\documentclass[a4paper]{article}
\usepackage[most]{tcolorbox}
\usepackage{amssymb}
\usepackage{amsmath}
\usepackage{algorithm,algpseudocode}
\usepackage{bm}

\usepackage{hyperref} %% inclusion after algorithm: mandatory
\usepackage{url}

\usepackage{boxedminipage}
\usepackage{graphicx}%%[draft] : do not embed figs/picts
\usepackage{float}
\usepackage{rotating}

\usepackage{xspace}
\usepackage{nicefrac}

\usepackage{comment}
\usepackage{soul}
\usepackage{verbatim}%%block comment

\usepackage{csquotes} % \textquote{}

\usepackage[T1]{fontenc}    % for accents
\usepackage[utf8]{inputenc} % for coding
\usepackage{xparse}   % \NewDocumentCommand
\usepackage{xifthen} % if-the-else see http://ctan.org/pkg/xifthen
\usepackage{pifont}
\usepackage{marvosym}

\usepackage{fullpage}
\usepackage{xstring}
\usepackage{placeins}

\newif\ifSLIDES
\SLIDESfalse

\newcommand{\stmath}[1]{\ifmmode\text{\sout{\ensuremath{#1}}}\else\sout{#1}\fi} % requires package ulem

\definecolor{darkgreen}{rgb}{0,0.7,0}
\definecolor{blue-violet}{rgb}{0.54, 0.17, 0.89}

\newcommand{\tored}{\color{red}\xspace}

\newcommand{\toblue}{\color{blue}\xspace}

\newcommand{\toblack}{\color{black}\xspace}
\newcommand{\togreen}{\color{green}\xspace}

\newcommand{\rred}[1]{{\textcolor{red}{#1}}}         %text added
\newcommand{\bblue}[1]{\textcolor{blue}{#1}}

\newcommand{\ppurple}[1]{\textcolor{purple}{#1}}
\newcommand{\ggreen}[1]{\textcolor{darkgreen}{#1}}

\NewDocumentCommand\sbold{g}{%
\IfNoValueTF{#1}
{$\bblue{\triangleright}$}
{$\bblue{\triangleright}$ {\bf #1}}
}

\NewDocumentCommand\sred{g}{%
\IfNoValueTF{#1}
{$\bblue{\triangleright}$}
{$\bblue{\triangleright}$ \textcolor{red}{#1}}
}
\NewDocumentCommand\sblue{g}{%
\IfNoValueTF{#1}
{$\bblue{\triangleright}$}
{$\bblue{\triangleright}$ \textcolor{blue}{#1}}
}
\NewDocumentCommand\sdgreen{g}{%
\IfNoValueTF{#1}
{$\ggreen{\triangleright}$}
{$\ggreen{\triangleright}$ \textcolor{darkgreen}{#1}}
}
\NewDocumentCommand\sgreen{g}{%
\IfNoValueTF{#1}
{$\ggreen{\triangleright}$}
{$\ggreen{\triangleright}$ \textcolor{darkgreen}{#1}}
}
\NewDocumentCommand\spurple{g}{%
\IfNoValueTF{#1}
{$\ppurple{\triangleright}$}
{$\ppurple{\triangleright}$ \textcolor{purple}{#1}}
}

\NewDocumentCommand\sbulem {g}{\noindent $\bullet$\IfNoValueTF{#1}{}{{\em #1}}}
\NewDocumentCommand\sbullet{g}{\noindent $\bullet$\IfNoValueTF{#1}{}{{#1}}}

\NewDocumentCommand\quoteen{o m}{\IfNoValueTF{#1}{}{(#1)}``{\em #2}''}
\NewDocumentCommand\citeen{o m}{\IfNoValueTF{#1}{}{(#1)}``{\em #2}''}

\newif\ifDRAFT
\DRAFTtrue     %%preparing the draft with comments
\NewDocumentCommand\bienfr{G{0} m m}{
\ifthenelse{\equal{#1}{0}}{\noindent{\bf En.} #2\newline}{}
\ifthenelse{\equal{#1}{2}}{\noindent{\bf En.} #2\newline}{}

\ifthenelse{\equal{#1}{1}}{\noindent{\bf Fr.} #3\newline}{}
\ifthenelse{\equal{#1}{2}}{\noindent{\bf Fr.} #3\newline}{}
}

\ifDRAFT %% Generic (i.e. non signed) comments in draft mode
\newcommand{\makeremark}[2]{
  \newcommand{#1}[1]
    {$\longrightarrow$\textcolor{red}{\sc #2: ##1}$\leftarrow$ \medskip}}

\newcommand{\makeremarkcolor}[3]{
  \newcommand{#1}[1]
    {$\longrightarrow$\textcolor{#3}{\sc #2: ##1}$\leftarrow$ \medskip}}

\newcommand{\added}[1]{{\textcolor{red}{#1}}}         %text added
\newcommand{\changed}[1]{{\textcolor{red}{\sc #1}}}   %text changed
\newcommand{\tcomment}[1]{{\textcolor{blue}{\sc #1}}} %comment in text

\newcommand{\makeremarkMargin}[2]{\marginpar{\tiny{\color{red}#1}:\color{blue}#2}}

\else %% Generic comments (empty macros i.e. hide everything) in final model
\newcommand{\makeremark}[2]{\newcommand{#1}[1]{}}

\newcommand{\makeremarkMargin}[2]{}

\newcommand{\added}[1]{{}}
\newcommand{\changed}[1]{{}}
\newcommand{\tcomment}[1]{{}}
\newcommand{\rem}[1]{{}}

\fi
\newcommand*\mnote[3][0pt]{%
  \if l#2\reversemarginpar\def\pointer{\triangleright}%
    \def\stackalignment{r}\fi%
  \if r#2\normalmarginpar\def\pointer{\triangleleft}%
    \def\stackalignment{l}\fi%
  \marginpar{%
    \topinset{%
      \scalebox{1.5}{\textcolor{blue}{$\pointer$}}}{%
      \belowbaseline[-1.5\baselineskip-#1]{%
        \stackengine%
          {-5pt}%
          {\fcolorbox{blue}{white}{\parbox{1.8cm}%
            {\vspace{3pt}\raggedright#3}}}%
          {~\colorbox{white}{\sffamily Note}}%
          {O}%
          {l}%
          {F}%
          {F}%
          {S}%
        }%
      }{%
      3ex+#1}{%
      -2ex}%
}
}

\makeremark{\vgsays}{Victor says}
\makeremark{\vmsays}{Vincent says}

\makeremark{\anoops}{Anoop says}
\makeremark{\arinjays}{Arinjay says}
\makeremark{\sikaos}{Sikao says}
\makeremark{\michals}{Michal says}
\makeremark{\nelsons}{Nelson says}

\makeremark{\fc}{Frederic says}
\makeremark{\FC}{Frederic says}
\makeremarkcolor{\fcb}{Frederic says}{blue}

\makeremark{\ac}{Antoine says}
\makeremark{\ah}{Antoine says}
\makeremark{\sq}{Simon says}
\makeremark{\gc}{Guillaume says}
\makeremark{\as}{Augusto says}
\makeremark{\gs}{Guilherme says}
\makeremark{\nmd}{Noel says}
\makeremark{\td}{Tom says}
\makeremark{\ar}{Andrea says}
\makeremark{\chr}{Charles says}
\makeremark{\da}{Deepesh says}
\makeremark{\al}{Alix says}
\makeremark{\sm}{Simon says}
\makeremark{\dc}{David says}
\makeremark{\db}{Denys says}
\makeremark{\ms}{Meline says}
\makeremark{\tod}{Timothee  says}
\makeremark{\lgold}{Louis says}

\makeremark{\dm}{Dorian says}
\makeremark{\pb}{PierreB says}

\makeremark{\ds}{Darsh says}
\makeremark{\rt}{Romain  says}

\makeremark{\edo}{Edoardo says}
\makeremark{\vla}{Vladimir says}
\makeremark{\mma}{Maximillien says}
\makeremark{\theos}{Theo says}

\ifx\question\undefined

\NewDocumentCommand\question{m+g}{%
\IfNoValueTF{#2}
{\bigskip\noindent \ding{43} \quoteen{\em #1}}
{\bigskip\noindent \ding{43} \quoteen{\textcolor{#1}{\em #2}}}
}

\definecolor{block-gray}{gray}{0.85}
\definecolor{amaranth}{rgb}{0.9, 0.17, 0.31}
\definecolor{candypink}{rgb}{0.89, 0.44, 0.48}

\newtcolorbox{qcolgray}{colback=block-gray,boxrule=0pt,boxsep=0pt,breakable}
\newtcolorbox{qcolcyan}{colback=cyan,boxrule=0pt,boxsep=0pt,breakable}
\newtcolorbox{qcolred}{colback=candypink,boxrule=0pt,boxsep=0pt,breakable}

\newcommand{\questionsymbol}{Q}
\newcounter{questiongcounter}
\fi

\newcommand{\beginsupplement}{
\setcounter{table}{0}
\renewcommand{\thetable}{S\arabic{table}}%
\setcounter{figure}{0}
\renewcommand{\thefigure}{S\arabic{figure}}%

\setcounter{section}{0}
\renewcommand{\thesection}{S\arabic{section}}               %Roman numeral title

}%% end supplement

\newcommand{\beginSI}{\beginsupplement}

\newcommand{\footurl}[1]{\footnote{\url{#1}}}

\ifSLIDES
\else
\newtheorem{conjecture}{Conjecture.}

\newtheorem{theorem}{Theorem.}

\newtheorem{example}{\noindent Example}{}
{}
{}

\newtheorem{remark}{Remark}

\fi
\newcounter{explecounter}

\NewDocumentEnvironment{proof-not-ams}{G{gray}}{%
  \noindent \color{#1}\emph{Proof.}%
}{%
  \hfill$\Box$\par\medskip
}

\newenvironment{proof-app}[1]
 {\noindent \emph{Proof. [#1]}}{$\Box$\par\medskip}

{}
{}
{}

\newcounter{pln}

\newenvironment{itemizep}
{ \begin{itemize}
    \setlength{\itemsep}{0pt}
    \setlength{\parskip}{0pt}
    \setlength{\parsep}{0pt}     }
{ \end{itemize}   }

\NewDocumentCommand\paragraphm{O{4pt} m}{%
{\vspace{#1}
\noindent{\bf #2}
}}
\NewDocumentCommand\paramini{O{4pt} m}{%
{\vspace{#1}
\noindent{\bf #2}
}}
\NewDocumentCommand\paragraphmini{O{4pt} m}{%
{\vspace{#1}
\noindent{\bf #2}
}}

\newcommand{\calA}{{\mathcal A}}

\newcommand{\calC}{{\mathcal C}}

\newcommand{\calF}{{\mathcal F}}

\newcommand{\calK}{{\mathcal K}}
\newcommand{\calL}{{\mathcal L}}

\newcommand{\calN}{{\mathcal N}}

\newcommand{\calS}{{\mathcal S}}

\newcommand{\calW}{{\mathcal W}}

\NewDocumentCommand\mmoA{g}{\IfNoValueTF{#1}{\text{\sffamily \bfseries A}}{\boldsymbol{a}_{#1}}}
\NewDocumentCommand\mmoB{g}{\IfNoValueTF{#1}{\text{\sffamily \bfseries B}}{\boldsymbol{b}_{#1}}}
\NewDocumentCommand\mmoC{g}{\IfNoValueTF{#1}{\text{\sffamily \bfseries C}}{\boldsymbol{c}_{#1}}}
\NewDocumentCommand\mmoD{g}{\IfNoValueTF{#1}{\text{\sffamily \bfseries D}}{\boldsymbol{d}_{#1}}}
\NewDocumentCommand\mmoE{g}{\IfNoValueTF{#1}{\text{\sffamily \bfseries E}}{\boldsymbol{e}_{#1}}}
\NewDocumentCommand\mmoF{g}{\IfNoValueTF{#1}{\text{\sffamily \bfseries F}}{\boldsymbol{f}_{#1}}}
\NewDocumentCommand\mmoG{g}{\IfNoValueTF{#1}{\text{\sffamily \bfseries G}}{\boldsymbol{g}_{#1}}}
\NewDocumentCommand\mmoH{g}{\IfNoValueTF{#1}{\text{\sffamily \bfseries H}}{\boldsymbol{h}_{#1}}}
\NewDocumentCommand\mmoI{g}{\IfNoValueTF{#1}{\text{\sffamily \bfseries I}}{\boldsymbol{i}_{#1}}}
\NewDocumentCommand\mmoJ{g}{\IfNoValueTF{#1}{\text{\sffamily \bfseries J}}{\boldsymbol{j}_{#1}}}
\NewDocumentCommand\mmoK{g}{\IfNoValueTF{#1}{\text{\sffamily \bfseries K}}{\boldsymbol{k}_{#1}}}
\NewDocumentCommand\mmoL{g}{\IfNoValueTF{#1}{\text{\sffamily \bfseries L}}{\boldsymbol{l}_{#1}}}
\NewDocumentCommand\mmoM{g}{\IfNoValueTF{#1}{\text{\sffamily \bfseries M}}{\boldsymbol{m}_{#1}}}
\NewDocumentCommand\mmoN{g}{\IfNoValueTF{#1}{\text{\sffamily \bfseries N}}{\boldsymbol{n}_{#1}}}
\NewDocumentCommand\mmoO{g}{\IfNoValueTF{#1}{\text{\sffamily \bfseries O}}{\boldsymbol{o}_{#1}}}
\NewDocumentCommand\mmoP{g}{\IfNoValueTF{#1}{\text{\sffamily \bfseries P}}{\boldsymbol{p}_{#1}}}
\NewDocumentCommand\mmoQ{g}{\IfNoValueTF{#1}{\text{\sffamily \bfseries Q}}{\boldsymbol{q}_{#1}}}
\NewDocumentCommand\mmoR{g}{\IfNoValueTF{#1}{\text{\sffamily \bfseries R}}{\boldsymbol{r}_{#1}}}
\NewDocumentCommand\mmoS{g}{\IfNoValueTF{#1}{\text{\sffamily \bfseries S}}{\boldsymbol{s}_{#1}}}
\NewDocumentCommand\mmoT{g}{\IfNoValueTF{#1}{\text{\sffamily \bfseries T}}{\boldsymbol{t}_{#1}}}
\NewDocumentCommand\mmoU{g}{\IfNoValueTF{#1}{\text{\sffamily \bfseries U}}{\boldsymbol{u}_{#1}}}
\NewDocumentCommand\mmoV{g}{\IfNoValueTF{#1}{\text{\sffamily \bfseries V}}{\boldsymbol{v}_{#1}}}
\NewDocumentCommand\mmoW{g}{\IfNoValueTF{#1}{\text{\sffamily \bfseries W}}{\boldsymbol{w}_{#1}}}
\NewDocumentCommand\mmoX{g}{\IfNoValueTF{#1}{\text{\sffamily \bfseries X}}{\boldsymbol{x}_{#1}}}
\NewDocumentCommand\mmoY{g}{\IfNoValueTF{#1}{\text{\sffamily \bfseries Y}}{\boldsymbol{y}_{#1}}}
\NewDocumentCommand\mmoZ{g}{\IfNoValueTF{#1}{\text{\sffamily \bfseries Z}}{\boldsymbol{z}_{#1}}}

\NewDocumentCommand\momA{g}{\IfNoValueTF{#1}{\text{\sffamily \bfseries A}}{\boldsymbol{a}_{#1}}}
\NewDocumentCommand\momB{g}{\IfNoValueTF{#1}{\text{\sffamily \bfseries B}}{\boldsymbol{b}_{#1}}}
\NewDocumentCommand\momC{g}{\IfNoValueTF{#1}{\text{\sffamily \bfseries C}}{\boldsymbol{c}_{#1}}}
\NewDocumentCommand\momD{g}{\IfNoValueTF{#1}{\text{\sffamily \bfseries D}}{\boldsymbol{d}_{#1}}}
\NewDocumentCommand\momE{g}{\IfNoValueTF{#1}{\text{\sffamily \bfseries E}}{\boldsymbol{e}_{#1}}}
\NewDocumentCommand\momF{g}{\IfNoValueTF{#1}{\text{\sffamily \bfseries F}}{\boldsymbol{f}_{#1}}}
\NewDocumentCommand\momG{g}{\IfNoValueTF{#1}{\text{\sffamily \bfseries G}}{\boldsymbol{g}_{#1}}}
\NewDocumentCommand\momH{g}{\IfNoValueTF{#1}{\text{\sffamily \bfseries H}}{\boldsymbol{h}_{#1}}}
\NewDocumentCommand\momI{g}{\IfNoValueTF{#1}{\text{\sffamily \bfseries I}}{\boldsymbol{i}_{#1}}}
\NewDocumentCommand\momJ{g}{\IfNoValueTF{#1}{\text{\sffamily \bfseries J}}{\boldsymbol{j}_{#1}}}
\NewDocumentCommand\momK{g}{\IfNoValueTF{#1}{\text{\sffamily \bfseries K}}{\boldsymbol{k}_{#1}}}
\NewDocumentCommand\momL{g}{\IfNoValueTF{#1}{\text{\sffamily \bfseries L}}{\boldsymbol{l}_{#1}}}
\NewDocumentCommand\momM{g}{\IfNoValueTF{#1}{\text{\sffamily \bfseries M}}{\boldsymbol{m}_{#1}}}
\NewDocumentCommand\momN{g}{\IfNoValueTF{#1}{\text{\sffamily \bfseries N}}{\boldsymbol{n}_{#1}}}
\NewDocumentCommand\momO{g}{\IfNoValueTF{#1}{\text{\sffamily \bfseries O}}{\boldsymbol{o}_{#1}}}
\NewDocumentCommand\momP{g}{\IfNoValueTF{#1}{\text{\sffamily \bfseries P}}{\boldsymbol{p}_{#1}}}
\NewDocumentCommand\momQ{g}{\IfNoValueTF{#1}{\text{\sffamily \bfseries Q}}{\boldsymbol{q}_{#1}}}
\NewDocumentCommand\momR{g}{\IfNoValueTF{#1}{\text{\sffamily \bfseries R}}{\boldsymbol{r}_{#1}}}
\NewDocumentCommand\momS{g}{\IfNoValueTF{#1}{\text{\sffamily \bfseries S}}{\boldsymbol{s}_{#1}}}
\NewDocumentCommand\momT{g}{\IfNoValueTF{#1}{\text{\sffamily \bfseries T}}{\boldsymbol{t}_{#1}}}
\NewDocumentCommand\momU{g}{\IfNoValueTF{#1}{\text{\sffamily \bfseries U}}{\boldsymbol{u}_{#1}}}
\NewDocumentCommand\momV{g}{\IfNoValueTF{#1}{\text{\sffamily \bfseries V}}{\boldsymbol{v}_{#1}}}
\NewDocumentCommand\momW{g}{\IfNoValueTF{#1}{\text{\sffamily \bfseries W}}{\boldsymbol{w}_{#1}}}
\NewDocumentCommand\momX{g}{\IfNoValueTF{#1}{\text{\sffamily \bfseries X}}{\boldsymbol{x}_{#1}}}
\NewDocumentCommand\momY{g}{\IfNoValueTF{#1}{\text{\sffamily \bfseries Y}}{\boldsymbol{y}_{#1}}}
\NewDocumentCommand\momZ{g}{\IfNoValueTF{#1}{\text{\sffamily \bfseries Z}}{\boldsymbol{z}_{#1}}}

\DeclareMathOperator*{\ArgminOp}{argmin}

\NewDocumentCommand{\argmin}{o}{%
  \IfNoValueTF{#1}%
    {\ArgminOp}%
    {\ArgminOp_{#1}}%
}

\NewDocumentCommand\dtms{O{k} g}{d^2_{#1}\IfNoValueTF{#2}{}{(#2)}}
\NewDocumentCommand\dtm{O{k} g}{d_{#1}\IfNoValueTF{#2}{}{(#2)}}
\NewDocumentCommand\dtmmed{g}{d_{DTMm}\IfNoValueTF{#1}{}{(#1)}}
\NewDocumentCommand\dtmemd{g}{d_{EMD}\IfNoValueTF{#1}{}{(#1)}}

\NewDocumentCommand\dwassk{O{k} G{}}{d_{\calW_{#1}}\IfNoValueTF{#2}{}{(#2)}}
\NewDocumentCommand\dpdzero{g}{d_{PD0}\IfNoValueTF{#1}{}{(#1)}}
\NewDocumentCommand\dpdone{g}{d_{PD1}\IfNoValueTF{#1}{}{(#1)}}
\NewDocumentCommand\dpdtwo{g}{d_{PD2}\IfNoValueTF{#1}{}{(#1)}}

\newcommand{\eg}{{\em e.g.}\xspace}

\NewDocumentCommand\atilde{g}{\IfNoValueTF{#1}{\tilde{a}}{\tilde{a_{#1}}}}
\NewDocumentCommand\ptilde{g}{\IfNoValueTF{#1}{\tilde{p}}{\tilde{p_{#1}}}}
\NewDocumentCommand\qtilde{g}{\IfNoValueTF{#1}{\tilde{q}}{\tilde{q_{#1}}}}

\NewDocumentCommand\degree{g}{%
\IfNoValueTF{#1}
{\ensuremath{^{\circ}}}
{\ensuremath{{#1}^{\circ}}}
}

\NewDocumentCommand\Rn{G{n}}{\mathbb{R}^{#1}}
\NewDocumentCommand\Rd{G{d}}{\mathbb{R}^{#1}}
\NewDocumentCommand\Zn{G{n}}{\mathbb{Z}^{#1}}
\NewDocumentCommand\Zd{G{d}}{\mathbb{Z}^{#1}}

\NewDocumentCommand\Euclidn{G{n}}{\mathbb{E}^{#1}}
\NewDocumentCommand\Euclidd{G{d}}{\mathbb{E}^{#1}}

\NewDocumentCommand\GLn{G{n}}{GL(#1)} % group:  GL(n) or GL(n, R): general linear group

\NewDocumentCommand\Ogro{G{3}}{O(#1)} % group: O(n)
\NewDocumentCommand\On{G{3}}{O(#1)} % group:  O(n)
\NewDocumentCommand\SOn{G{3}}  {SO(#1)} % group: SO(n)
\NewDocumentCommand\SOnlie{G{3}}{\mathfrak{so}(#1)} % associated Lie algebra

\NewDocumentCommand\En{G{3}}{E(#1)} % group: SE(n)
\NewDocumentCommand\SEn{O{} G{3}}{SE(#2) \IfNoValueTF{#1}{}{^{#1}} }

\NewDocumentCommand\SEnp{G{3}}{E^{+}(#1)} % group: SE(n) denote E^+(n)
\NewDocumentCommand\SEnlie{G{3}}{\mathfrak{se}(#1)} % associated Lie algebra

\NewDocumentCommand\TpM{O{p} m}{T_{#1}{#2}} % Tangent space

\NewDocumentCommand\UFmakeset{g}{\text{UF.make\_set} \IfNoValueTF{#1}{}{(#1)}}
\NewDocumentCommand\UFunion{g}{\text{UF.union} \IfNoValueTF{#1}{}{(#1)}}
\NewDocumentCommand\UFfind{g}{\text{UF.find} \IfNoValueTF{#1}{}{(#1)}}
\NewDocumentCommand\UFnumcc{g}{\text{UF.num\_cc} \IfNoValueTF{#1}{}{(#1)}}
\NewDocumentCommand\UFnumnodes{g}{\text{UF.num\_nodes} \IfNoValueTF{#1}{}{(#1)}}

\newcommand{\size}[1]{\lvert #1 \rvert}

\NewDocumentCommand\expL{O{p} m}{
\exp
\IfEqCase{#1}{
{p}{(}      % p for parenthesis
{P}{\big(}
{b}{[}      % b/B for brackets
{B}{\big[}
{c}{\{}     % c/C for braces/curly brackets
{C}{\big\{}
}[\PackageError{expL}{Undefined option to expL: #1}{}]
#2
\IfEqCase{#1}{
{p}{)}
{P}{\big)}
{b}{]}
{B}{\big]}
{c}{\{}
{C}{\big\{}
}[\PackageError{expL}{Undefined option to expL: #1}{}]
}
\NewDocumentCommand\expl{O{p} m}{\expL[#1]{#2}}

\NewDocumentCommand\erf{O{p} m}{
\text{erf}
\IfEqCase{#1}{
{p}{(}      % p for parenthesis
{P}{\big(}
{b}{[}      % b/B for brackets
{B}{\big[}
{c}{\{}     % c/C for braces/curly brackets
{C}{\big\{}
}[\PackageError{expL}{Undefined option to expL: #1}{}]
#2
\IfEqCase{#1}{
{p}{)}
{P}{\big)}
{b}{]}
{B}{\big]}
{c}{\{}
{C}{\big\{}
}[\PackageError{expL}{Undefined option to expL: #1}{}]
}

\newcommand{\stirlingnumfk}[2]{} % Stirling number of the first kind

\NewDocumentCommand\nabladec{O{} G{}}{\nabla_{#1} #2}

\NewDocumentCommand{\divergence}{O{} g}{\nabla_{#1}\!\bullet\IfNoValueTF{#2}{}{\left(#2\right)}}
\NewDocumentCommand\laplacian{O{} G{}}{\bigtriangleup_{#1} #2}
\NewDocumentCommand\gradient{O{} G{}}{\nabla_{#1}#2}

\NewDocumentCommand{\partiald}{s m m} {
\IfBooleanTF{#1}
{\frac{\partial }{\partial #3} #2} % partialdl
{\frac{\partial #2}{\partial #3}}
}

\newcommand{\partialD}[2] {\frac{d #1}{d #2}} % operand in the numerator
\NewDocumentCommand\partialDe{g g G{2}}{(\partialD{#1}{#2})^{#3}} % power

\NewDocumentCommand\partialDk{g g G{2}}   {\frac{d^{#3} {#1}}{d {#2}^{#3}} }      % kth order deriv

\NewDocumentCommand\partialdk  {g g G{2}}   {\frac{\partial^{#3} {#1}}{\partial {#2}^{#3}} }      % kth order deriv
\NewDocumentCommand\partialdkl{g g G{2}}    {\frac{\partial^{#3} }{\partial {#2}^{#3}} #1}
\NewDocumentCommand\partialdkat{g g G{2} g} {\frac{\partial^{#3} {#1}}{\partial {#2}^{#3}}_{|#4}}
\NewDocumentCommand\partialdklat{g g G{2} g}{ \frac{\partial^{#3} }{\partial {#2}^{#3}} #1 _{|#4}}

\NewDocumentCommand\partialdtwo{g g g}{ \frac{\partial^{2} {#1}}{\partial {#2} \partial {#3}}}
\NewDocumentCommand\partialdtwoat{g g g g}{ \frac{\partial^{2} {\tored #1}}{\partial {#2} \partial {#3}} _{| #4}}

\NewDocumentCommand\distproj{O{} g g}{d_{#2}^{#1}\IfNoValueTF{#3}{}{(#3)}}
\NewDocumentCommand\distHAOS{g g}{\overrightarrow{d}_{#1}\IfNoValueTF{#2}{}{(#2)}} % Hausdorff one-sided
\NewDocumentCommand\distHA{g g}{d_{HA}\IfNoValueTF{#1}{}{(#1, #2)}}                 % Hausdorff one-sided

\NewDocumentCommand\distMHAOS{g g}{\overrightarrow{\overline{d}}_{#1}\IfNoValueTF{#2}{}{(#2)}} % Hausdorff one-sided
\NewDocumentCommand\distMHA{g g}{\overline{d}_{HA}\IfNoValueTF{#1}{}{(#1, #2)}}                 % Hausdorff one-sided

\NewDocumentCommand\command{m+g}{
\IfNoValueTF{#2}
{command-1}
{command-2 }%
}

\NewDocumentCommand\vvnorm{O{} m g}{\left\lVert {#2} \right\rVert_{#1}\IfNoValueTF{#3}{}{^{#3}}}

\NewDocumentCommand\vvnormi{G{\cdot}} {{\vvnorm[\infty]{#1}}}
\NewDocumentCommand\vvnormtv{G{\cdot}}{{\| #1 \|_{TV}}}

\NewDocumentCommand\dist{g}{%
\IfNoValueTF{#1}
{\text{dist}}
{\text{dist}(#1)}
}

\NewDocumentCommand\demd{g+g}{%
\IfNoValueTF{#1}{d_{\text{EMD}}}
{d_{\text{EMD}}(#1,#2)}}

\NewDocumentCommand\wasser{O{} m G{}}{
W_{#1} 
\IfNoValueTF{#3}{}{^{#3}} % exponent 
\IfNoValueTF{#2}{}{(#2)}
}

\NewDocumentCommand\PSD{G{d}}{S^+_{#1}}
\NewDocumentCommand\PSDcor{G{d}}{\calC^+_{#1}}

\NewDocumentCommand\onevector{G{}}{ {{\bf 1_{#1}}} }
\NewDocumentCommand\lavectorone{G{}}{ {{\bf 1_{#1}}} }

\NewDocumentCommand\identitymatrix{G{}}{ {{\bf I}_{#1}} }
\NewDocumentCommand\idmatrix{G{}}{ {{\bf I}_{#1}} }
\NewDocumentCommand\Imatrix{G{}}{ {\rred{\bf I}_{#1}} } %% fix me -- pdt concours crcn 2025

\newcommand{\ladiag}[1]{\text{Diag}(#1)}

\NewDocumentCommand\transpose{m G{}}{{#1}^{\mathsf{#2 T}}}
\NewDocumentCommand\latrans{m G{}}{{#1}^{\mathsf{#2 T}}}

\newcommand{\interior}[1]{ {\kern0pt#1}^{\mathrm{o}}}

\NewDocumentCommand\Tn{G{n}}{ \mathbb{T}^{#1} }
\NewDocumentCommand\Td{G{d}}{ \mathbb{T}^{#1} }
\NewDocumentCommand\torusd{G{d}}{ \mathbb{T}^{#1} }
\NewDocumentCommand\torusdc{O{c} G{d}}{\mathbb{T}_{#1}^{#2} } % T_c^d
\NewDocumentCommand\torustri{O{{\bm{2\pi}}} G{d}}{\torusdc[#1]{#2}}

\NewDocumentCommand\diam{g}{%
\IfNoValueTF{#1}
{\text{diam}}
{\text{diam}(#1)}
}

\NewDocumentCommand\dotpn{o m m}{%
\langle #2, #3 \rangle>
\IfNoValueTF{#1}
{}
{_{#1}}
}

\NewDocumentCommand\Volume{O{} g}{{\tt Volume}_{#1}\IfNoValueTF{#2}{}{(#2)}}
\NewDocumentCommand\Area  {O{} g}{{\tt Area}_{#1} \IfNoValueTF{#2}{}{(#2)}}

\NewDocumentCommand\Sd{G{d}}{S^{#1}}
\NewDocumentCommand\sphered{G{d}}{S^{#1}}

\NewDocumentCommand\Bd{G{d}}{B^{#1}}
\NewDocumentCommand\Bdcr{O{d} g} { B^{#1} \IfNoValueTF{#2}{}{(#2)} }

\NewDocumentCommand\unitSA{g}{A \IfNoValueTF{#1}{}{(#1)}}
\NewDocumentCommand\unitSV{g}{V \IfNoValueTF{#1}{}{(#1)}}

\NewDocumentCommand\spheredr{m+g}{\IfNoValueTF{#2}{S^{#1}}{S^{#1}(#2)}}
\NewDocumentCommand\balldr{m+g}{\IfNoValueTF{#2}{B^{#1}}{B^{#1}(#2)}}
\NewDocumentCommand\chord{O{} m}{\text{\tt chord}_{#1}(#2)} % may be subscripted by the body

\NewDocumentCommand\spheredrarea{m+g}{\IfNoValueTF{#2}{\text{Area}_{#1}}{\text{Area}_{#1}(#2)}}
\NewDocumentCommand\balldrvol{m+g}{\IfNoValueTF{#2}{\text{Vol}_{#1}}{\text{Vol}_{#1}(#2)}}

\NewDocumentCommand\coneDARvol{m+m+g}{
\IfNoValueTF{#3}
{V^{\text{Cone}}_{#1, #2}}
{V^{\text{Cone}}_{#1, #2}(#3)}
}

\NewDocumentCommand\ballcapDARvol{m+m+g}{
\IfNoValueTF{#3}
{V^{\text{Cap}}_{#1, #2}}
{V^{\text{Cap}}_{#1, #2}(#3)}
}
\NewDocumentCommand\sphericalcapDARarea{m+m+g}{
\IfNoValueTF{#3}
{A^{\text{Cap}}_{#1, #2}}
{A^{\text{Cap}}_{#1, #2}(#3)}
}
\NewDocumentCommand\RatioCut{g}{\IfNoValueTF{#1}{\text{RatioCut}}{\text{RatioCut}[#1]}}
\NewDocumentCommand\Ncut{g}{\IfNoValueTF{#1}{\text{NCut}}{\text{NCut}[#1]}}

\NewDocumentCommand\giv{O{} g}{%
\text{name}
\IfNoValueTF{#1}{V_{G}(#2)}V_{#1}(#2){}
}

\NewDocumentCommand\NNG{O{} g}{
\text{NNG}_{#1}
\IfNoValueTF{#2}{}{(#2)}
}

\NewDocumentCommand\MST{O{} g}{
\text{MST}_{#1}
\IfNoValueTF{#2}{}{(#2)}
}

\NewDocumentCommand\normalD{m g}{\calN(#1 \IfNoValueTF{#2}{)}{\mid #2)}}

\NewDocumentCommand\dopearl{G{X=x}}{\text{do}(#1)} %% Pearl's do operator
 
\NewDocumentCommand\probX{O{} g}{
\mathbb{P}_{#1}\IfNoValueTF{#2}{}{\left[{#2} \right]}}

\NewDocumentCommand\probXP{m+g}{%
\IfNoValueTF{#2}
{\ensuremath{ {\mathbb{P}_{#1}}}}
{\ensuremath{ {\mathbb{P}_{#1}}\left[ {#2} \right]}}
}

\NewDocumentCommand\expX{O{} g}{%
\IfNoValueTF{#2} 
{ {\mathbb{E}_{#1}} }
{ {\mathbb{E}_{#1}}\left[ {#2} \right] }
}

\NewDocumentCommand\expXB{O{} m}{
{\langle #2 \rangle_{#1} }}

\NewDocumentCommand\expXu{O{} g}{%
\mathbb{E}
\IfNoValueTF{#1}
{}
{_{#1}}
\IfNoValueTF{#2}
{}
{[#2]}
}
\NewDocumentCommand\varX{g}{\mathrm{Var} \IfNoValueTF{#1}{}{\left[ {#1} \right]}}

\NewDocumentCommand\sigX{g}{%
\IfNoValueTF{#1}
{\ensuremath{ {\mathbb{\sigma}}}}
{\ensuremath{ {\mathbb{\sigma}}\left[ {#1} \right]}}
}
\NewDocumentCommand\mcflow{g}{\varphi \IfNoValueTF{#1}{}{(#1)}} % MC: ergodic flow
\NewDocumentCommand\mcconduct{g}{\phi \IfNoValueTF{#1}{}{(#1)}} % MC: conductance

\NewDocumentCommand\fisheremp{g}{F \IfNoValueTF{#1}{}{(#1)}}

\NewDocumentCommand\entH{g}{H\IfNoValueTF{#1}{}{(#1)}} % entropy H

\NewDocumentCommand\fisherinfo{O{} G{\theta}}{I_{#1}(#2)}

\NewDocumentCommand\fishermat{G{\theta}}{I_{#1}}

\NewDocumentCommand\priorjef{g} {J\IfNoValueTF{}{}{(#1)}}
\NewDocumentCommand\jefprior{g} {J\IfNoValueTF{}{}{(#1)}}

\NewDocumentCommand\divKL{g g}{D_{\mathrm{KL}} \IfNoValueTF{#1}{}{\left(#1 \Vert #2\right)}}
\NewDocumentCommand\KLdiv{g g}{D_{\mathrm{KL}} \IfNoValueTF{#1}{}{\left(#1 \Vert #2\right)}}

\NewDocumentCommand\divJS{g g}{D_{\mathrm{JS}} \IfNoValueTF{#1}{}{\left(#1 \Vert #2\right)}}
\NewDocumentCommand\JSdiv{g g}{D_{\mathrm{JS}} \IfNoValueTF{#1}{}{\left(#1 \Vert #2\right)}}

\NewDocumentCommand\emcres{O{t} G{ij}}{r_{#2}^{(#1)}}  % EM component rEsponsibility
\NewDocumentCommand\emcsum{O{t} G{j}}{n_{#2}^{(#1)}}  % EM component: sum for compo j at iter t
\NewDocumentCommand\emcmu{O{t} G{j}}{\mu_{#2}^{(#1)}} % EM component: mu_j at iter t
\NewDocumentCommand\emcw{O{t} G{j}}{w_{#2}^{(#1)}}    % EM component: weight w_j at iter t
\NewDocumentCommand\emisum{O{t}}{N^{(#1)}}           % EM  sum of emcsum at iter t

\NewDocumentCommand\lacovar{G{}}{\Sigma_{#1}}
\NewDocumentCommand\lacorr{g G{}}{\Omega_{#1}}

\NewDocumentCommand\pearson{g}{\IfNoValueTF{#1}{\text{Pear.}}{\text{Pear.}(#1)}}

\NewDocumentCommand\spear{g}{\IfNoValueTF{#1}{\text{Spear.}}{\text{Spear.}(#1)}}
\NewDocumentCommand\mic{g}{\IfNoValueTF{#1}{\text{MIC}}{\text{MIC}(#1)}}
\NewDocumentCommand\micd{O{d} g}{
\text{MIC}_{#1}
\IfNoValueTF{#2}{}{ (#2)}}

\NewDocumentCommand\pvalue{o}{\text{p}\!-\!\text{val}\IfNoValueTF{#1}{}{_{#1}}}

\NewDocumentCommand\fTheta{g}{\IfNoValueTF{#1}{f_{\Theta}}{f_{\Theta}(#1)}}

\NewDocumentCommand\Xsample{g}{\IfNoValueTF{#1}{\ensuremath{x^{(n)}}\xspace}{\ensuremath{x^{(#1)}}\xspace}}
\NewDocumentCommand\Ysample{g}{\IfNoValueTF{#1}{\ensuremath{y^{(n)}}\xspace}{\ensuremath{y^{(#1)}}\xspace}}

\NewDocumentCommand\mean{O{} g}{{\text{mean}}_{#1}\IfNoValueTF{#2}{}{(#2)}}
\NewDocumentCommand\average{O{} g}{{\text{mean}}_{#1}\IfNoValueTF{#2}{}{(#2)}}

\NewDocumentCommand\median{O{} g}{{\text{median}_{#1}}\IfNoValueTF{#2}{}{(#2)}}
\NewDocumentCommand\minimum{O{} g}{{\text{minimum}_{#1}}\IfNoValueTF{#2}{}{(#2)}}

\NewDocumentCommand\likeli{G{X} G{\Theta}}{L(#1\mid #2)}
\NewDocumentCommand\loglik{G{X} G{\Theta}}{\calL(#1\mid #2)}

\NewDocumentCommand\mmd{O{} g}{%
\text{MMD}
\IfNoValueTF{#1}
{}
{_{#1}}
\IfNoValueTF{#2}
{}
{[#2]}
}

\NewDocumentCommand\mmds{O{} g}{%
\text{MMD}^2
\IfNoValueTF{#1}
{}
{_{#1}}
\IfNoValueTF{#2}
{}
{[#2]}
}

\NewDocumentCommand\normat{O{r} g}{\text{\codecx{Norm.}}_{#1} \IfNoValueTF{#2}{} {(#2)} } % normalize matrix: row or column
\NewDocumentCommand\lanorm{O{r} g}{\text{\codecx{Norm.}}_{#1} \IfNoValueTF{#2}{} {(#2)} } % same
\NewDocumentCommand\softmax{O{r} g}{\text{\codecx{softmax}}_{#1} \IfNoValueTF{#2}{} {(#2)} }
 
\NewDocumentCommand\sinkhorn{g}{\codecx{Sinkhorn}\IfNoValueTF{#1}{}{(#1)}} % Sinkhorn operator applied to a matrix

\NewDocumentCommand\concat{g}{{\codecx{Concat}}{\codecx{Concat}(#1)}}

\NewDocumentCommand\attn{g}{\codecx{Attn} \IfNoValueTF{#1}{}{(#1)}}
\NewDocumentCommand\selfattn{g}{\codecx{SelfAttn} \IfNoValueTF{#1}{}{(#1)}}

\NewDocumentCommand\Katt{O{1} G{}} {K^{(#1)}_{#2}} % Katt coeff ij
\NewDocumentCommand\Kasink{G{}}{\Katt{#1}}

\NewDocumentCommand\attQW{s O{} g}{\IfBooleanTF{#1}{\prescript{h}{}{W_{Q_{#2}}}}{W_{Q_{#2}}} \IfNoValueTF{#3}{}{(#3)}}
\NewDocumentCommand\attWQ{s O{} g}{\IfBooleanTF{#1}{\prescript{h}{}{W_{Q_{#2}}}}{W_{Q_{#2}}} \IfNoValueTF{#3}{}{(#3)}}

\NewDocumentCommand\attKW{s O{} g}{\IfBooleanTF{#1}{\prescript{h}{}{W_{K_{#2}}}}{W_{K_{#2}}} \IfNoValueTF{#3}{}{(#3)}}
\NewDocumentCommand\attWK{s O{} g}{\IfBooleanTF{#1}{\prescript{h}{}{W_{K_{#2}}}}{W_{K_{#2}}} \IfNoValueTF{#3}{}{(#3)}}

\NewDocumentCommand\attVW{s O{} g}{\IfBooleanTF{#1}{\prescript{h}{}{W_{V_{#2}}}}{W_{V_{#2}}} \IfNoValueTF{#3}{}{(#3)}}
\NewDocumentCommand\attWV{s O{} g}{\IfBooleanTF{#1}{\prescript{h}{}{W_{V_{#2}}}}{W_{V_{#2}}} \IfNoValueTF{#3}{}{(#3)}}

\NewDocumentCommand\attQ{s O{} g}{\IfBooleanTF{#1}{\prescript{h}{}{Q_{#2}}}{Q_{#2}} \IfNoValueTF{#3}{}{(#3)}}
\NewDocumentCommand\attK{s O{} g}{\IfBooleanTF{#1}{\prescript{h}{}{K_{#2}}}{K_{#2}} \IfNoValueTF{#3}{}{(#3)}}
\NewDocumentCommand\attV{s O{} g}{\IfBooleanTF{#1}{\prescript{h}{}{V_{#2}}}{V_{#2}} \IfNoValueTF{#3}{}{(#3)}}

\NewDocumentCommand\roformer{g}{\codecx{RoFormer}\IfNoValueTF{#1}{}{(#1)}}
\NewDocumentCommand\alibi{g}{\codecx{ALiBi}\IfNoValueTF{#1}{}{(#1)}}

\NewDocumentCommand\sijref{G{ij}}{\overset{\circ}{s}_{#1}}
\NewDocumentCommand\dijref{G{ij}}{\overset{\circ}{d}_{#1}}

\NewDocumentCommand\Vgnm{g}{V_{GNM}\IfNoValueTF{#1}{}{(#1)}}
\NewDocumentCommand\Vanm{g}{V_{ANM}\IfNoValueTF{#1}{}{(#1)}}

\NewDocumentCommand\wienerW{s O{t} g}{
\IfBooleanTF{#1}{\bar{W}_{#2}}{W_{#2}}
\IfNoValueTF{#3}{}{(#3)}}

\NewDocumentCommand\wienerB{s O{t} g}{
\IfBooleanTF{#1}{\bar{B}_{#2}}{B_{#2}}
\IfNoValueTF{#3}{}{(#3)}}

\NewDocumentCommand\gdct{O{} g}{\text{GDCT}_{#1}\IfNoValueTF{#2}{}{(#2)}}

\NewDocumentCommand\chemeq{ m m O{} O{}} {
{#1 \underset{#4}{\stackrel{#3}{\rightleftharpoons}} #2}
}

\NewDocumentCommand\brackets{m+g}{\IfNoValueTF{#2}{[#1]}{[#1]_{#2}}}

\NewDocumentCommand\deltagd{s}{\IfBooleanTF{#1}{\Delta G_d^{\circ}}{\Delta G_d}}
\NewDocumentCommand\deltaga{s}{\IfBooleanTF{#1}{\Delta G_a^{\circ}}{\Delta G_a}}

\NewDocumentCommand\ddeltagd{s}{\IfBooleanTF{#1}{\deltagd*}{\deltagd}}

\NewDocumentCommand\deltaGexp{g}{%
  \IfNoValueTF{#1}
  {\ensuremath{\Delta G}\xspace}
  {\ensuremath{\Delta G_{#1}}\xspace}
}

\NewDocumentCommand\bsa{g}{%
\IfNoValueTF{#1}
{\ensuremath{\text{BSA\xspace}}}
{\ensuremath{\text{BSA}(#1)}}
}

\NewDocumentCommand\Calpha{O{} g}{\IfNoValueTF{#2}{C_{\alpha}^{#1}}{C_{\alpha;#2}^{#1}}}
\NewDocumentCommand\Cbeta{O{} g} {\IfNoValueTF{#2}{C_{\beta}^{#1}}{C_{\beta;#2}^{#1}}}
\NewDocumentCommand\Cgamma{O{} g}{\IfNoValueTF{#2}{C_{\gamma}^{#1}}{C_{\gamma;#2}^{#1}}}
\NewDocumentCommand\Cdelta{O{} g} {\IfNoValueTF{#2}{C_{\delta}^{#1}}{C_{\delta;#2}^{#1}}}
\NewDocumentCommand\Cepsilon{O{} g} {\IfNoValueTF{#2}{C_{\epsilon}^{#1}}{C_{\epsilon;#2}^{#1}}}

\newcommand{\msa}[1]{MSA\xspace}
\NewDocumentCommand\SASA{g}{%
\IfNoValueTF{#1}
{\ensuremath{\text{SASA}}}
{\ensuremath{\text{SASA}(#1)}}
}

\NewDocumentCommand\hemogflp{g}{%
\IfNoValueTF{#1}
{FL+}{FL+[#1]}
}
\NewDocumentCommand\hemoghlm{g}{%
\IfNoValueTF{#1}
{HL-}{HL-[#1]}
}
\NewDocumentCommand\hemoghlp{g}{%
\IfNoValueTF{#1}
{HL+}{HL+[#1]}
}

\NewDocumentCommand\dlactiv{O{} g}{\sigma_{#1} \IfNoValueTF{#2}{}{(#2)}} % activation function
\NewDocumentCommand\dlsoftmax{O{} g}{\codecx{softmax}_{#1} \IfNoValueTF{#2}{}{(#2)}} % activation function
\NewDocumentCommand\dlonehot{g}{\codecx{onehot} \IfNoValueTF{#1}{}{(#1)}} % relu
\NewDocumentCommand\dlconcat{O{} g}{\codecx{concat}_{#1} \IfNoValueTF{#2}{}{(#2)}} % relu

\NewDocumentCommand\dllinearnb{g}{\codecx{LinearNoBias} \IfNoValueTF{#1}{}{(#1)}} % linear no bias
\NewDocumentCommand\dllinear{g}{\codecx{Linear} \IfNoValueTF{#1}{}{(#1)}} % linear
\NewDocumentCommand\dlrelu{O{} g}{\codecx{relu}_{#1} \IfNoValueTF{#2}{}{(#2)}} % relu
\NewDocumentCommand\dlmlpex{g}{\dlrelu{\dllinear{\IfNoValueTF{#1}{}{#1}}}}   % explicit/expanded; NB: dllinear provides ()

\NewDocumentCommand\dlmlp{O{} g}  {\codecx{MLP}_{#1}  \IfNoValueTF{#2}{}{(#2)}} % multilayer perceptron
\NewDocumentCommand\dlmlpone{O{} g}{\codecx{MLP}_{#1} \IfNoValueTF{#2}{}{(#2)}} % multilayer perceptron
\NewDocumentCommand\dlmlptwo{O{} g}{\codecx{MLP2}_{#1}\IfNoValueTF{#2}{}{(#2)}} % multilayer perceptron

\NewDocumentCommand\dllayernorm{g}{\codecx{LayerNorm} \IfNoValueTF{#1}{}{(#1)}} % layer norm
\NewDocumentCommand\dldropout{O{} g}{\codecx{Dropout}_{#1} \IfNoValueTF{#2}{}{(#2)}} % dropout

\NewDocumentCommand\pointj{g}{%
\IfNoValueTF{#1}
{p^{(j)}}{p^{(#1)}}
}

\NewDocumentCommand\alphaij{O{} G{ij}}{\alpha_{#2}^{#1}}
\NewDocumentCommand\betaij{O{} G{ij}} {\beta_{#2}^{#1}}
\NewDocumentCommand\sigmaij{O{} G{ij}}{\sigma_{#2}^{#1}}

\NewDocumentCommand\phiij{O{} G{i}}{\phi_{#2}^{#1}}
\NewDocumentCommand\varphiij{O{} G{i}}{\varphi_{#2}^{#1}}
\NewDocumentCommand\tauij{O{} G{ij}}{\tauij_{#2}^{#1}}
\NewDocumentCommand\deltaij{O{} G{ij}}{\delta_{#2}^{#1}}

\NewDocumentCommand\gammaij{O{}  G{ij}}{\gamma_{#2}^{#1}}
\NewDocumentCommand\Lambdaij{O{} G{ij}}{\Lambda_{#2}^{#1}}
\NewDocumentCommand\Lambdadec{O{} G{}}{\Lambda_{#2}^{#1}}

\NewDocumentCommand\thetai{O{} G{i}}{\theta_{#2}^{#1}}
\NewDocumentCommand\Thetai{O{} G{i}}{\Theta_{#2}^{#1}}

\NewDocumentCommand\lambdarg{g}{\IfNoValueTF{#1}{\lambda}{\lambda_{#1}}}

\NewDocumentCommand\aij{O{} G{ij}}{a_{#2}^{#1}}
\NewDocumentCommand\Aij{O{} G{ij}}{A_{#2}^{#1}}
\NewDocumentCommand\aijvec{O{} G{ij}}{\overrightarrow{a}_{#2}^{#1}}
\NewDocumentCommand\aijt{G{i}G{j}} {a_{#1,#2}}
\NewDocumentCommand\Aijt{G{i}G{j}} {A_{#1,#2}}

\NewDocumentCommand\bij{O{} G{ij}}{b_{#2}^{#1}}
\NewDocumentCommand\Bij{O{} G{ij}}{B_{#2}^{#1}}
\NewDocumentCommand\bijvec{O{} G{ij}}{\overrightarrow{b}_{#2}^{#1}}

\NewDocumentCommand\bijt{G{i}G{j}} {b_{#1,#2}}
\NewDocumentCommand\Bijt{G{i}G{j}} {B_{#1,#2}}

\NewDocumentCommand\cij{G{ij}} {c_{#1}}
\NewDocumentCommand\Cij{G{ij}} {C_{#1}}
\NewDocumentCommand\cijk{G{ijk}} {c_{#1}}
\NewDocumentCommand\Cijk{G{ijk}} {C_{#1}}

\NewDocumentCommand\cijt{G{i}G{j}} {c_{#1,#2}}
\NewDocumentCommand\Cijt{G{i}G{j}} {C_{#1,#2}}
\NewDocumentCommand\calCijt{G{i}G{j}} {\calC_{#1,#2}}

\NewDocumentCommand\dij{O{} G{ij}} {d_{#2}^{#1}}
\NewDocumentCommand\Dij{O{} G{ij}} {D_{#2}^{#1}}

\NewDocumentCommand\dijt{G{i}G{j}} {d_{#1,#2}}
\NewDocumentCommand\Dijt{G{i}G{j}} {D_{#1,#2}}

\NewDocumentCommand\dijts{G{i}G{j}}{d_{#1,#2}^2}
\NewDocumentCommand\dijs{G{ij}}{d_{#1}^2}

\NewDocumentCommand\eij{G{ij}} {e_{#1}}
\NewDocumentCommand\eijt{G{i}G{j}} {e_{#1,#2}}
\NewDocumentCommand\Eij{G{ij}} {E_{#1}}
\NewDocumentCommand\Eijt{G{i}G{j}} {E_{#1,#2}}

\NewDocumentCommand\eijs{G{ij}}{e_{#1}^2}
\NewDocumentCommand\eijts{G{i}G{j}}{e_{#1,#2}^2}

\NewDocumentCommand\fij{G{ij}}{f_{#1}}
\NewDocumentCommand\Fij{G{ij}}{f_{#1}}
\NewDocumentCommand\fijt{G{i}G{j}}{f_{#1,#2}}
\NewDocumentCommand\Fijt{G{i}G{j}}{F_{#1,#2}}

\NewDocumentCommand\gij{G{ij}}{g_{#1}}
\NewDocumentCommand\Gij{G{ij}}{G_{#1}}
\NewDocumentCommand\gijt{G{i}G{j}}{g_{#1,#2}}
\NewDocumentCommand\Gijt{G{i}G{j}}{G_{#1,#2}}

\NewDocumentCommand\hij{G{ij}}{h_{#1}}
\NewDocumentCommand\Hij{G{ij}}{H_{#1}}
\NewDocumentCommand\hijt{G{i}G{j}}{h_{#1,#2}}
\NewDocumentCommand\Hijt{G{i}G{j}}{H_{#1,#2}}

\NewDocumentCommand\iij{G{ij}}{i_{#1}}
\NewDocumentCommand\iijt{G{i}G{j}}{i_{#1,#2}}
\NewDocumentCommand\Iij{G{ij}}{I_{#1}}
\NewDocumentCommand\Iijt{G{i}G{j}}{I_{#1,#2}}

\NewDocumentCommand\jij{G{ij}}{j_{#1}}
\NewDocumentCommand\jijt{G{i}G{j}}{j_{#1,#2}}
\NewDocumentCommand\Jij{G{ij}}{J_{#1}}
\NewDocumentCommand\Jijt{G{i}G{j}}{J_{#1,#2}}

\NewDocumentCommand\kij{O{} G{ij}}{k_{#2}^{#1}}
\NewDocumentCommand\Kij{O{} G{ij}}{K_{#2}^{#1}}
\NewDocumentCommand\kijvec{O{} G{ij}}{\overrightarrow{k}_{#2}^{#1}}

\NewDocumentCommand\mij{G{ij}}{m_{#1}}
\NewDocumentCommand\Mij{G{ij}}{M_{#1}}
\NewDocumentCommand\mijt{G{i}G{j}}{m_{#1,#2}}
\NewDocumentCommand\Mijt{G{i}G{j}}{M_{#1,#2}}

\NewDocumentCommand\oij{O{} G{ij}}{o_{#2}^{#1}}
\NewDocumentCommand\Oij{O{} G{ij}}{O_{#2}^{#1}}
\NewDocumentCommand\oijvec{O{} G{ij}}{\overrightarrow{o}_{#2}^{#1}}
\NewDocumentCommand\oijtilde{O{} G{ij}}{\tilde{o}_{#2}^{#1}}

\NewDocumentCommand\pij{O{} G{ij}}{p_{#2}^{#1}}
\NewDocumentCommand\Pij{O{} G{ij}}{P_{#2}^{#1}}
\NewDocumentCommand\pijvec{O{} G{ij}}{\overrightarrow{p}_{#2}^{#1}}

\NewDocumentCommand\rij{O{} G{ij}}{r_{#2}^{#1}}
\NewDocumentCommand\Rij{O{} G{ij}}{R_{#2}^{#1}}

\NewDocumentCommand\qij{O{} G{ij}}{q_{#2}^{#1}}
\NewDocumentCommand\Qij{O{} G{ij}}{Q_{#2}^{#1}}
\NewDocumentCommand\qijvec{O{} G{ij}}{\overrightarrow{q}_{#2}^{#1}}

\NewDocumentCommand\qbarfunc{O{} g}{\bar{q}_{#1}  \IfNoValueTF{#2}{}{(#2)}}

\NewDocumentCommand\sij{O{} G{ij}}{s_{#2}^{#1}}
\NewDocumentCommand\Sij{O{} G{ij}}{S_{#2}^{#1}}
\NewDocumentCommand\sijvec{O{} G{ij}}{\overrightarrow{s}_{#2}^{#1}}
\NewDocumentCommand\sijtilde{O{} G{ij}}{\tilde{s}_{#2}^{#1}}

\NewDocumentCommand\tij{G{ij}}{t_{#1}}
\NewDocumentCommand\Tij{G{ij}}{T_{#1}}

\NewDocumentCommand\uij{G{ij}}{u_{#1}}
\NewDocumentCommand\Uij{G{ij}}{U_{#1}}
\NewDocumentCommand\uijt{G{i}G{j}}{u_{#1,#2}}
\NewDocumentCommand\Uijt{G{i}G{j}}{U_{#1,#2}}

\NewDocumentCommand\vij{O{} G{ij}}{v_{#2}^{#1}}
\NewDocumentCommand\Vij{O{} G{ij}}{V_{#2}^{#1}}
\NewDocumentCommand\vijvec{O{} G{ij}}{\overrightarrow{v}_{#2}^{#1}}
\NewDocumentCommand\vijt{G{i}G{j}}{v_{#1,#2}}
\NewDocumentCommand\Vijt{G{i}G{j}}{V_{#1,#2}}

\NewDocumentCommand\wij{O{} G{ij}}{w_{#2}^{#1}}
\NewDocumentCommand\Wij{O{} G{ij}}{W_{#2}^{#1}}
\NewDocumentCommand\wijvec{O{} G{ij}}{\overrightarrow{w}_{#2}^{#1}}

\NewDocumentCommand\xij{O{} G{ij}}{x_{#2}^{#1}}
\NewDocumentCommand\Xij{O{} G{ij}}{X_{#2}^{#1}}

\NewDocumentCommand\yij{G{ij}}{y_{#1}}
\NewDocumentCommand\Yij{G{ij}}{Y_{#1}}

\NewDocumentCommand\zij{G{ij}}{z_{#1}}
\NewDocumentCommand\Zij{G{ij}}{Z_{#1}}

\NewDocumentCommand\piij{G{ij}}{\pi_{#1}}
\NewDocumentCommand\thetaij{G{ij}}{\theta_{#1}}
\NewDocumentCommand\Thetaij{G{ij}}{\Theta_{#1}}

\NewDocumentCommand\costpc{g}{\Delta\IfNoValueTF{#1}{}{(#1)}} % procruste cost

\NewDocumentCommand\eigenvec{O{v} G{v}}{
\IfNoValueTF{#2}{#2}{#2_{#1}}}

\NewDocumentCommand\FMgeodesic{O{} g}{\gamma_{#1} \IfNoValueTF{#2}{}(#2)}

\NewDocumentCommand\OTCPL{g}{\pi \IfNoValueTF{#1}{}{(#1)}}  %% OT coupling
\NewDocumentCommand\otcpl{g}{\pi \IfNoValueTF{#1}{}{(#1)}}  %% OT coupling
\NewDocumentCommand\otcost{g}{c\IfNoValueTF{#1}{}{(#1)}}  %% OT cost

\NewDocumentCommand\RectFlow{g}{\text{RectFlow}\IfNoValueTF{#1}{}{(#1)}}

\NewDocumentCommand{\verblinett}{v}{{\tt #1}}
\NewDocumentCommand{\verblineem}{v}{{\em #1}}
\NewDocumentCommand{\verblinebf}{v}{{\bf #1}}

\newcommand{\codec}[1]{{\tt \text{#1}}} %code convention: definition
\newcommand{\codecx}[1]{{\tt \text{#1}}\xspace} %code convention codecx def codecx-def codecxdef
\newcommand{\ucainria}{Centre Inria at Université Côte d'Azur, France\xspace}

\newcounter{classd}%class description

\NewDocumentCommand\lloyd{g}{\codecx{\toblue Lloyd}\IfNoValueTF{#1}{}{(#1)}}
\NewDocumentCommand\lloyditers{g}{\codecx{\toblue Lloyd}\IfNoValueTF{#1}{}{(#1)}}

\NewDocumentCommand\Distproba{g}{D_{\text{COST}}\IfNoValueTF{#1}{}{(#1)}}

\NewDocumentCommand\smartseeding{g}{\codecx{\toblue SmartSeeding}\IfNoValueTF{#1}{}{(#1)}}
\NewDocumentCommand\kmeanspp{O{blue} g}{\codecx{\textcolor{#1}{k-means++}}\IfNoValueTF{#2}{}{(#2)}}

\NewDocumentCommand\Dsquare{g}{D^2\IfNoValueTF{#1}{}{(#1)}}
\NewDocumentCommand\Dsquaremean{g}{\overline{D}^2\IfNoValueTF{#1}{}{(#1)}}

\NewDocumentCommand\onehot{g}{\codecx{One\_hot}\IfNoValueTF{#1}{}{(#1)}}

\NewDocumentCommand\kmeansfunc{o G{K}}{\IfNoValueTF{#1}{\Phi_{#2}}{\Phi_{#2,#1}}}

\NewDocumentCommand\kmeansfun {G{K}}{\Phi_{#1}} % kmeans functional
\NewDocumentCommand\kmeansfuncmean{G{K}}{\overline{\Phi}_{#1}} % kmeans functional
\NewDocumentCommand\kmeansfunmean {G{K}}{\overline{\Phi}_{#1}} % kmeans functional

\NewDocumentCommand\kmeansinertia{G{S} G{s}}{\varphi_{#1}(#2)}
\NewDocumentCommand\kmeansfunS {O{K}}{\Phi_{#1}^\text{D}} % wrt Seeds
\NewDocumentCommand\kmeansfunCOM {G{K}}{\Phi_{#1}^\text{S-COM}} % wrt COM induced by the seeds

\NewDocumentCommand\kmeansopt{G{K}}{\Phi_{#1,OPT}}
\newcommand{\frederic}{Fr\'ed\'eric\xspace}

\newcommand{\lheritier}{Lh\'eritier\xspace}

\newcommand{\carriere}{Carri\`ere\xspace}

\NewDocumentCommand\gmpfr{g}{%
\IfNoValueTF{#1}
{\codecx{Gmpfr}}
{\codecx{Gmpfr}[#1]}
}

\NewDocumentCommand\scoreDFM{g}{%
\IfNoValueTF{#1}
{\Phi(\calS)}
{\Phi(#1)}
}

\NewDocumentCommand\varinf{O{} g}{
\text{VI}_{#1}
\IfNoValueTF{#2}{}{(#2)}
}

\NewDocumentCommand\pibest{g}{%
\IfNoValueTF{#1}
{{\pi_{best}}}
{{\pi_{best}}(#1)}
}

\NewDocumentCommand\blosum{g}{%
\IfNoValueTF{#1}
{{\codecx{BLOSUM}}}
{{\codecx{BLOSUM#1}}}
}

\NewDocumentCommand\alphafold{G{2}}{\codecx{AlphaFold#1}}

\newcommand{\phisical}{\codecx{PhiSiCal}}

\NewDocumentCommand\sblwebhref{O{black}}{\href{https://sbl.inria.fr}{\textcolor{#1}{Structural Bioinformatics Library}}}

\NewDocumentCommand\powerdist{g}{\pi \IfNoValueTF{#1}{}{(#1)}}

\NewDocumentCommand\vorr{g}{%
\IfNoValueTF{#1}
{\text{Vor}}
{\text{Vor}(#1)
}}

\NewDocumentCommand\kalp{G{\alpha}}{\calK_{#1}}
\NewDocumentCommand\kalpha{G{\alpha}}{\calK_{#1}}
\NewDocumentCommand\Kalpha{G{\alpha}}{\calK_{#1}}
\NewDocumentCommand\acomplex{G{\alpha}}{\calK_{#1}}
\NewDocumentCommand\ashape{G{\alpha}}{\calW_{#1}}
\NewDocumentCommand\ballrestriction{O{R} G{}}{{#1}_{#2}}
\NewDocumentCommand\vorregion{G{i}} { V_{#1}}

\NewDocumentCommand\sfm{g}{\IfNoValueTF{#1}{\calF_{\alpha}}{\calF_{#1}}}
\NewDocumentCommand\sfmA{g}{\IfNoValueTF{#1}{\calF^(A)_{\alpha}}{\calF^{A}_{#1}}}
\NewDocumentCommand\sfmB{g}{\IfNoValueTF{#1}{\calF^(B)_{\alpha}}{\calF^{B}_{#1}}}

\NewDocumentCommand\spectraldom{g}{\IfNoValueTF{#1}{\codecx{SPECTRALDOM}}{\codecx{SPECTRALDOM/#1}}}

\NewDocumentCommand\stiffc{G{}}{ \gamma^\text{Cov}_{#1} }
\NewDocumentCommand\stiffnc{G{}}{ \gamma^\text{NCov}_{#1} }
\NewDocumentCommand\stiffhbond{G{}}{ \gamma^\text{HB}_{#1} }
\NewDocumentCommand\stiffSB{G{}}{  \gamma^\text{SB}_{#1} }

\NewDocumentCommand\tmalign{g}{\codecx{TM}\IfNoValueTF{#1}{}{(#1)}}
\NewDocumentCommand\TMXref{O{A}}{{#1}^\text{ref}} % default letter:now A

\newcommand{\ang}{\textup{\AA}\xspace} % angstrom short ie A^\circ
\NewDocumentCommand\lrmsdpw{g}{\text{lRMSDpw}\IfNoValueTF{#1}{}{(#1)}}

\NewDocumentCommand\rmsd{gg}{%
\IfNoValueTF{#2}
{\ensuremath{\text{RMSD}}\xspace }
{\ensuremath{\text{RMSD}(#1,#2)}\xspace}
}
\NewDocumentCommand\rmsdw{gg}{%
\IfNoValueTF{#2}
{\ensuremath{\text{RMSD}_\text{w}}\xspace }
{\ensuremath{\text{RMSD}_\text{w}(#1,#2)}\xspace}
}

\NewDocumentCommand\gopt{g+g}{%
\IfNoValueTF{#2}
{\ensuremath{g^{\text{OPT}}}\xspace}
{\ensuremath{g^{\text{OPT}}(#1, #2)}\xspace}
}

\NewDocumentCommand\lrmsdoptrm{gg}{%
\IfNoValueTF{#2}
{\ensuremath{g_{#1}^{\text{OPT}}}}
{\ensuremath{g_{#1}^{\text{OPT}}}(#2)}
}

\NewDocumentCommand\lrmsd{gg}{%
\IfNoValueTF{#2}
{\ensuremath{\text{lRMSD}}\xspace}
{\ensuremath{\text{lRMSD}(#1,#2)}\xspace}
}
\NewDocumentCommand\lrmsds{gg}{%
\IfNoValueTF{#2}
{\ensuremath{\text{lRMSD}^2}\xspace}
{\ensuremath{\text{lRMSD}^2(#1,#2)}\xspace}
}
\NewDocumentCommand\lrmsdvw{gg}{%
\IfNoValueTF{#2}
{\ensuremath{\text{lRMSD}_{\text{vw}}}\xspace}
{\ensuremath{\text{lRMSD}_{\text{vw}}(#1,#2)}\xspace}
}

\NewDocumentCommand\lrmsdvws{gg}{%
\IfNoValueTF{#2}
{\ensuremath{\text{lRMSD}^2_{\text{vw}}}\xspace}
{\ensuremath{\text{lRMSD}^2_{\text{vw}}(#1,#2)}\xspace}
}

\NewDocumentCommand\lrmsdew{gg}{%
\IfNoValueTF{#2}
{\ensuremath{\text{lRMSD}_{\text{ew}}}\xspace}
{\ensuremath{\text{lRMSD}_{\text{ew}}(#1,#2)}\xspace}
}

\NewDocumentCommand\lrmsdews{gg}{%
\IfNoValueTF{#2}
{\ensuremath{\text{lRMSD}^2_{\text{ew}}}\xspace}
{\ensuremath{\text{lRMSD}^2_{\text{ew}}(#1,#2)}\xspace}
}

\NewDocumentCommand\lrmsdfinal{gg}{%
\IfNoValueTF{#2}
{\ensuremath{\text{lRMSD}_{\text{final}}}\xspace}
{\ensuremath{\text{lRMSD}_{\text{final}}(#1,#2)}\xspace}
}

\NewDocumentCommand\rmsdcomb{gg}{%
\IfNoValueTF{#2}
{\ensuremath{\text{RMSD}_{\text{Comb.}}}\xspace}
{\ensuremath{\text{RMSD}_{\text{Comb.}}(#1,#2)}\xspace}
}
\NewDocumentCommand\rmsdcombs{gg}{%
\IfNoValueTF{#2}
{\ensuremath{\text{RMSD}^2_{\text{Comb.}}}\xspace}
{\ensuremath{\text{RMSD}^2_{\text{Comb.}}(#1,#2)}\xspace}
}

\NewDocumentCommand\dCalC{gg}{%
\IfNoValueTF{#2}
{d_{\calC}}
{d_{\calC}(#1,#2)}
}

\NewDocumentCommand\extendConf{g}{
\IfNoValueTF{#1}
{\codec{ExtendConformation}}
{\codec{ExtendConformation}(#1)}
}
\NewDocumentCommand\selectConf{g}{
\IfNoValueTF{#1}
{\codec{SelectConformationToExtend}}
{\codec{SelectConformationToExtend}(#1)}
}
\NewDocumentCommand\acceptConf{g}{
\IfNoValueTF{#1}
{\codec{AcceptConformation}}
{\codec{AcceptConformation}(#1)}
}
\NewDocumentCommand\recordConf{g}{
\IfNoValueTF{#1}
{\codec{RecordNewConformation}}
{\codec{RecordNewConformation}(#1)}
}

\NewDocumentCommand\energyP{gg}{
\IfNoValueTF{#1}
{\codec{E}}
{\codec{E}(#1)}
}

\NewDocumentCommand\sampleConfUnif{gg}{
\IfNoValueTF{#1}
{\codec{SampleConformationUniformly}}
{\codec{SampleConformationUniformly}(#1)} }

\NewDocumentCommand\sampleConfMoveSet{gg}{
\IfNoValueTF{#1}
{\codec{SampleConformationWithMoveSet}}
{\codec{SampleConformationWithMoveSet}(#1,#2)}}

\NewDocumentCommand\algoHY{g}{%
\IfNoValueTF{#1}
{\codecx{Hybrid}}
{\codecx{Hybrid-switch}-{#1}}
}
\NewDocumentCommand\algohybrid{g}{%
\IfNoValueTF{#1}
{\codecx{Hybrid}}
{\codecx{Hybrid-switch}-{#1}}
}

\NewDocumentCommand\distij{g}{%
\IfNoValueTF{#1}
{d_{ij}}
{d_{#1}}
}

\newcommand{\distik}{\distik{ik}}

\NewDocumentCommand\flowij{g}{%
\IfNoValueTF{#1}
{f_{ij}}
{f_{#1}}
}

\NewDocumentCommand\motifAB{g}{\IfNoValueTF{#1}{M^{(AB)}}{M^{(AB)}_{#1}}}
\NewDocumentCommand\motifab{g}{\IfNoValueTF{#1}{M^{(AB)}}{M^{(AB)}_{#1}}}

\NewDocumentCommand\motifa{g}{\IfNoValueTF{#1}{M^{(A)}}{M^{(A)}_{#1}}}
\NewDocumentCommand\motifb{g}{\IfNoValueTF{#1}{M^{(B)}}{M^{(B)}_{#1}}}

\NewDocumentCommand\pconsSeq{gg}{%
  \IfNoValueTF{#1}
  {PCS}
  {PCS_{\leq #2}^{#1}}
}

\NewDocumentCommand\consSeq{g}{%
\IfNoValueTF{#1}{CS}
{CS^{(#1; D)}}
}

\NewDocumentCommand\consSS{g}{%
\IfNoValueTF{#1}
{CSS}
{CSS^{(#1, D)}}
}

\NewDocumentCommand\pconsSS{ggg}{%
\IfNoValueTF{#1}
{PCSS}
{PCSS_{\leq #2, \geq #3}^{(#1; D)}
}}

\NewDocumentCommand\scoreij{g+g}{%
\IfNoValueTF{#2}
{s_{ij}}
{s_{#1, #2}}
}

\NewDocumentCommand\dihedralangles{g}{\IfNoValueTF{#1}{\text{DihedralAngles}}{\text{DihedralAngles}(#1)}}

\NewDocumentCommand\cposeA{G{i}}{A^{(#1)}}
\NewDocumentCommand\cposeB{G{i}}{B^{(#1)}}

\NewDocumentCommand\Kdist{G{i} G{j}}{R_{#1, #2}}

\NewDocumentCommand\msafreq{G{} G{}}{f_{#1}(#2)}
\NewDocumentCommand\msafreqab{G{} G{}}{f_{#1}(#2)}

\NewDocumentCommand\oneseq{O{}}{ {a^{#1}}_1,\dots,a^{#1}_L }
\NewDocumentCommand\seqone{O{}}{ a^{#1}_1,\dots, a^{#1}_L }
\NewDocumentCommand\sequniverse{G{a}}{ \calA^L}

\NewDocumentCommand\dcaHam{G{\cdot}}{H(#1)}

\NewDocumentCommand\dcaone{O{} G{i}} {\bblue{h_{#2}^{ a^{#1}_{#2}} }}  
\NewDocumentCommand\dcatwo{O{} G{i} G{j}} {\bblue{J_{#2#3}^{a^{#1}_{#2}a^{#1}_{#3}  } }} % O{}: exponent
\NewDocumentCommand\artZi{G{} G{i}}{z_{#2}(a_{#2-1}^{#1},\dots,a_1^{#1})}

\NewDocumentCommand\seqall{G{m}}{a^{#1}}
\NewDocumentCommand\seqsym{O{} G{i}}{a^{#1}_{#2}}                % seq input symbol
\NewDocumentCommand\seqsympre{O{} G{i}}{\underline{a}^{#1}_{#2}} % and its prefix
\NewDocumentCommand\seqmod{g}{\bblue{a\IfNoValueTF{#1}{}{_{\backslash #1}}}}

\NewDocumentCommand\MMLlength{G{\theta} G{X}}{\text{Length}(#1, #2)} % MML

\newcommand{\cbmd}{CBMD\xspace}
\newcommand{\cbmds}{CBMDs\xspace}

\NewDocumentCommand\Tdrvec{G{n}}{\Theta}       % random vector
\NewDocumentCommand\Tdrvar{G{i}}{{\Theta}_{#1}} % random variable
\NewDocumentCommand\Tdobs {G{}}{\theta_{#1}}

\NewDocumentCommand\Tddatum{G{n}}{{\bm \theta}^{(#1)}} % datum / entire dataset
\NewDocumentCommand\Tdptij{G{i}}{{\bm \theta}_{#1}} % Td pt / sample point
\NewDocumentCommand\tdptij{G{i}}{{\bm \theta}_{#1}} % Td pt / sample point

\NewDocumentCommand\CMM{g}{CMM\IfNoValueTF{#1}{}{(#1)}}
\renewcommand\torustri{\mathbb{T}_{2\pi}^d}

\NewDocumentCommand\vmoned{O{} g}{\mathrm{vm}_{#1}\IfNoValueTF{#1}{}{(#2)}}
\NewDocumentCommand\VMoned{O{} g}{\mathrm{VM}_{#1}\IfNoValueTF{#1}{}{(#2)}}
\NewDocumentCommand\VMonedinv{O{} g}{\mathrm{VM}_{#1}^{-1}\IfNoValueTF{#1}{}{(#2)}}

\NewDocumentCommand\wconed{O{} g}{\mathrm{wc}_{#1}\IfNoValueTF{#1}{}{(#2)}}
\NewDocumentCommand\WConed{O{} g}{\mathrm{WC}_{#1}\IfNoValueTF{#1}{}{(#2)}}

\NewDocumentCommand\vMvM{g}{\text{vM-vM}\IfNoValueTF{#1}{}{(#1)}}
\RenewDocumentCommand\vMvM{g}{\ifmmode \text{vM-vM}\IfNoValueTF{#1}{}{(#1)} \else  \text{vM-vM}\IfNoValueTF{#1}{}{(#1)}\xspace \fi}

\NewDocumentCommand\wCwC{g}{\text{wC-wC}\IfNoValueTF{#1}{}{(#1)}}
\RenewDocumentCommand\wCwC{g}{\ifmmode \text{wC-wC}\IfNoValueTF{#1}{}{(#1)} \else  \text{wC-wC}\IfNoValueTF{#1}{}{(#1)}\xspace \fi}

\NewDocumentCommand\vMwC{g}{\text{vM-wC}\IfNoValueTF{#1}{}{(#1)}}
\RenewDocumentCommand\vMwC{g}{\ifmmode \text{vM-wC}\IfNoValueTF{#1}{}{(#1)} \else  \text{vM-wC}\IfNoValueTF{#1}{}{(#1)}\xspace \fi}

\NewDocumentCommand\wCvM{g}{\text{wC-vM}\IfNoValueTF{#1}{}{(#1)}}
\RenewDocumentCommand\wCvM{g}{\ifmmode \text{wC-vM}\IfNoValueTF{#1}{}{(#1)} \else  \text{wC-vM}\IfNoValueTF{#1}{}{(#1)}\xspace \fi}

\newif\ifLONG
\LONGfalse

\title{Circula-based multivariate distributions on the flat torus, with applications in structural biology}
\author{Guillaume \carriere\thanks{\ucainria; Guillaume.Carriere@inria.fr}
 and Alix \lheritier\thanks{Amadeus, Sophia Antipolis, France; alherit@gmail.com}
and \frederic Cazals\thanks{\ucainria; Frederic.Cazals@inria.fr}
}

\newcommand{\wdir}{./}
\begin{document}
\maketitle

%\input{cmm-main.tex}
%%===Inclusion starts for file cmm-main.tex

\begin{abstract}
Modeling dependencies between random variables independently from
their marginals is fundamental in applications ranging from finance to
(structural) biology.  In this work, we undertake this problem using
circula to model data living on the $d$-dimensional flat torus
$\torusd$, making two contributions. First, using a low rank covariance structure to define circulae based
on a latent variable model, we design the first closed-form normalized
distribution on the flat torus $\torusd$--with covariance structure. Second, building on this framework, we propose the first models for joint
distributions of torsion angles (backbone and side-chains) for
neighboring amino-acids in proteins.  In practice, we fit mixtures on
flat torii from $\torusd{2}$ to $\torusd{14}$, and show they are SOTA
in terms of likelihood and sparsity. We anticipate that these models will prove fundamental to move from discrete
structural studies like in \alphafold, to thermodynamics and kinetics,
which are the ultimate goals in theoretical biophysics.
\end{abstract} 

\section{Introduction}
%%i%%%%%%%%%%%%%%%%%%%%%%%%%%%%%%%%%%%%%%%%%%%%%%%%%%%%%%%%%%%%%%%%%%%%%%%%%%%%%%%

\subsection{Previous work}
%%ii-%-%-%-%-%-%-%-%-%-%-%-%-%-%-%-%-%-%-%-%-%-%-%-%-%-%-%-%-%-%-%-%-%-%-%-%-%-%-%

\paramini{Copulae and circulae.}
Copulae are functions that allow us to model the dependence
structure between random variables independently of their marginal
distributions~\cite{nelsen2006introduction}. This makes them
especially useful for settings where complex dependencies matter, \eg in
finance and risk management, hydrology, or (structural) biology.  A
central result is Sklar's theorem~\cite{sklar1959fonctions}, which
states that any multivariate joint distribution can be written as the
product of the marginals and a copula fully capturing their
dependence.
In practice, fitting copulae typically involves a two-step approach:
first estimating the marginal distributions, then selecting and
calibrating a copula family (e.g., Gaussian, t, Archimedean) using
methods such as maximum likelihood or inference functions for margins
(IFM)~\cite{joe1996estimation}. For elliptical copulae--which generalize multivariate normal distributions, an estimate of
the covariance or correlation matrix plays a central role, as it
directly parameterizes the dependence structure. In such cases, one
often fits the copula by estimating this matrix from
observations. When data are high-dimensional or noisy,
low-rank approximations of the covariance matrix can be used to
stabilize estimation, sometimes followed by a projection onto the
nearest valid correlation matrix using the Higham correction to ensure
positive semi-definiteness~\cite{borsdorf2010computing}.
%%
%%\cite{borsdorf2010computing} Computing a nearest correlation matrix with factor structure
%%\cite{duan2014generalized} On the generalized low rank approximation of the correlation matrices arising in the asset portfolio
%%
%% While classical covariance captures only linear dependence, copulae
%% provide a much richer description, including nonlinear and tail
%% relationships. 
Extensions like vine copulae (pair-copula
constructions)~\cite{czado2022vine} allow flexible high-dimensional
modeling by decomposing complex dependence structures into cascades of
bivariate copulae.

Circulae are a class of copulae designed to capture dependence
structures with circular or periodic features arising in
directional data~\cite{mardia2009directional}.  Circulae are
especially relevant to handle periodic data found meteorology,
oceanography, finance, or (structural) biology.  Construction of
circulae typically involves adapting copula definitions to circular
domains or using transformations of linear copulae onto the unit
circle~\cite{jones2015class}.
In this work, we are concerned with 
circulae defined on the $d$-dimensional flat torus $\torusd$.

\paramini{Applications in structural biology.}
Polypeptide chains, which are polymers of amino acids, are the
building blocks of proteins. There are 20 natural amino acids (a.a.) sharing the
same backbone (which involves the three atoms $N-\Calpha-C$), and
differing by their so-called side-chains.
The intrinsic geometry of a protein is modeled by its internal
coordinates, namely bond lengths, valence angles, and torsion angles.
Of fundamental importance are the two torsion angles $(\phi, \psi)$
found before and after the $\Calpha$ carbon, and those of the
side-chains -- denoted $\chi_i$s.  The angles $(\phi_i, \psi_i)$
indeed define the Ramachandran diagram which encodes secondary
structures \cite{ramachandran1963stereochemistry}, while the $\chi$
angles define the so-called rotational isomers of side-chains, namely
the {\em preferred} low energy side-chain
conformations~\cite{shapovalov2011smoothed,hintze2016molprobity}.
These angles have been used for decades in structural
analysis\cite{williams2018molprobity,amarasinghe2023getting,amarasinghe2025phisical},
structure prediction~\cite{jumper2021highly}, protein
design~\cite{hallen2019protein}, generative
models~\cite{paluszewski2010mocapy++}, loop closure
reconstructions\cite{coutsias2004kinematic,odonnell2022tripeptide}, etc.

More recently, torsion angles of individual a.a. played a pivotal role
in particular in \alphafold~\cite{jumper2021highly}, the protein
structure prediction program for which the 2024 Nobel prize in
chemistry was co-awarded.
Indeed, the geometry inference in the structure module in \alphafold
starts by pulling a tuple of torsion angles $\phi, \psi, \chi_i$s for
each a.a.  from a high dimensional latent space.  These tuples define the
individual a.a. geometries, and are iteratively adjusted by a {\em
  message passing} like algorithm also resorting to {\em invariant
  point attention}~\cite{jumper2021highly}.

\ifLONG
%%ii-%-%-%-%-%-%-%-%-%-%-%-%-%-%-%-%-%-%-%-%-%-%-%-%-%-%-%-%-%-%-%-%-%-%-%-%-%-%-%
%% The second one is \proteinmpnn~\cite{dauparas2022robust}
Stat models for couplings:
\cite{gonzalez2025dependence}

\cite{gonzalez2022statistical}
Statistical proofs of the interdependence between nearest neighbor effects on polypeptide backbone conformations
\fi

\subsection{Contributions}
%%ii-%-%-%-%-%-%-%-%-%-%-%-%-%-%-%-%-%-%-%-%-%-%-%-%-%-%-%-%-%-%-%-%-%-%-%-%-%-%-%

Our work unifies and expands previous work in two directions.

\begin{figure}[h]
\centerline{\includegraphics[width=\linewidth]{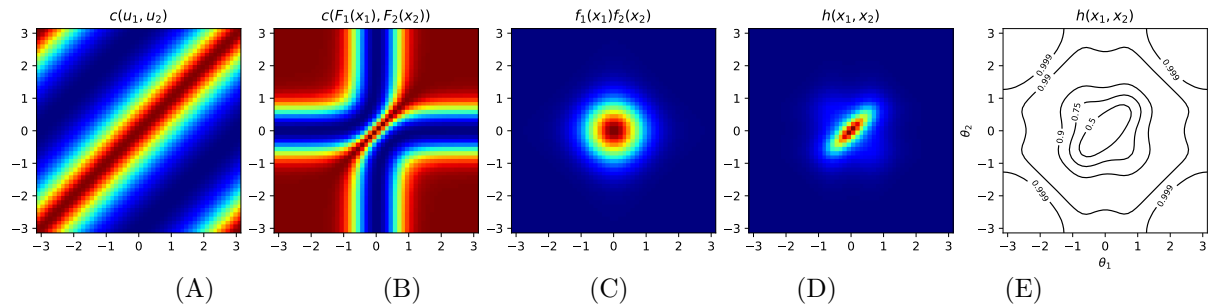}}
%\makebox[\linewidth]{A\hfill B\hfill C\hfill D\hfill E}
\vspace{-.5cm}
\makebox[\linewidth]{\hspace{\fill}(A)\hspace{\fill}(B)\hspace{\fill}(C)\hspace{\fill}(D)\hspace{\fill}(E)\hspace{\fill}}
\caption{{\bf Modeling covariance on the flat torus:  illustration with a 2D von Mises-wrapped Cauchy  distribution (\vMwC).}
{\bf (A)} The wrapped Cauchy circula correlates circular uniform variables.
{\bf (B)} The wrapped Cauchy circula is applied to variables of interest passed through the probability integral transform.
{\bf (C)} The variables of interest follow their marginal von Mises densities.
{\bf (D-E)}  The vM-wC distribution is the product of (B) and (C).
% {The \vMwC distribution shown uses $\mu = 0$, $\rho = 0.8$ and $\rho_c = 0.3$ on both dimensions.}  
}
\label{fig:circula-decomposition}
\end{figure}

Mathematically, we design the first closed-form normalized
distribution on the hypertorus, also embarking an explicit covariance
structure (Fig.~\ref{fig:circula-decomposition}).
%%
%% These distributions are based on particular cases of a multivariate
%% extension of circulae, for which we derive novel closed-form
%% expressions.
%%
These distributions are essentially based on applying a multivariate
extension of the 2D circula of \cite{jones2015class} using von Mises or wrapped Cauchy as {\em binding
densities}. We combine these circula densities with von Mises marginal
distributions, which results in tractable models with useful
characteristics. We refer to these models as the multivariate von
Mises-von Mises (\vMvM) distribution and the multivariate von
Mises-wrapped Cauchy (\vMwC) distribution.

Application-wise, we extend and unify previous work focusing on
torsion angles in proteins, fitting circula mixtures on flat torii
from dimension $d=2$ to $d=14$. The long term goal driving this
initiative is to move from structural studies like in \alphafold, to
thermodynamics and kinetics, which are the
ultimate open problems in theoretical biophysics~\cite{lelievre2010free}.
More specifically, we make two fundamental contributions.
First, for single a.a., we expand the mixture model \phisical which
models joint distributions of all torsion angles in a given a.a.
\cite{amarasinghe2023getting}.  We add to the product of
univariate von Mises a circula with covariance structure, obtaining
mixtures which are SOTA in terms of model selection criteria
(likelihood vs model complexity).  Practically,
we fit 20 mixture models--one per a.a., using flat torii $\torusd{2}$
(a.a.: alanine) to $\torusd{7}$ (a.a.: arginine).
Second, correlations between the torsion angles $\psi_i$ and
$\phi_{i+1}$ of two consecutive a.a.  have recently been observed in
the {\em cross-landscape} Ramachandran
diagram~\cite{rosenberg2023amino}.
We provide the first quantitative model for cross-landscapes,
computing joint distributions of all torsion angles for a pair of
consecutive a.a. -- $20\times 20$ of them.

All proofs are given in the Supporting Information.

\section{Background for multivariate models on the flat torus}
%%i%%%%%%%%%%%%%%%%%%%%%%%%%%%%%%%%%%%%%%%%%%%%%%%%%%%%%%%%%%%%%%%%%%%%%%%%%%%%%%%

\ifLONG
\subsection{Notations and terminology}
%%ii-%-%-%-%-%-%-%-%-%-%-%-%-%-%-%-%-%-%-%-%-%-%-%-%-%-%-%-%-%-%-%-%-%-%-%-%-%-%-%

We work on the $d$-dimensional flat torus $\torusd$.
A random vector is denoted $\torusd: \Tdrvec = (\Tdrvar{1},\dots,\Tdrvar{d})$,
and a realization of this vector $\Tdobs = (\Tdobs{1}, \dots, \Tdobs{d} )$.
A dataset of $n$ observations is written $\Tddatum = \{ \Tdptij{1}, \dots, \Tdptij{n}\}$.
Finally, we access coordinates as  $\Tdptij = (\Tdptij{ij})$.
\fi

\ifLONG
\begin{remark}
NB: as in the literature we distinguish
\begin{itemize}
\item multivariate von Mises-wrapped Cauchy (\vMwC) distribution: the final product,
\item multivariate von Mises-wrapped Cauchy (\vMwC) circula: the {\em coupling/binding} density.
\end{itemize}
\end{remark}
\fi
\subsection{Copula models}
%%ii-%-%-%-%-%-%-%-%-%-%-%-%-%-%-%-%-%-%-%-%-%-%-%-%-%-%-%-%-%-%-%-%-%-%-%-%-%-%-%

\ifLONG
\else
We work on the $d$-dimensional flat torus $\torusd$.
A random vector is denoted $\torusd: \Tdrvec = (\Tdrvar{1},\dots,\Tdrvar{d})$,
and a realization of this vector $\Tdobs = (\Tdobs{1}, \dots, \Tdobs{d} )$.
A dataset of $n$ observations is written $\Tddatum = \{ \Tdptij{1}, \dots, \Tdptij{n}\}$.
Finally, we access coordinates as  $\Tdptij = (\Tdptij{ij})$.
\fi
A copula $c(\cdot)$ is a multivariate distribution whose marginals are uniform on
the interval  $[0,1]$. Copulae are used to describe the dependence
structure of a multivariate distribution independently of its
marginals. Indeed,  Sklar's theorem~\cite{sklar1959fonctions} states that  any multivariate
distribution $h$ can be decomposed in terms of its marginals $f_i$ and
a copula $c$ modeling the coupling, namely
\begin{align*}
h(x_1, ..., x_d) = f_1(x_1) \cdot ... \cdot f_d(x_d) \cdot c(F_1(x_1), ..., F_d(x_d)),
\end{align*}
with $f_i$ (resp. $F_i$) are the marginal densities (resp. cumulative
distribution functions) and $c$ is the copula density. Note that the
copula is applied to the observed vector $(F_1(x_1), ..., F_d(x_d))$,
whose marginals are uniformly distributed on $[0,1]$ as a result of
the probability integral transform~\cite{jacod2004probability}.

\subsection{Circula models}
%%ii-%-%-%-%-%-%-%-%-%-%-%-%-%-%-%-%-%-%-%-%-%-%-%-%-%-%-%-%-%-%-%-%-%-%-%-%-%-%-%

\ifLONG
\else

\paramini{Circulae.}
Circulae are a circular analogue of copulae, namely distributions whose marginals
are uniform on $[0, 2\pi)$. 
We focus on a circula construction  based on the latent variable
model from~\cite{jones2015class} (Fig.~\ref{fig:latent-variable-model}), in
which each RV of interest is conditioned to a uniform random variable
$\phi$ on $[0, 2\pi)$.
\ifLONG Integrating out $\phi$, which amounts to saying that $\phi$ is
unobserved, leaves the residual correlations, yielding a multivariate
circula which can be regarded as a latent variable model.  This is
akin to factor copula for linear multivariate data
\cite{krupskii2013factor}, adapted to the hypertorus.  
\fi
Formally, each uniform marginal of the circula is defined as $\Theta_i
= (\Omega_i + q_i \phi)(\bmod 2\pi)$ where $\phi$ is circular
uniformly distributed, $\Omega_i$ is a RV with density $g_i$, and $q_i
= \pm 1$ gives the sign of the dependency on $\phi$.  It follows that
$\Theta_i$ is a uniform variable obtained by shifting $\phi$ by an
angle $\omega_i \sim g_i$, in the following we refer to $g_i$ as {\em  binding densities}.
\ifLONG
In this construct, $\phi$ is the unobserved variable explaining the
dependency structure of the full model.
\fi
The complete-data joint density of the circula $c(\cdot)$ is the product of the density on $\phi$ and conditional densities on $\theta_i$ given $\phi$:
\begin{equation}
\label{eq:circula-density-full}
c(\phi, \theta_1, ..., \theta_d) = \frac{1}{2\pi} \prod_{i=1}^d g_i(\theta_i - q_i\phi).
\end{equation}
The circula is then obtained by integrating out the latent variable $\phi$:
\begin{equation}
\label{eq:multivar-circula}
c(\theta_1, ..., \theta_d) = \frac{1}{2\pi} \int_0^{2\pi} \prod_{i=1}^d g_i(\theta_i - q_i\phi)d\phi.
\end{equation}
We assume the binding densities $g_i$ are centered with mean resultant length $\rho_i$.
(NB: the length of the {\em mean resultant vector} acts as concentration
parameter \cite{mardia2009directional}). We also assume they are
symmetric about 0 -- see Rmk \ref{rmk:mu-zero}.
%%
%\footnote{Mean resultant vector, which serves as concentration parameter.}
%%
The mean resultant length of the $(k,l)$th marginal copula
density, which acts as dependence parameter for variables $(\theta_k,
\theta_l)$, satisfies~\cite{jones2015class}:
\begin{equation}
\label{eq:rhokl-dependence}
\rho_{kl} =  \rho_k\rho_l.
\end{equation}

\paramini{Circula-based multivariate distributions (CBMDs)}.
We focus on CBMDs, namely distributions of the form
\begin{equation}
\label{eq:full-density}
h(\theta_1, ..., \theta_d) = 
 (2\pi)^d \cdot f_1(\theta_1) \cdot ... \cdot f_d(\theta_d) \cdot c(2\pi F_1(\theta_1), ..., 2\pi F_D(\theta_d)),
\end{equation}
with  $f_i(\cdot)$ (resp. $F_i(\cdot)$) are the marginal densities
(resp. marginal cumulative distribution functions), and $c(\cdot)$ is
a circula. 
%% Note the normalization term $(2\pi)^d$ due to circulae having uniform marginals on $[0, 2\pi)$.  %% as in (\ref{eq:multivar-circula}). 
Circulas'  uniform marginals on $[0, 2\pi)$ require the normalization term $(2\pi)^d$.

The CBMD construction can be used to generate various distributions by
selecting different distributions as marginals/circula. In the
following, we use the notation f-c to denote instances of CBMDs in
which f specifies the marginal distributions and c specifies the
circula.

\fi

\ifLONG

\begin{figure}[htbp]
\centerline{\includegraphics[width=.4\linewidth]{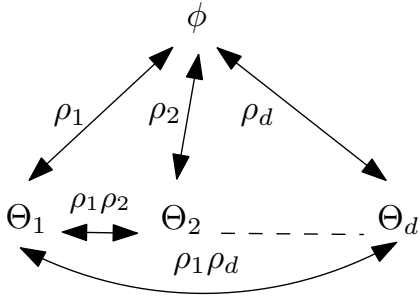}}
\caption{{\bf Circula as a latent variable
    model~\cite{jones2015class}.} Arrows
    represent modeled correlations.
$\rho$ is the length of the the mean resultant vector, which acts as a
  concentration parameter~\cite{mardia2009directional}.
The product $\rho_k\rho_l$ is the dependence value for the pair $(k,l)$.
}
\label{fig:latent-variable-model}
\end{figure}

\else

\begin{figure}[htbp]
\centering
\begin{minipage}{0.35\linewidth}
  \includegraphics[width=\linewidth]{\wdir/latent-variable-model.pdf}
\end{minipage}
\hfill
\begin{minipage}{0.5\linewidth}
  \caption{{\bf Circula as a latent variable      model~\cite{jones2015class}.}  Arrows
    represent modeled correlations.
  $\rho$ is the length of the mean resultant vector, which acts as a
    concentration parameter~\cite{mardia2009directional}.
  The product $\rho_k\rho_l$ is the dependence value for the pair $(k,l)$.}
  \label{fig:latent-variable-model}
\end{minipage}
\end{figure}

\fi

\section{Circula-based multivariate distributions: generic model}
\label{sec:vMwC}
%%i%%%%%%%%%%%%%%%%%%%%%%%%%%%%%%%%%%%%%%%%%%%%%%%%%%%%%%%%%%%%%%%%%%%%%%%%%%%%%%%

We first analyze the dependency structure of circula-based multivariate
distributions, and proceed with practical methods for sampling and
parameter estimation of CBMDs.

\subsection{Dependency structure analysis}
\label{seq:dependency-structure}
%%ii-%-%-%-%-%-%-%-%-%-%-%-%-%-%-%-%-%-%-%-%-%-%-%-%-%-%-%-%-%-%-%-%-%-%-%-%-%-%-%

\ifLONG
To provide insight on the dependency structures that can be capture
through CBMDs as in Eq.~\ref{eq:full-density}, we consider the
dependency measure of Jammalamadaka and Sarma
\cite[Sec. 8.2]{jammalamadaka2001topics} which we denote JS correlation
coefficient\footnote{Note that the correlation is positive iff 
$\theta_1\mu_1$ and $\theta_2-\mu_2$ have the same sign, that is, the
two angles $\theta_1$ and $\theta_2$ are on the direction with respect to their means}:
\else
With $\mu_i$ the circular mean of $\theta_i$, we characterize the dependency structure in \cbmds
(Eq.~\ref{eq:full-density}) using the so-called Jammalamadaka and
Sarma (JS) correlation coefficient~\cite[Sec. 8.2]{jammalamadaka2001topics}:
\fi 
\begin{equation}
\rho_{JS}(\theta_1, \theta_2) = \frac{\mathbb{E} [\sin(\theta_1 - \mu_1)\sin(\theta_2 - \mu_2)]}{\sqrt{\mathbb{E}[\sin^2(\theta_1 - \mu_1)] \mathbb{E}[\sin^2(\theta_1 - \mu_1)]}}
\end{equation}
Since $E[sin(\theta - \mu)] = 0$, the JS correlation corresponds to the
pearson correlation applied to variables $Z_i = \sin(\theta_i -
\mu_i)$.
%% , and thus all properties of the pearson correlation coefficient are retained. 
%%
%% JS correlation can be seen as an analogue of pearson correlation on
%% circular variables, in which $\sin(\theta - \mu)$ is taken as the
%% deviation of $\theta$ to the mean direction $\mu$. Indeed, because
%% $E[sin(\theta - \mu)] = 0$ then JS correlation corresponds to the
%% pearson correlation applied to variables $Z_i = \sin(\theta_i -
%% \mu_i)$, and thus all properties of the pearson correlation
%% coefficient are retained. 
%%
In a multivariate setting one may
consequently be interested in the JS correlation matrix, that is the
matrix $R_{JS}$ with typical element $[R_{JS}]_{i,j} =
\rho_{JS}(\theta_i, \theta_j)$.
\medskip

\ifLONG
We now characterize the space of JS correlation matrices spanned by
populations of circula-based multivariate distributions. In the 2D
case, Jones et al.~\cite{jones2015class} underline the role of the
concentration $\rho_g$ of the binding density $g$ as a dependence
parameter in the circula model, such that $\rho_{JS}(\theta_1,
\theta_2) = q \rho_g$. In the multivariate extension, Jones et
al.~\cite{jones2015class} also show that, under the condition the
binding densities $g_i$ are symmetric about 0 with mean resultant
length $\rho_i$, the mean resultant length of the $(k,l)$th marginal
copula density is equal to $\rho_k\rho_l$ and acts as dependence
parameter for variables $(\theta_k, \theta_l)$. Therefore, the JS
correlation matrix $R_c$ of populations of such a circula
reads as
\else
Recall that the dependence parameter for the pair $(k,l)$ is
the sign-less number $\rho_{kl} = \rho_k\rho_l$, with $\rho_{kl}\in [0,1]$ -- Eq.~\ref{eq:rhokl-dependence}.
Therefore, taking into account the sign of correlations as encoded by
the $q_i =\pm 1$ values, the JS correlation matrix $R_c(\{q_i, \rho_i\})$ of populations of
such a circula reads as 
 \fi
\begin{equation}
\label{eq:Rc}
R_c(\{q_i, \rho_i\}) = 
\begin{bmatrix}
1  & q_1  \rho_1 q_2 \rho_2 & \cdots & q_1  \rho_1 q_d \rho_d \\
q_2  \rho_2 q_1 \rho_1 & 1 & \cdots & q_2  \rho_2 q_d \rho_d \\
\vdots & \vdots & \ddots & \vdots \\
q_d  \rho_d q_1 \rho_1 & q_d  \rho_d q_2 \rho_2 & \cdots & 1
\end{bmatrix}
\end{equation}
Let $w$ be a vector with element $w_i = q_i\rho_i$ with $w_i\in [-1,1]$, $R_c$ can also be
expressed as
\begin{equation}
\label{eq:Rc-structure}
R_c(\{q_i, \rho_i\}) =  G(w) := w\latrans{w} - \ladiag{w\latrans{w}} + \idmatrix{d}.
\end{equation}
The previous equation shows that $R_c$ exhibits a {\em rank-one
  off-diagonal} structure, namely a rank-one matrix with its diagonal
replaced by ones. This stems from the construction of the multivariate
circula as a latent variable model: the dependency structure is
determined by a single explanatory factor (the latent variable), with
only $d$ degrees of freedom.

\subsection{Marginal and conditional distributions}
%%ii-%-%-%-%-%-%-%-%-%-%-%-%-%-%-%-%-%-%-%-%-%-%-%-%-%-%-%-%-%-%-%-%-%-%-%-%-%-%-%

A useful property of CBMDs, arising from their construction as a
one-factor latent variable model, is that they are closed under
marginalization. That is, the marginal distribution of a CBMD over a
subset of marginal variables is itself a CBMD.
One has:
\begin{theorem}
\label{thm:conditioning}
Let $h(\theta)$ be a CBMD.  and $u_i = 2\pi F_i(\theta_i)$.  The
marginal distribution of $h(\theta)$ with respect to the subset of
marginal variables $\theta_A$ satisfies:
\begin{equation} 
h(\theta_A) = (2\pi)^{|\theta_A|  -1}  \prod_{i \in A} f_i(\theta_i) \int_0^{2\pi} \prod_{i\in A} g_i(u_i - q_i\phi)d\phi.
\end{equation}
Let $\theta_A$ (resp. $\theta_B$) denote conditioned
(resp. conditioning) variables, then the conditional distribution
$h(\theta_A | \theta_B)$ is simply expressed as
\begin{equation}
h(\theta_A | \theta_B) = \frac{h(\theta_A, \theta_B)}{h(\theta_B)} = (2\pi)^{|\theta_A|} \prod_{i\in A}f_i(\theta_i) \frac{ c(u_A, u_B) }{c(u_B)}
\end{equation}
with $c(u_B)$ the marginal distribution of the circula $c(\cdot)$ over variables $u_B$:
\begin{equation}
c(u_B) = \frac{1}{2\pi} \int_0^{2\pi} \prod_{i \in B} g_i(u_i - q_i\phi)d\phi
\end{equation}
\end{theorem}

\subsection{Sampling}
\label{sec:sampling}
%%ii-%-%-%-%-%-%-%-%-%-%-%-%-%-%-%-%-%-%-%-%-%-%-%-%-%-%-%-%-%-%-%-%-%-%-%-%-%-%-%

The simplest method for generating values from a CBMD is to first sample the latent variable, and
then sample the remaining variables conditioned by the latent
variable, that is:
\ifLONG
\begin{itemizep}
\label{alg:sampling}
\item Sample $\phi \sim U[0,2\pi)$
\item For each $i=1,...,d$, sample $\omega_i \sim g_i$ and compute $u_i = (\omega_i + q_i\phi)(\bmod 2\pi)$
\item For each $i=1,...,d$, compute $\theta_i = F_i^{-1}(u_i/2\pi)$
\end{itemizep}
Sampling from the circular uniform density is straightforward.
\else
(i)  Sample $\phi \sim U[0,2\pi)$;
(ii) For each $i=1,...,d$, sample $\omega_i \sim g_i$ and compute $u_i = (\omega_i + q_i\phi)(\bmod 2\pi)$;
(iii) For each $i=1,...,d$, compute $\theta_i = F_i^{-1}(u_i/2\pi)$.
\fi 
Sampling from the binding densities $g_i$ and evaluating the
quantile functions $F_i^{-1}$ depends on their analytical form.
See  Sec. \ref{sec:si:von-mises-background}
and Sec.~\ref{sec:si:wrapped-cauchy-background} for the von Mises and
wrapped Cauchy distributions.

\begin{comment} Difficulties may arise when evaluating the von Mises quantile function, as it does not admit a closed-form expressions. 
However its evaluation is computed efficiently by means of the Newton-raphson method, which iterative updates are of the form:

\begin{align*}
\theta_{i+1} = \theta_i - \frac{VM_i(\theta_i) - u_i}{\vmoned[i]{\theta_i}}
\end{align*}
\end{comment}

\ifLONG
\begin{remark}
In some applications, sampling from conditional distributions can be
of interest. For a CBMD,
the process is only slightly more complicated, and involves sampling
the latent variable in accordance with the conditioning variables. Let
$\theta_b$ denote the subset of conditioning variables and let $u_i = F_i(\theta_i)$. Then the conditional
density of $\phi$ becomes
\begin{align*}
c(\phi | u_b) = \frac{c(u_b | \phi)c(\phi)}{c(u_b)} = \frac{\prod_{i \in b} g_i(u_i - q_i \phi)}{2\pi c(u_b)}
\end{align*}
where $c(u_b)$ is the marginal distribution of the circula $c(\cdot)$
over variables $u_b$.  Sampling from this distribution is best done
with Monte-Carlo methods. The remaining variables $\theta_a = \theta
\setminus \theta_b$ are then generated as described above.
\end{remark}
\fi

\subsection{Parameter estimation}
\label{sec:parameter-estimation}
%%ii-%-%-%-%-%-%-%-%-%-%-%-%-%-%-%-%-%-%-%-%-%-%-%-%-%-%-%-%-%-%-%-%-%-%-%-%-%-%-%

CBMDs requires estimating two types of parameters: the
location and concentration parameters of the marginal and binding densities, and the
coupling indicator variables $q_i$ in Eq.~\ref{eq:multivar-circula}.
For a dataset of $n$ observations $\Tddatum$, the log likelihood of density (\ref{eq:full-density}) is equal to :
\begin{equation*}
\resizebox{\columnwidth}{!}{$\displaystyle
\log(h(\Tddatum)
= 
n d \log(2\pi) 
+  \sum_{i=1}^n \log(f_1(\Tdptij{i1}))
+ \dots
+ \sum_{i=1}^n  \log(f_d(\Tdptij{id}))
+ \sum_{i=1}^n \log(c(2\pi F_1(\Tdptij{i1}), ..., 2\pi F_d(\Tdptij{id} )))
$}
\end{equation*}

There is generally no closed form solution for the maximum likelihood
estimator (MLE) because the marginal summations are coupled to the circula
summation in terms of parameters~\cite[Section 4.1]{jones2015class}.
Therefore, numerical methods such as L-BFGS-B are used to estimate
continuous parameters.
In contrast, $q_i = \pm 1$ yields $2^d$ possibilities for the
coupling indicators, whence $2^d$ runs of  L-BFGS-B.
This prohibitive complexity for large values of $d$ is dealt with 
in Alg.~\ref{alg:estimate-circula} using approximation techniques.

\paramini{IFM method for continuous parameters inference.}
%% The problem of estimating the continuous parameters can be simplified
%% by estimating marginal parameters independently, followed by
%% estimating the circula parameters conditional on the marginal
%% parameter estimates. This approach was proposed as a computationally
%% efficient approximation for the bivariate von Mises MLE
%% \cite{shieh2005inferences}\footnote{Here we refer to the bVM variant
%% from Shieh \& Johnson later generalized by circula models, to be
%% distinguished from the more popular bVM \textit{sine} variant
%% \cite{singh2002} and related distributions.}.
%%
%%
%% More generally, for multivariate models on which marginals are
%% specified, this is known as the Inference Function for Margins (IFM)
%% method \cite{joe1996estimation}. Although IFM does not yield an exact
%% MLE, it provides accurate estimates in practice, while significantly
%% reducing the original optimization problem on $3d$ parameters to $d$
%% separate optimization problems on $2$ parameters (marginal parameters)
%% plus a single optimization problem on $d$ parameters
%% (circula). Moreover, estimation of the individual marginal parameters
%% can often be performed using specialized efficient methods -- for
%% example, von Mises parameters can be estimated as described in
%% appendix \ref{sec:si:vm-mle} -- rather than generic numerical methods.
%%
Originating with the fit of bivariate von Mises
MLE~\cite{shieh2005inferences}, the estimation of continuous
parameters can be be simplified by estimating the marginal parameters
independently, and then the circula parameters conditional on the
marginal parameter estimates.
\ifLONG
\begin{remark}
Here we refer to the bVM variant
from Shieh \& Johnson later generalized by circula models, to be
distinguished from the more popular bVM \textit{sine} variant
\cite{singh2002} and related distributions.
\end{remark}
\fi
This method is known as the Inference Function for Margins
(IFM)~\cite{joe1996estimation} for multivariate models on which
marginals are specified. IFM does not yield an exact
MLE but provides accurate estimates in practice.
\ifLONG
It reduces  the original optimization problem on $3d$ parameters to $d$
separate optimization problems on $2$ parameters (marginal parameters)
plus a single optimization problem on $d$ parameters
(circula). 
\fi
Moreover, estimation of the individual marginal parameters
can often be performed using specialized efficient methods -- for
example, von Mises parameters can be estimated as described in
appendix \ref{sec:si:vm-mle} -- rather than generic numerical methods.

\paramini{1-factor approximation for discrete parameters inference.}
\label{sec:1-factor-approx}
To avoid estimating the aforementioned $2^d$ models, we infer the best
dependency structure directly from the data  using the sample
estimate of the correlation matrix $\hat{R}$ (Eq.~\ref{eq:Rc}).
Note that the one factor structure (Eq.~\ref{eq:Rc-structure})
of correlation matrices $R_c$ of CBMDs prevents representing all correlation matrices.
The classical one-factor approximation is usually undertaken using the
so-called Higham correction~\cite{higham2002computing,borsdorf2010computing,duan2014generalized}.
We revisit this problem and provide a constructive algebraic
characterization for the rank-one case:
\begin{theorem}
\label{thm:rank-one-approx}
The best one factor approximation for the Frobenius norm of a $d\times d$ 
correlation matrix can be found by solving an algebraic system
involving $d+1$ equations of degree five in $d+1$ variables.
\end{theorem}
Practically, this algebraic system can be solved
using tools from real algebraic geometry--\eg Gr\"obner basis 
and the real univariate representation.
In practice and as suggested by our analysis which exhibits a
functional to be optimized involving a Rayleigh-Ritz ratio, we simply
re-scale the top eigenvector of the $\hat{R}$ (or $\hat{R}-\idmatrix$), 
and clip its entries to $[-1,1]$.
See Sec.~\ref{sec:rank-one-approx}, Alg. \ref{alg:estimate-circula}, Sec.~\ref{sec:circula-estimation}.
This 1-factor approximation yields estimates for the discrete sign
parameters $q$, and provides initial estimates for correlation
magnitude parameters $\rho_{\text{init}}$, which are then refined numerically.

%% We determine the set of $q$ values that enable
%% the population of the distribution to reproduce this matrix. This is
%% akin to moment matching estimation conceptually, although in this case
%% the procedure selects one model from $2^d$ discrete candidates.

\ifLONG
\begin{remark}
 Although heuristic, aligning  the $q$ estimates with the principal
direction displayed by the data correlation matrix does not provide
an exact MLE for $q$ in general.  Experiments assessing the practical
effectiveness of this heuristic are given in
Sec.~\ref{sec:parameter-estimation-results}.
\end{remark}
\fi

\section{Circula-based multivariate distributions: instantiations}
\label{sec:instantiations}
%%i%%%%%%%%%%%%%%%%%%%%%%%%%%%%%%%%%%%%%%%%%%%%%%%%%%%%%%%%%%%%%%%%%%%%%%%%%%%%%%%

\subsection{von Mises circula}
%%ii-%-%-%-%-%-%-%-%-%-%-%-%-%-%-%-%-%-%-%-%-%-%-%-%-%-%-%-%-%-%-%-%-%-%-%-%-%-%-%

The von Mises circula distribution is expressed as the
circula from Eq.~\ref{eq:multivar-circula} with centered von
Mises distributions ($\mu = 0$) as binding densities $g_i$, namely:
\begin{align*}
c(u_1, ..., u_d) &= \frac{1}{2\pi} \int_0^{2\pi} \prod_{i=1}^d \vmoned[i]{\theta_i - q_i\phi} d\phi
\end{align*}
The free parameters of this distribution are the concentrations
parameters $\kappa_i \in (0,\infty)$ of the von Mises binding
densities, and the discrete indicators $q_i =\pm 1$ for the sign 
of
correlations with the latent variable. Let $I_0(\cdot)$ denote
the modified Bessel function of the first kind and order
0~\cite{abramowitz1988handbook}. Plugging the expression of
$\vmoned{\cdot}$ from Eq.~\ref{eq:von-mises-1D} into the above formula
yields:

\begin{equation}
\label{eq:integrate_form_mises_circula}
c(u_1, ..., u_d) = \frac{1}{(2\pi)^{d+1}} \int_0^{2\pi} \prod_{i=1}^d \frac{\exp\{\kappa_i\cos(u_i - q_i\phi)\}}{I_0(\kappa_i)} d\phi
\end{equation}

\begin{theorem}
\label{thm:vMvM-closed-form}
The von Mises circula density of Eq.~\ref{eq:integrate_form_mises_circula} admits the following closed form expression:
\ifLONG
\begin{align*}
&c(u_1, ..., u_d) = \frac{I_0(R)}{(2\pi)^{d}\prod_{i=1}^d I_0(\kappa_i)},\\
&\\
&R = \sqrt{(\sum_{i=1}^d \kappa_i \cos(\theta_i))^2 + (\sum_{i=1}^d \kappa_i q_i \sin(\theta_i))^2}
\end{align*}
\else
\begin{equation}
c(u_1, ..., u_d) = \frac{I_0(R)}{(2\pi)^{d}\prod_{i=1}^d I_0(\kappa_i)},
\text{ with }
R = \sqrt{(\sum_{i=1}^d \kappa_i \cos(\theta_i))^2 + (\sum_{i=1}^d \kappa_i q_i \sin(\theta_i))^2}
\end{equation}
\end{theorem}
\fi

\subsection{Wrapped Cauchy circula}
%%ii-%-%-%-%-%-%-%-%-%-%-%-%-%-%-%-%-%-%-%-%-%-%-%-%-%-%-%-%-%-%-%-%-%-%-%-%-%-%-%

The wrapped Cauchy circula distribution is expressed as the circula from Eq.~\ref{eq:multivar-circula} with centered wrapped
Cauchy distributions ($\mu = 0$) as binding densities $g_i$. This is
equal to
\begin{align*}
c(u_1, ..., u_d) &= \frac{1}{2\pi} \int_0^{2\pi} \prod_{i=1}^d \wconed[i]{\theta_i - q_i\phi} d\phi.
\end{align*}
The free parameters of this model are concentration parameters $\rho_i
\in (0,1)$ of the wrapped Cauchy binding densities, and again the
discrete indicators variables $q_i = \pm 1$. Plugging the expression of $\wconed{\cdot}$ from Eq.~\ref{eq:wrapped-cauchy-1D} into the above formula yields

\begin{equation}
\label{eq:integrate_form_cauchy_circula}
c(u_1, ..., u_d) = \frac{1}{(2\pi)^{d+1}} \int_0^{2\pi} \prod_{i=1}^d \frac{1-\rho_i^2}{1+\rho_i^2 - 2\rho_i\cos(u_i-q_i\phi)} d\phi
\end{equation}

\begin{theorem}
\label{thm:vMwC-closed-form}
Under the condition $\rho_k \neq \rho_l, (k\neq l)$, the wrapped Cauchy circula density of Eq.~\ref{eq:integrate_form_cauchy_circula} admits the following closed form expression
\begin{equation}
\label{eq:vMwC-closed-form}
c(u_1, ..., u_d) = \frac{1}{(2\pi)^d} \sum_{j=1}^d (\rho_je^{iq_ju_j})^{d-1} \prod_{\substack{k=1 \\ k\neq j}}^d \frac{1-\rho_k^2}{(\rho_j e^{iq_ju_j}-\rho_k e^{iq_ku_k})(1-\rho_j\rho_ke^{i(q_ju_j - q_ku_k)})}
\end{equation}
\end{theorem}

\ifLONG
\begin{remark}
In cases when some $\rho$ parameters are equal, the expression obtained differs from Eq.~\ref{eq:vMwC-closed-form} and its computation becomes significantly more expensive, although theoretically tractable. In practice, it is simpler to apply small perturbations to the $\rho$ parameters in order to ensure the condition for Eq.~\ref{eq:vMwC-closed-form}.
\end{remark}

\begin{remark}
It is important to note that in practice, sampling from wrapped Cauchy is very efficient, comparable to sampling from the uniform distribution. As such, the cost of sampling from CBMD constructed using a wrapped Cauchy circula using the method described in Sec.~\ref{sec:sampling} is driven by the evaluation of marginal quantile functions.
\end{remark}
\fi

\begin{comment}
From the perspective of practical utility, we
then compare four resulting instances of CBMDs induced by these
multivariate circulae : the von Mises-von Mises (\vMvM), the von
Mises-wrapped Cauchy (\vMwC), the wrapped Cauchy-wrapped Cauchy
(\wCwC) and the wrapped Cauchy-von Mises (\wCvM).
\end{comment}

\subsection{Instantiations}
The circula framework decouples the marginals and the circula itself.
We provide two instantiations of Eq. (\ref{eq:multivar-circula}), namely the von Mises circula and the
wrapped Cauchy circula. 
Using von Mises and wrapped Cauchy distributions as marginals yields the following four instantiations:
\begin{itemize}
\item \cbmds with von Mises marginals: \vMwC and \vMvM;
\item \cbmds with wrapped Cauchy marginals: \wCwC and \wCvM.
\end{itemize}
A comparison of these instantiations is provided in Section~\ref{sec:shape-analysis}. 
In the sequel, we focus on \vMwC and \vMvM which use von
Mises marginals, due to their Gaussian-like behavior on the circle.

\section{Mixture modeling with circulae}
%%i%%%%%%%%%%%%%%%%%%%%%%%%%%%%%%%%%%%%%%%%%%%%%%%%%%%%%%%%%%%%%%%%%%%%%%%%%%%%%%%

\ifLONG
\subsection{Circula mixture models (CMM): definition and fitting}
%%ii-%-%-%-%-%-%-%-%-%-%-%-%-%-%-%-%-%-%-%-%-%-%-%-%-%-%-%-%-%-%-%-%-%-%-%-%-%-%-%
\fi

\paramini{Mixture.}
Given a component distribution $f_k$ the density of a mixture model is expressed as
\begin{align*}
p(x) = \sum_{k=1}^K w_k f_k(x), \quad \sum_{k=1}^K w_k = 1.
\end{align*}
Estimating a mixture model requires estimating the parameters of each
component, as well as the component weights $w_k$, a task usually
undertaken using the Expectation-Maximization (EM)
algorithm~\cite{mclachlan2008algorithm}.
Inference of CMMs via the EM algorithm is enabled by the
tractable likelihood maximization of each CBMD -- Sec.~\ref{sec:parameter-estimation}. 

\paramini{Fitting.}
As detailed in Sec.~\ref{sec:fitting-mixtures}, fitting a mixture consists of two steps:
\ifLONG
\begin{itemize}
\item Initialization using  \kmeanspp[black] on the flat torus~\cite{carriere2026clustering};
\item Unsupervised mixture learning using the MML-EM algorithm~\cite{figueiredo2002unsupervised}.
\end{itemize}
\else
(i)  Initialization using  \kmeanspp[black] on the flat torus;
(ii) Unsupervised mixture learning using the MML-EM algorithm~\cite{figueiredo2002unsupervised}.
\fi
We note in passing that the learning method from
\cite{figueiredo2002unsupervised} combines EM and the {\em Minimum
  Message Length} (MML), instead of BIC or AIC to promote
sparsity. MML provides the 
length of the shortest code (message) needed to describe both the model and the data, and 
has the units of \textit{bits} -- see also Rmk \ref{rmk:mml-length}.
This is important in our mixture fitting setting for several
reasons:
(i) MML does not rely on parameter identifiability (permutating the
labels of components), as opposed to BIC and AIC; (ii) MML uses an
adaptive penalty, which is important to penalize components poorly
supported by data; and (iii) MML accounts for the precision with
which parameters are stated (instead of treating them as point
estimates as in BIC/AIC). 
%%

\begin{comment}

\subsection{Comparing two models: sparsity and the compensation metric}

\label{sec:compensation-metric}
When comparing two models $\theta_A$ and $\theta_B$ of the same
dataset $X$, the one achieving the shorter Message Length is
preferred. However, as Message Length scaling  with the dataset size,
its significance is hard to quantify. 
%%
Let $N_A$ be the number of parameters per component of $\theta_A$ in
the message length formula.  We define the {\em compensation metric}
of $\theta_A$ w.r.t.  $\theta_B$ as a dataset-size invariant
alternative, defined as the real number $\alpha$ solution of
\begin{equation}
\label{eq:compensation-metric}
\text{Length}(\theta_A, X) |_{N_A \gets \alpha * N_A} = \text{Length}(\theta_B, X).
\end{equation}
  Intuitively, $\alpha$ is
the factor by which one must scale the per-component complexity of
$\theta_A$ to match the Message Length of $\theta_B$. A value $\alpha
>> 1$ indicates that $\theta_A$ is robustly superior, while $\alpha <
1$ indicates that $\theta_B$ is in fact the better model.
\end{comment}

\section{Experiments and results}
\label{sec:experiments-results}
%%i%%%%%%%%%%%%%%%%%%%%%%%%%%%%%%%%%%%%%%%%%%%%%%%%%%%%%%%%%%%%%%%%%%%%%%%%%%%%%%%

\subsection{Dataset}
%%ii-%-%-%-%-%-%-%-%-%-%-%-%-%-%-%-%-%-%-%-%-%-%-%-%-%-%-%-%-%-%-%-%-%-%-%-%-%-%-%

Protocol and data availability are discussed in Sec.~\ref{sec:protocal-availability}.

The torsion angle values of individual amino acid / amino acid pairs
were collected from the PDB50 collection, used to train \phisical~\cite{amarasinghe2023getting}.
This dataset includes 38895 structures with non-redundant amino
acid sequences from the Protein Data Bank \cite{berman2000protein},
and contains only structures with an R-factor cut-off at 0.3 and
resolution cut-off at 3.5\ang or better.
For individual amino acids, PDB50 yields $N=22,177,093$
observations $\{\phi, \psi, \chi_1,\dots,\chi_d\}$~\cite{amarasinghe2023getting},
ascribed to 20 individual dataset--one per a.a. type
(NB: smallest one: CYS, $296,547$ observations; largest one: LEU,
$2,171,630$ observations.)
\ifLONG
These observations are publicly available at \url{https://lcb.infotech.monash.edu.au/phisical/}.
(NB: $\omega$ angle excluded since it is almost always $\sim
180$\textdegree.)
\fi
For amino acid pairs, we consider the $20*20=400$ pairs of (ordered)
a.a. types with the first (resp. second) a.a. on the N-ter
(resp. C-ter) side.  A given pair is denoted XXX-YYY.
The PDB50 collection yields $22,191,566$ observations, the most
abundant pair being LEU-LEU ($200,982$ observations), and the least
one TRP-CYS ($4458$ observations).

\subsection{Circula mixture models for torsion angles of single amino acids}
%%ii-%-%-%-%-%-%-%-%-%-%-%-%-%-%-%-%-%-%-%-%-%-%-%-%-%-%-%-%-%-%-%-%-%-%-%-%-%-%-%

\begin{comment}

mean bit gain VMVM
Model=3558.01, Data=-22422.87, Total=-18864.86

mean bit gain VMWC
Model=3576.85, Data=-24083.71, Total=-20506.86

total bit gain (min, mean, median, max)
VMVM : -92804.59, -18864.86, -8055.7, -985.95
VMWC : -91499.52, -20506.86, -8142.68, -1369.13

PER OBSERVATION : 

mean bit gain VMVM
Model=0.003, Data=-0.020, Total=-0.017

mean bit gain VMWC
Model=0.003, Data=-0.022, Total=-0.018

total bit gain (min, mean, median, max)
VMVM : -0.084, -0.017, -0.007, -0.001
VMWC : -0.083, -0.018, -0.007, -0.001
\end{comment}

\begin{comment}
The \phisical model provides a mixture for all torsion angles of each
a.a. type~\cite{amarasinghe2023getting}, the mixture component being a
product of univariate von Mises. This corresponds to a particular case of \cbmd where
binding densities are circular uniform distributions.
%%
The \phisical mixture is fitted by combining an exhaustive search of
the mixture space using split/merge/delete
operations~\cite{kasarapu2015minimum}, and the model selection uses
MML with specialized parameter priors--rather than generic Jeffrey
priors.
\end{comment}
%%

The \phisical model--a particular case of \cbmd where  binding densities are circular uniform distributions,
provides a mixture for all torsion angles of each
a.a. type~\cite{amarasinghe2023getting}.
The mixture space is thoroughly explored using split/merge/delete
operations between components~\cite{kasarapu2015minimum}, and the
model selection uses MML with specialized parameter priors--rather
than generic Jeffrey priors.

To assess the usefulness of our CBMDs, we add circulae to the the
components from \phisical and re-estimate the full set of parameters.
We do so using the MML-EM algorithm~\cite{figueiredo2002unsupervised}
launched with $k_{\textrm{max}}$ equal to the initial number of
components, and also limiting the maximal concentration of von Mises
distributions to $700$--as for
\phisical. Tab.~\ref{tab:phisical-results} reports the resulting
Message Lengths obtained by using either von Mises circula or wrapped
Cauchy circula. The message length difference in bits per observation yields the
following min median values: (min=-0.084, mean=-0.017, median=-0.007,
max=-0.001) for \vMvM; (min=-0.083, mean=-0.018, median=-0.007,
max=-0.001) for \vMwC. 
The following observations are made :

\sbulem All mixtures kept the same number of components.  We
conjecture this is because the \phisical models are already well
optimized in terms of component placement and concentrations, under
the assumption of no component covariance. Incorporating covariance
thus reduces message length, but does not warrant component
annihilation.\\
\sbulem Mixtures with \vMwC components yields slightly better results
than those using \vMvM components, consistent with their more practical
unimodality -- see Sec.~\ref{sec:shape-analysis}. The modest difference owes to similar shapes in the
presence of moderate correlation--which is the case on this dataset.\\
\sbulem Our circula-based models consistently outperforms \phisical in
terms of message length, confirming that the  gain in likelihood
brought by circulae outweighs the cost of the additional parameters.

\subsection{Circula mixture models for torsion angles of pairs of amino acids}
%%ii-%-%-%-%-%-%-%-%-%-%-%-%-%-%-%-%-%-%-%-%-%-%-%-%-%-%-%-%-%-%-%-%-%-%-%-%-%-%-%

%% On amino acid pair data, we estimate models at two levels of
%% specificity : (i) models that operates solely on the backbone torsion
%% angles -- the precise cross-landscape as introduced in
%% \cite{rosenberg2023amino} -- and (ii) residue-specific models for each
%% of the 400 amino acid pair combinations, with side chain angles
%% included. In both cases, we use only vM-wC components, following the
%% observation in sec. that these provided slightly better performance on
%% single amino acid data.

In the sequel, we refer to all torsion angles of two consecutive
a.a. as the {\em cross landscape}. (NB: in \cite{rosenberg2023amino},
the term simply refers to pairs $(\phi_i, \psi_{i+1})$.)  For the sake
of conciseness, the three Ramachandran diagrams associated to two
consecutive a..a, namely $(\phi_1,\psi_1)$, $(\psi_1, \phi_2)$ (cross
landscape), and $(\phi_2,\psi_2)$, is termed the {\em Ramachandran
  trinity}.
We define two types of mixtures on the cross landscape: one for the backbone--regardless
of the two a.a. types; and one for each of the $20\times 20$ oriented pairs.
In both cases, we use CMM with \vMwC components which proved slightly more
efficient on single amino acid data.

\paramini{Backbone cross landscape model.}  We compare a CMM with our
\vMwC components against a baseline whose components are those of
\phisical. Both mixtures are estimated using the MML-EM algorithm with
$k_{\textrm{max}}$ set to $600$. Given the large volume of data
available (20M+ observations), we perform each MML-EM iteration on a
randomly drawn batch of 1\% of the data, and estimate Message Length
on the full dataset by scaling the batch log-likelihood accordingly.
The baseline and CMM models achieve Message Lengths of
$67,562,262$ bits (with $k=575$) and $67,070,004$ bits (with $k=563$), combining model lengths of $35,479$ bits and $49,137$ bits, and data lengths of $67,526,784$ bits and $67,020,868$ bits, respectively.
The $505,916$ bit reduction in data length far outweighs the $13,658$ bit increase in model complexity confirming that — as one might 
expect — a heavier, more expressive model becomes increasingly advantageous 
as the dataset grows larger.

%yielding a  compensation metric of  $\alpha=118.75$. While the raw message 
%lengths indicate the improvement over the baseline is consistent with the 
%other experiments, the compensation metric confirms that — as one might 
%expect — a heavier, more expressive model becomes increasingly advantageous 
%as the dataset grows larger.

\begin{figure}[ht]
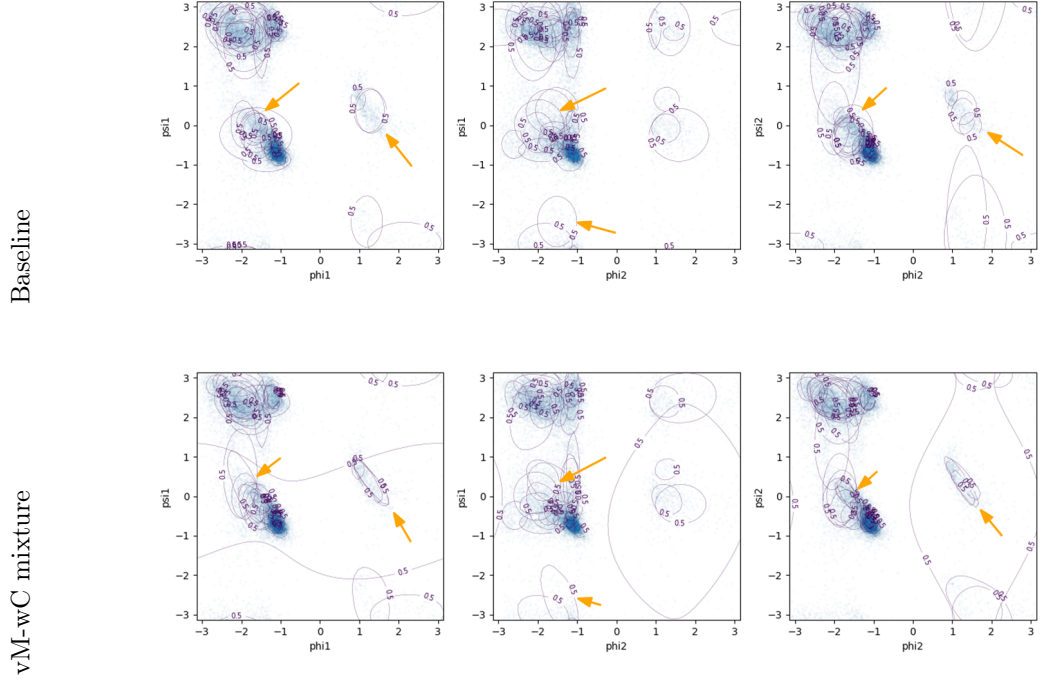

\begin{center}
\begin{tabular}{cc}
%\rotatebox{90}{Baseline} &      \includegraphics[width=0.8\linewidth]{\wdir/vmm_backbone.png}\\
%\rotatebox{90}{\vMwC mixture} & \includegraphics[width=0.8\linewidth]{\wdir/vmwcm_backbone.png}
\rotatebox{90}{Baseline} &      \includegraphics[width=0.9\linewidth]{\wdir/backbone-trinity-vmm-baseline-montage.png}\\
\rotatebox{90}{\vMwC mixture} & \includegraphics[width=0.9\linewidth]{\wdir/backbone-trinity-vMwC-montage.png}
\end{tabular}
\end{center}
\caption{{\bf The Ramachandran trinity for a  mixture with 50 components: credible region contours
at $50\%$ level  for  each unweighted component of the baseline and \vMwC mixture.}
{\small For the latter, orange arrow indicate slanted components due to the covariance structure.}
}
\label{fig:k50-per-component-credible}
\end{figure}

\begin{figure}[htbp]
\centerline{\includegraphics[width=\linewidth]{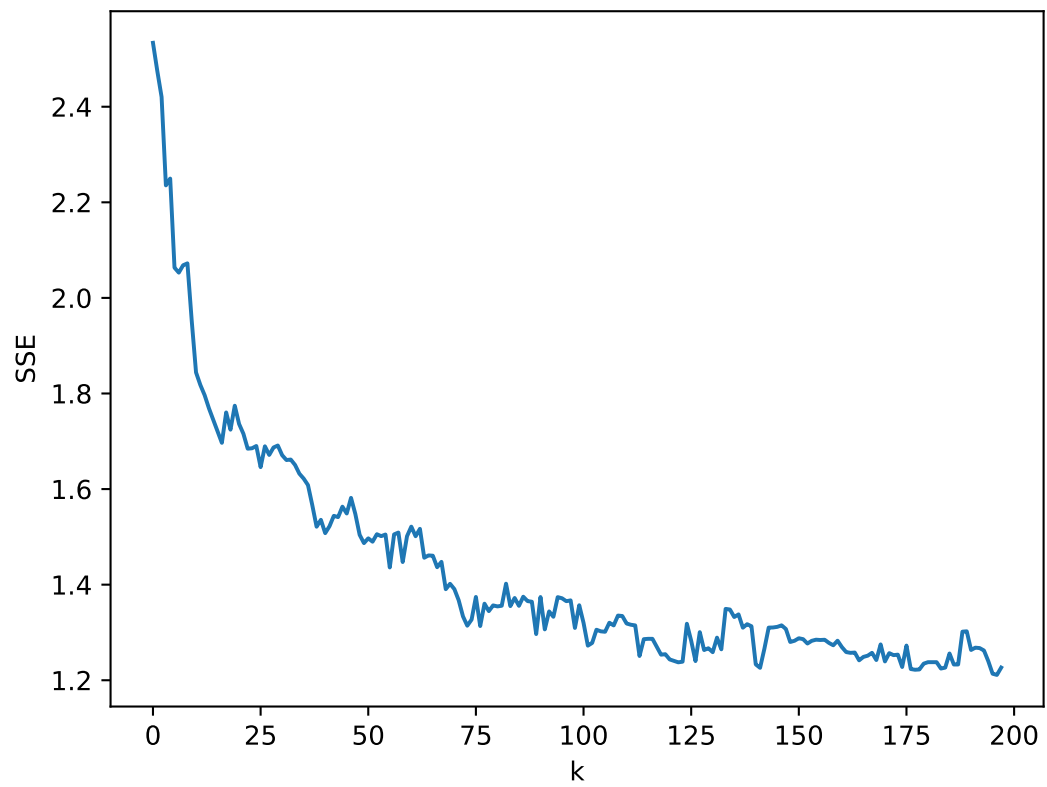}}
\caption{{\bf Sum of square error (SSE) from k-means clustering of cross-landscape data, for varying value of k} Additional cluster bring diminishing reductions of SSE around k=50}
\label{fig:scree-plot}
\end{figure}

\begin{figure}[ht]
\begin{center}
\begin{tabular}{cc}
\rotatebox{90}{Baseline}      & \includegraphics[width=0.9\linewidth]{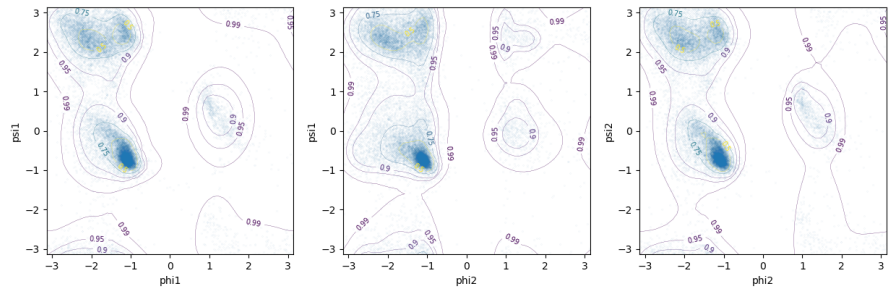}\\
\rotatebox{90}{\vMwC mixture} &  \includegraphics[width=0.9\linewidth]{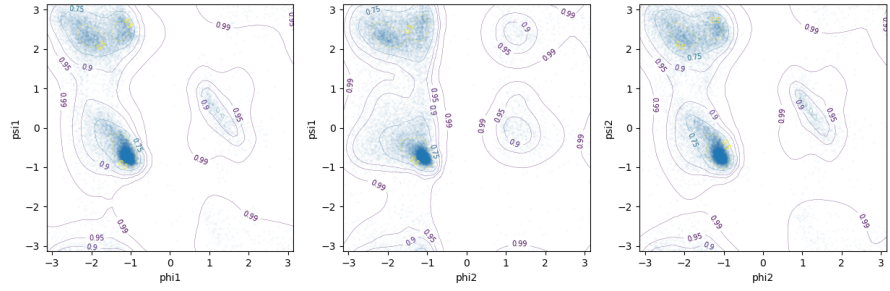}
\end{tabular}
\end{center}
\caption{ {\bf The Ramachandran trinity for a  mixture with 50 components: contour plots of the overall density.} Contours at thresholds $\{99\%, 95\%, 90\%, 75\%, 50\%\}$ levels.}
\label{fig:k50-full-density-contours}
\end{figure}

While our reference model for practical use remains the MML estimated
mixture presented above, we also estimate a simpler model for
comparison with the cross-landscape analysis of
\cite{rosenberg2023amino}, in which k-means clustering with $k=20$ is
applied to the data. Our scree plot analysis (Fig.~\ref{fig:scree-plot}) suggest that $\sim k=50$ clusters is more
appropriate, beyond which additional clusters brings diminishing
returns in SSE reduction.  Consequently, we estimate baseline and CMM
mixture models using the classical EM algorithm with $k=50$
(Fig.~\ref{fig:k50-per-component-credible},
Fig.~\ref{fig:k50-full-density-contours}).
Direct comparison illustrates how modeling covariance yields tighter
densities around the mass in correlated data. The baseline and CMM
models achieve log-likelihoods of $-50,738,249$ and $-50,065,195$ respectively. 

%A more thorough analysis of the
%insights from \cite{rosenberg2023amino} through our statistical models
%is provided in Sec.\ref{sec:cross-landscape-analysis}.

\begin{comment}
AA1 median extremes: (min, mean, median, max)
Min median for aa1: ('CYS', '0.50', '1.20', '1.27', '1.81')
Max median for aa1: ('PRO', '1.67', '2.08', '2.02', '2.74')

AA2 median extremes: (min, mean, median, max)
Min median for aa2: ('TRP', '0.07', '1.22', '1.28', '1.96')
Max median for aa2: ('PRO', '1.29', '1.97', '1.96', '2.80')

Number of values < 1: 18 out of 400

\paramini{Pair specific cross landscape models.}
We also compute the compensation metric for the 400 CMMs with  \vMwC  components (Fig.~\ref{fig:aa-pairs-results-matrix}).
%%
The compensation metric is $>1$ in 382 cases out of 400,
with   (min=0.07, mean=1.66, median=1.67, max=2.80).
%%
The min median values are obtained for cystein for the first a.a. (Min
median for a.a. one: (CYS, min=0.50, mean=1.20, median=1.27, max=1.81), and for
tryptophan for the second a.a.  (Min median for a.a. two: (TRP,
min=0.07, mean=1.22, median=1.28, max=1.96).  

These results indicate that the vM-wC CMM is worthwhile even of smaller datasets. The CMM may become too expensive
when dimensionality is high relative to dataset size, as exemplified
by the worst observed score of 0.07 on MET-TRP (5,020 observations,
$d=9$).
\end{comment}

\begin{figure}[htbp]
\centerline{\includegraphics[width=.6\linewidth]{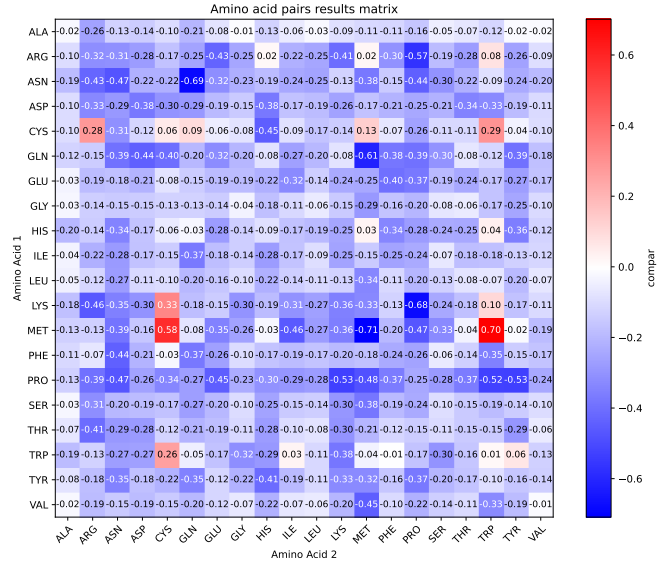}}
\caption{{\bf Difference in bits per observation between the baseline and \vMwC mixtures, for the 
$20\times 20$ pairs of amino acids.}
$(
\MMLlength{\theta_{\vMwC}}{X}-
\MMLlength{\theta_{\text{baseline}}}{X})/\size{X}$}
\label{fig:aa-pairs-results-matrix}
\end{figure}

\paramini{Pair specific cross landscape models.}
Fig.~\ref{fig:aa-pairs-results-matrix} shows the difference in bits per observation for the 400 CMMs with  \vMwC  components.
The bit difference is negative in 382 cases out of 400,
with (min=-0.71, mean=-0.19, median=-0.18, max=0.70).
The min median values are obtained for cystein for the first a.a. (Min
median for a.a. one: (CYS, min=-0.45, mean=-0.07, median=-0.09, max=0.29), and for
tryptophan for the second a.a.  (Min median for a.a. two: (TRP,
min=-0.38, mean=-0.13, median=-0.15, max=0.26).  
These results indicate that the vM-wC CMM is
worthwhile even on smaller datasets. The CMM may become too expensive
when dimensionality is high relative to dataset size, as exemplified
by the worse observed score of 0.70 bit gain per observation on MET-TRP (5,020 observations,
$d=9$).
The complete message lengths for all 400 datasets are published on an external webpage -- see Sec.~\ref{sec:protocal-availability}.

\section{Discussion and outlook}
\label{sec:outlook}
%%i%%%%%%%%%%%%%%%%%%%%%%%%%%%%%%%%%%%%%%%%%%%%%%%%%%%%%%%%%%%%%%%%%%%%%%%%%%%%%%%

Our work introduces a new type of normalized multivariate
distributions with covariance structure on the flat torus $\torusd$.
This unlocks mixture modeling with \vMwC / \vMvM components, which is
not practical using classical toroidal distributions,
as most of them are not normalized. To the best of our knowledge, the
only exception is the multivariate wrapped Normal distribution, used
in mixture modeling in \cite{greco2023finite}. However, von Mises and
wrapped Cauchy distributions are generally preferred to wrapped Normal
distribution, as they are more tractable and use a more practical
cosine based measure of distance, instead of the geodesic distance.
An appealing further extension would be expand the circula model,
using multiple latent variables and/or Vine copulae.

Application-wise, we focus on the problem of modeling joint
distributions of torsion angles in proteins (backbone+side chains) 
which live in $\torusd{2}$ to $\torusd{14}$.
On the one hand and as recalled in Introduction, these angles are
ubiquitous in structural modeling.  We show that our mixture models
are SOTA in two respect: they improve on previous models for single
amino acids, and for the first time, they enable modeling joint
distributions for pairs of amino acids.
On the other hand, these angles  play a pivotal role in the recent Nobel prize winning
program \alphafold, which solely focuses on structure prediction, via  a hybrid message passing
like architecture operating in latent and physical space.
Also combined with message passing like algorithms,
we  anticipate that our mixtures will play a fundamental role to make a stride
towards the efficient prediction of thermodynamics and kinetics, the overarching goals
in theoretical biophysics.

%% moving from structure prediction \cite{jumper2021highly} to thermodynamics \cite{kamisetty2008free}

%\clearpage
%\FloatBarrier

\clearpage

%%===Inclusion end for file cmm-main.tex

\noindent{\bf Acknowledgments.}  This work has been supported by the French government, through the projects
3IA C\^ote d’Azur Investments (ANR-23-IACL-0001), 
ANR project Innuendo (ANR-23-CE45-0019).

\bibliographystyle{unsrt}
%\bibliography{\wmybib/biogeom,\wmybib/mcs,\wmybib/abs,\wmybib/abs-sub,local.bib}
%%===Inclusion starts for file cmm.bbl

%%===Inclusion end for file cmm.bbl

\beginSI

\clearpage
%\input{cmm-background.tex}
%%===Inclusion starts for file cmm-background.tex
\section{Background}
%%ii-%-%-%-%-%-%-%-%-%-%-%-%-%-%-%-%-%-%-%-%-%-%-%-%-%-%-%-%-%-%-%-%-%-%-%-%-%-%-%

\subsection{von Mises distribution}
%%ii-%-%-%-%-%-%-%-%-%-%-%-%-%-%-%-%-%-%-%-%-%-%-%-%-%-%-%-%-%-%-%-%-%-%-%-%-%-%-%

\label{sec:si:von-mises-background}

The {\em von Mises distribution} $\vmoned{\mu, \kappa}$ (sometimes referred to as the {\em Circular Normal distribution}) is the most commonly used analogue of the normal distribution on the unit circle. The von Mises distribution density function is

\begin{equation}
\label{eq:von-mises-1D}
\vmoned{\theta; \mu, \kappa} = \frac{1}{2\pi I_0(\kappa)} \exp \left\{\kappa  \cos(\theta-\mu) \right\}
\end{equation}
where $\mu$ is the mean parameter, $\kappa$ is the concentration parameter and $I_0(\cdot)$ stands for the modified Bessel function
of the first kind and order 0~\cite{abramowitz1988handbook}.
The mean resultant length $\rho$ of the von Mises distribution is then given by $\rho = A(\kappa)$ where
\begin{equation}
A(\kappa) = I_1(\kappa) / I_0(\kappa)
\end{equation}
As it is defined on the circle, the cumulative distribution function of a von Mises distribution can only be expressed analytically by integrating the pdf between $\theta$ and a lower limit $\theta_0$, corresponding to our "frame of reference" on the circle.
As such, the cdf of a univariate von Mises marginal distribution is defined by~\cite{jammalamadaka2001topics}[2.2.4]:
\begin{align*}
\mathrm{VM}(\theta) = \int_{\theta_0}^{\theta} \vmoned{t} dt
\end{align*}
with 
\begin{align*}
\int \vmoned{t}dt = \frac{1}{2 \pi} \left( \theta + \frac{2}{I_0(\kappa)} \sum_{j=1}^{\infty} I_j(\kappa) \frac{\sin [j(\theta - \mu)]}{j} \right),
\end{align*}
where $I_j(\cdot)$ is modified bessel function of the first king and order $j$.

\paragraph{Sampling}

The von Mises distribution can be sampled using a standard algorithm by Best \& Fisher \cite{best1979efficient}, summarized in \cite[Section 3.5.4]{mardia2009directional}. This algorithms implements an accept-reject scheme using a wrapped Cauchy distribution as envelope, with achieves a high acceptance ratio for most values of $\kappa$. Due to its simple implementation and good performance, this algorithm remains the dominant approach in for most scientific libraries (numpy, scipy, R package \textit{CircStats}...)

\paragraph{Quantile function evaluation}
The von Mises quantile function $\mathrm{VM}^{-1}(q)$ does not admit a closed-form expression. 
However, it can be evaluated efficiently using standard root-finding schemes -- most commonly the Newton-Raphson method, whose iterative updates are of the form:

\begin{align*}
\theta_{i+1} = \theta_i - \frac{\mathrm{VM}(\theta_i) - q}{\vmoned{\theta_i}}
\end{align*}

\subsection{Wrapped Cauchy distribution}
%%ii-%-%-%-%-%-%-%-%-%-%-%-%-%-%-%-%-%-%-%-%-%-%-%-%-%-%-%-%-%-%-%-%-%-%-%-%-%-%-%
\label{sec:si:wrapped-cauchy-background}
The Cauchy distribution on the real line $C(\mu, \gamma)$ has density

\begin{equation}
C(x; \mu, \gamma) = \frac{1}{\pi} \frac{\gamma}{\gamma^2 + (x - \mu)^2}
\end{equation}

The corresponding {\em wrapped Cauchy distribution} $\wconed{\mu, \rho}$ is obtained by wrapping a Cauchy distribution on the circle with the map $x_w = x \mod 2 \pi$. Through its characteristic function \cite{mardia2009directional}, it is one of the rare wrapped distributions that yields a closed form expression

\begin{equation}
\label{eq:wrapped-cauchy-1D}
\wconed{\theta; \mu, \rho} = \sum_{k=-\infty}^{\infty} C(\theta + 2 k \pi; \mu, \gamma) = \frac{1}{2\pi} \frac{1-\rho^2}{1+ \rho^2 - 2 \rho \cos(\theta - \mu)}
\end{equation}

where $\rho$ is the mean resultant length, with $\rho = e^{-\gamma}$.

With the lower limit set at $\theta_0 = \mu$, the cumulative distribution function is expressed with the following closed form \cite[Section 3.5.7]{mardia2009directional}

\begin{align*}
WC(\theta) = \int_{\mu}^{\theta} \wconed{t} dt = \frac{1}{2\pi}\arccos \left( \frac{(1+\rho^2)\cos(\theta -\mu) - 2\rho}{1+\rho^2 - 2\rho\cos(\theta - \mu)}  \right)
\end{align*}

\paragraph{Sampling}

Sampling can be performed by generating a linear Cauchy random variable and wrapping it onto the circle. If $u \sim U[0,1]$ and as $\gamma = - \log(\rho)$, samples for the wrapped Cauchy distribution are then generated as 

\begin{align*}
\theta = \left(\mu + -\log(\rho) \tan \big(\pi  (u - \tfrac{1}{2})\big)\right) \mod 2\pi
\end{align*}

\paragraph{Quantile function evaluation}
The wrapped Cauchy quantile function is obtained by inverting the CDF, which yields 

\begin{align*}
WC^{-1}(q) = \begin{cases}
			r(q), & \  0 \leq q \leq \frac{1}{2}\\
            2\pi - r(1 - q), &\  \frac{1}{2} \leq q \leq 1
		 \end{cases} 
\end{align*}
where 
\begin{align*}
r(q) = 2 \arctan \left( \frac{1 - \rho}{1 + \rho} \tan(\pi q \right)
\end{align*}

\subsection{von Mises distribution MLE}
\label{sec:si:vm-mle}
%%ii-%-%-%-%-%-%-%-%-%-%-%-%-%-%-%-%-%-%-%-%-%-%-%-%-%-%-%-%-%-%-%-%-%-%-%-%-%-%-%

The log likelihood of an individual von Mises marginal is equal to 

\begin{align*}
\log(f(\mathbf{\theta})) &= n \log 2\pi + \kappa \sum_{i=1}^n \cos(\theta_i - \mu) - n\log I_0(\kappa)\\
&= n\{ \log 2\pi + \kappa \bar{R} \cos(\bar{\theta} - \mu) - \log I_0(\kappa)  \}
\end{align*}

with $\bar{\theta}$ the mean direction and $\bar{R}$ the mean resultant length.

\paragraph{$\mu$ estimator}
Since $\cos(x)$ is maximal at $x=0$ the MLE of $\mu$ is $\hat{\mu} = \bar{\theta}$.

\paragraph{$\kappa$ estimator}
Differentiating the log likelihood with respect to $\kappa$ results in 

\begin{align*}
\frac{\partial}{\partial \kappa} \log(f(\mathbf{\theta})) = n \{ \bar{R}\cos(\bar{\theta} - \mu) - \frac{I_1(\kappa)}{I_0(\kappa)} \}
\end{align*}

Substituting $\hat{\mu} = \bar{\theta}$, the MLE of $\kappa$ is the solution of

\begin{align*}
\frac{I_1(\kappa)}{I_0(\kappa)} - \bar{R} = 0
\end{align*}

This non linear equation is difficult to solve analytically but a good approximation can be obtained efficiently using the \textit{Sra's truncated newton approximation} method \cite{sra2012short}, as summarized in \cite[Section 2.2]{kasarapu2015minimum}.

%%===Inclusion end for file cmm-background.tex

\clearpage
%\input{cmm-theory-proofs.tex}
%%===Inclusion starts for file cmm-theory-proofs.tex

\section{Circula: theory}
\label{sec:si:theory}
%%i%%%%%%%%%%%%%%%%%%%%%%%%%%%%%%%%%%%%%%%%%%%%%%%%%%%%%%%%%%%%%%%%%%%%%%%%%%%%%%%

\subsection{Marginal and conditional distributions}
\label{sec:conditioning}
%%ii-%-%-%-%-%-%-%-%-%-%-%-%-%-%-%-%-%-%-%-%-%-%-%-%-%-%-%-%-%-%-%-%-%-%-%-%-%-%-%

\begin{proof-not-ams}{black}[Thm.~\ref{thm:conditioning}]
Let $h(\theta)$ be a CBMD, $\theta_A$ denote the marginal variables and $\theta_B = \theta \setminus \theta_A$ denote the remaining variables, the marginal distribution of $h(\theta)$ on $\theta_A$ is equal to

\begin{align*}
h(\theta_A) &= \int_{\theta_B} (2\pi)^d  c(2\pi F_1(\theta_1), ..., 2\pi F_d(\theta_d)) \left( \prod_{i=1}^d f_i(\theta_i) \right) d\theta_B  \\
&=(2\pi)^{d} \prod_{i \in A} f_i(\theta_i) \int_{\theta_B}  c(2\pi F_1(\theta_1), ..., 2\pi F_d(\theta_d)) \left( \prod_{i \in B} f_i(\theta_i) \right) d\theta_B 
\end{align*}

Let $u_i= 2\pi F(\theta_i)$, such that $du_i = 2\pi f_i(\theta_i)d\theta_i$. Plugging the density of the circula from Eq.~\ref{eq:multivar-circula} yields

\begin{align*}
h(\theta_A) &=(2\pi)^{d - |\theta_B|}  \prod_{i \in A} f_i(\theta_i) \int_{u_B} c(u_1, ..., u_d) du_B  \\
&=(2\pi)^{|\theta_A|}  \prod_{i \in A} f_i(\theta_i) \int_{u_B} \frac{1}{2\pi} \int_0^{2\pi}  \prod_{i=1}^d g_i(u_i - q_i\phi) d\phi du_B  \\
&=(2\pi)^{|\theta_A| - 1} \prod_{i \in A} f_i(\theta_i) \int_0^{2\pi} \left( \int_{u_B}  \prod_{i\in B} g_i(u_i - q_i\phi) du_B \right) \prod_{i\in A} g_i(u_i - q_i\phi) d\phi   \\
&=(2\pi)^{|\theta_A| - 1} \prod_{i \in A} f_i(\theta_i) \int_0^{2\pi} \left(\prod_{i\in B} \int_0^{2\pi} g_i(u_i - q_i\phi) du_i \right) \prod_{i\in A} g_i(u_i - q_i\phi) d\phi   \\
&=(2\pi)^{|\theta_A| - 1} \prod_{i \in A} f_i(\theta_i)  \int_0^{2\pi} \prod_{i\in A} g_i(u_i - q_i\phi) d\phi 
\end{align*}
\end{proof-not-ams}

\subsection{von Mises circula: closed form expression}
\label{sec:si:vMvM}
%%ii-%-%-%-%-%-%-%-%-%-%-%-%-%-%-%-%-%-%-%-%-%-%-%-%-%-%-%-%-%-%-%-%-%-%-%-%-%-%-%

\begin{proof-not-ams}{black}[Thm.~\ref{thm:vMvM-closed-form}]
Let the binding density $g$ be a von Mises parameterized by concentration $\kappa$ and mean direction $\mu=0$:

$$g(\theta) = \frac{\exp\{\kappa \cos (\theta)\}}{2\pi I_0(\kappa)}$$

Then the equation for the circula in integrate form is

\begin{align*}
c(\theta_1, ..., \theta_d) &= \frac{1}{2\pi} \int_0^{2\pi} \prod_{i=1}^d  \frac{\exp\{\kappa_i\cos(\theta_i-q_i \phi)\}}{2\pi I_0(\kappa_i)} d\phi \\
&= \frac{1}{(2\pi)^{d+1}\prod_{i=1}^d I_0(\kappa_i)} \int_0^{2\pi} \exp\left\{\sum_{i=1}^d \kappa_i\cos(\theta_i-q_i \phi)\right\} d\phi
\end{align*}

Focusing on the term in exponent we have

\begin{align*}
\sum_{i=1}^d \kappa_i\cos(\theta_i-q_i\phi) = \sum_{i=1}^d \kappa_i \left(\cos(\theta_i)\cos(q_i\phi) + \sin(\theta_i)\sin(q_i\phi)\right)
\end{align*}
Because $q_i \in \{-1,1\}$ this is equal to

\begin{align*}
\left(\sum_{i=1}^d \kappa_i \cos(\theta_i)\right)\cos(\phi) + \left(\sum_{i=1}^d \kappa_i q_i \sin(\theta_i)\right)\sin(\phi)
\end{align*}

Now recall the trigonometric identity

\begin{align*}
a\cos(x)+b\sin(x) = R\cos(x-\alpha)
\end{align*}

where $R=\sqrt{a^2+b^2}$ and $\alpha = \arctan(b/a)$

Set $a=\sum_{i=1}^d \kappa_i \cos(\theta_i)$ and $b=\sum_{i=1}^d \kappa_i q_i \sin(\theta_i)$ we have

\begin{align*}
&\left(\sum_{i=1}^d \kappa_i \cos(\theta_i)\right)\cos(\phi) + \left(\sum_{i=1}^d \kappa_i q_i \sin(\theta_i)\right)\sin(\phi) \\
=&a\cos(\phi) + b\sin(\phi) = R\cos(\phi-\alpha)
\end{align*}

we can now write the integrate form as

\begin{align*}
&\int_0^{2\pi} \exp\left\{\sum_{i=1}^d \kappa_i\cos(\theta_i-q_i \phi)\right\} d\phi\\
=&\int_0^{2\pi} \exp\{R\cos(\phi-\alpha)\} d\phi\\
=& 2\pi I_0(R)
\end{align*}

\end{proof-not-ams}

\subsection{wrapped Cauchy circula: closed form expression}
\label{sec:si:vMwC}
%%ii-%-%-%-%-%-%-%-%-%-%-%-%-%-%-%-%-%-%-%-%-%-%-%-%-%-%-%-%-%-%-%-%-%-%-%-%-%-%-%

The following proof of Thm.~\ref{thm:vMwC-closed-form} was claimed in
\cite{jones2015class} and detailed in \cite{kato2024pc}:
\begin{proof-not-ams}{black}[Thm.~\ref{thm:vMwC-closed-form}]
Let the binding density $g$ be a wrapped Cauchy parameterized by mean resultant length $\rho$ and mean direction $\mu=0$:

$$g(\theta) = \frac{1}{2\pi} \frac{1 - \rho^2}{1+\rho^2-2\rho\cos\theta}$$

Then the equation for the circula in integrate form is

$$c(\theta_1, ..., \theta_d) = \frac{1}{(2\pi)^{d+1}} \int_0^{2\pi} \prod_{k=1}^d \frac{1-\rho_k^2}{1+\rho_k^2 - 2\rho_k\cos(\theta_k-q_k\phi)} d\phi$$

Set $z=e^{i\phi}$ and $\eta_k=\rho_ke^{iq_k\theta_k}$ the integral part becomes

\begin{align*}
&\int_0^{2\pi} \prod_{k=1}^d \frac{1-\rho_k^2}{1+\rho_k^2-2\rho_k\cos(\theta_k-q_k\phi)} d\phi\\
=&\int_0^{2\pi} \prod_{k=1}^d \frac{1-\rho_k^2}{1+\rho_k^2-2\rho_k\cos(\phi-q_k\theta_k)} d\phi\\
=&\int_C \left( \prod_{k=1}^d \frac{1-|\eta_k|^2}{|z-\eta_k|^2} \right) \frac{1}{iz}dz\\
=&\int_C \left( \prod_{k=1}^d \frac{1-|\eta_k|^2}{(z-\eta_k)(\overline{z-\eta_k})} \right) \frac{1}{iz}dz\\
=&\frac{1}{i}\int_C z^{d-1} \prod_{k=1}^d \frac{1-|\eta_k|^2}{(z-\eta_k)(1-\overline{\eta_k}z)} dz
\end{align*}

where $C=\{z\in \mathbb{C} ; |z|=1\}$. This integral can be calculated with the residue theorem which states:

$$\int_\gamma f(z)dz = 2\pi i\sum \textrm{Res}(f, a_k)$$

where $\gamma$ is a closed curve on the complex plane, $\{a_1, ..., a_n\}$ are points on the interior of $\gamma$ which are singularities of $f$ and $Res(\cdot)$ is defined as :
$$ \textrm{Res}(f,c) = \lim_{z\rightarrow c} (z-c)f(z)$$

The list of points $\{\eta_1, ..., \eta_d\}$ are singularities of the integral and are in the interior of $C$ because $\rho_j \in (0,1)$. The residue for $\eta_j$ is equal to

\begin{align*}
&lim_{z\rightarrow \eta_j} (z - \eta_j) z^{d-1} \prod_{k=1}^d \frac{1-|\eta_k|^2}{(z-\eta_k)(1-\overline{\eta_k}z)}\\
=&lim_{z\rightarrow \eta_j} \frac{1-|\eta_j|^2}{1-\overline{\eta_j}z} z^{d-1} \prod_{\substack{k=1 \\ k\neq j}}^d \frac{1-|\eta_k|^2}{(z-\eta_k)(1-\overline{\eta_k}z)}\\
=& \eta_j^{d-1} \prod_{\substack{k=1 \\ k\neq j}}^d \frac{1-|\eta_k|^2}{(\eta_j-\eta_k)(1-\overline{\eta_k}\eta_j)}
\end{align*}

such that, under the condition $\rho_j \neq \rho_k \quad (j \neq k)$ the integral becomes

\begin{align*}
&\frac{1}{i}\int_C z^{d-1} \prod_{k=1}^d \frac{1-|\eta_k|^2}{(z-\eta_k)(1-\overline{\eta_k}z)} dz\\
=&2\pi \sum_{j=1}^d \eta_j^{d-1} \prod_{\substack{k=1 \\ k\neq j}}^d \frac{1-|\eta_k|^2}{(\eta_j-\eta_k)(1-\overline{\eta_k}\eta_j)}
\end{align*}

Resulting in the following closed form expression for the circula:

\begin{align*}
c(\theta_1, ..., \theta_d) =& \frac{1}{(2\pi)^d} \sum_{j=1}^d \eta_j^{d-1} \prod_{\substack{k=1 \\ k\neq j}}^d \frac{1-|\eta_k|^2}{(z-\eta_k)(1-\overline{\eta_k}z)}\\
=&\frac{1}{(2\pi)^d} \sum_{j=1}^d (\rho_je^{iq_j\theta_j})^{d-1} \prod_{\substack{k=1 \\ k\neq j}}^d \frac{1-\rho_k^2}{(\rho_j e^{iq_j\theta_j}-\rho_k e^{iq_k\theta_k})(1-\rho_j\rho_ke^{i(q_j\theta_j - q_k\theta_k)})}
\end{align*}
\end{proof-not-ams}

\begin{remark}
If $\rho_j = \rho_k$ for some $(j,k)$, the singularity on point $\eta_j=\rho_j e^{iq_j\theta_j}$ is a pole of order 2. More generally for a set of indices $P$ of size $p$ such that all $\rho_k$ parameters with $k$ in $P$ are equal, the corresponding $\eta_p$ point is a pole of order $p$. In such cases, the formula for the residue becomes

$$\textrm{Res}(f, c)=\frac{1}{(p-1)!} \lim_{z \rightarrow c} \frac{d^{p-1}}{dz^{p-1}} ((z-c)^pf(z))$$

Such that the residue at point $\eta_p$ becomes

\begin{align*}
&\frac{1}{(p-1)!} \lim_{z \rightarrow \eta_p} \frac{d^{p-1}}{dz^{p-1}} \left( \frac{1-|\eta_p|^2}{1-\overline{\eta_p}z} \right)^p z^{d-1} \prod_{\substack{k=1 \\ k\not\in P}}^d \frac{1-|\eta_k|^2}{(z-\eta_k)(1-\overline{\eta_k}z)}
\end{align*}
\end{remark}
%%===Inclusion end for file cmm-theory-proofs.tex

\clearpage
%\input{cmm-algorithms.tex}
%%===Inclusion starts for file cmm-algorithms.tex

\section{1-factor approximations for correlations matrices \toblack}
\label{sec:rank-one-approx}
%%i%%%%%%%%%%%%%%%%%%%%%%%%%%%%%%%%%%%%%%%%%%%%%%%%%%%%%%%%%%%%%%%%%%%%%%%%%%%%%%%

\subsection{Previous work}
%%ii-%-%-%-%-%-%-%-%-%-%-%-%-%-%-%-%-%-%-%-%-%-%-%-%-%-%-%-%-%-%-%-%-%-%-%-%-%-%-%

%% \begin{itemize}
%% \item \cite{higham2002computing} (higham first paper)
%% \item \cite{borsdorf2010computing} (higham)
%% \item \cite{duan2014generalized} (chinese)
%% \item \cite{deutsch2001best}: std convergence theory (cited higham 2009)
%% \end{itemize}

Approximating a given covariance matrix is a classical topic in numerical analysis,
and no less than 15 methods are discussed in ~\cite{simon2010majorization}.
The approximation is said to have a {\em $k$ factor structure} when the off diagonal terms
agree with that of a rank $k$ matrix. Properties of such approximation
are studied in detail~\cite{higham2002computing,borsdorf2010computing,duan2014generalized},
with particular solvers based on {\em alternating projections} \cite{borsdorf2010computing}.

In the sequel, we derive a novel (to the best of our knowledge) characterization
of solutions of the 1-factor approximation problem, and use an insight
obtained along the calculation to initialize our circula fitting procedure.

%% ~\cite{higham2002computing,borsdorf2010computing,duan2014generalized}.

\subsection{Exact 1-factor approximation of the correlation matrix $R$}
%%ii-%-%-%-%-%-%-%-%-%-%-%-%-%-%-%-%-%-%-%-%-%-%-%-%-%-%-%-%-%-%-%-%-%-%-%-%-%-%-%

Let $R$ be the correlation matrix.  We want to approximate it using the following map -- Eq.~\ref{eq:Rc-structure}:
\begin{equation*}
G(w) =  w\latrans{w} - \ladiag{w\latrans{w}} + \idmatrix{d}.
\end{equation*}
The following is, to the best of our knowledge, a novel algebraic
characterization of the 1-factor approximation. It will be used in 
our circula estimation algorithm -- Alg.~\ref{alg:estimate-circula},
see Section \ref{sec:circula-estimation}.

Since we aim at providing a one factor approximating the matrix $R_c$,
it is tempting to invoke the Eckart-Young-Mirsky Theorem and using the
eigenvector associated to the leading singular value of $R_c$.

However, as the following analysis shows, extra terms appear and
require solving a more elaborate problem, which still admits the
algebraic solution claimed in the theorem.

\begin{proof-not-ams}{black}[Thm.~\ref{thm:rank-one-approx}]

%%ii-%-%-%-%-%-%-%-%-%-%-%-%-%-%-%-%-%-%-%-%-%-%-%-%-%-%-%-%-%-%-%-%-%-%-%-%-%-%-%
\paragraph{Step 1: expressing the functional.}
We have
\begin{equation}
R-G(w) = R- \idmatrix{d} -( w\latrans{w} - \ladiag{  w\latrans{w} } ).
\end{equation}
Define 
\begin{equation}
\label{eq:def-A}
A=R-\idmatrix{d}.  
\end{equation}
We have
\begin{align}
f(w) = \vvnorm[F]{R-G(w)}{2} 
&= \sum_{i\neq j} (\Aij - w_iw_j)^2\\
&= \sum_{i\neq j}  \Aij[2] + \sum_{i\neq j} w_i^2 w_j^2 -2\sum_{i\neq j} \Aij w_iw_j\\
&= \vvnorm[F]{A}{2} -2 \latrans{w} A w + \vvnorm{w}{4} - \sum_i w_i^4\\
&=  \vvnorm[F]{A-w\latrans{w}}{2}  - \sum_i w_i^4.
\end{align}
We use polar coordinates to decouple the magnitude and the direction of $w$,
writing  $w=rv$ with $v\in \Sd{d-1}$. We get
\begin{align}
\label{eq:frv}
f(r,v) &= \vvnorm[F]{A}{2}  -2r^2 \latrans{v}Av + r^4(1 -\sum_i v_i^4)\\
     &= \vvnorm[F]{A}{2}  -2r^2 \alpha(v) + r^4\beta(v),
\end{align}
with 
\begin{equation}
\alpha(v)= \latrans{v}Av, \beta(v) = (1 -\sum_i v_i^4).  
\end{equation}

%%ii-%-%-%-%-%-%-%-%-%-%-%-%-%-%-%-%-%-%-%-%-%-%-%-%-%-%-%-%-%-%-%-%-%-%-%-%-%-%-%
\paragraph{Step 2: minimizing $f(r,v)$.}
In seeking critical points of $f(r,v)$, let us start with the variable $r$. 
The following is a necessary optimality condition:
\begin{equation}
\partiald{f}{r}  = -4r\alpha(v) + 4r^3\beta(v) = 4r(-\alpha(v) + r^2\beta(v)) = 0,
\end{equation}
whence the conditions
\begin{equation}
\label{eq:rz-r2}
\begin{cases}
r&=0\\
r^2&=\nicefrac{\alpha(v)}{\beta(v)}.
\end{cases}
\end{equation}
Let us focus on the non trivial solution.  Substituting
$r^2=\nicefrac{\alpha(v)}{\beta(v)}$ into the expression of
Eq.~\ref{eq:frv} yields
\begin{equation}
f(r,v) =    \vvnorm[F]{A}{2} -\frac{\alpha(v)^2}{\beta(v)}.
\end{equation}
From which we get that minimizing $f(r,v)$ is equivalent to maximizing 
\begin{equation}
\label{eq:Phiv2}
\Phi(v) =  \frac{\alpha(v)^2}{\beta(v)}, \vvnorm{v} = 1.
\end{equation}
One has $\sum_i v_i^2=1$; and  since $0\leq v_i^2\leq 1$, we get $v_i^4 \leq v_i^2$. Summing
the latter yields $\sum_i v_i^4 \leq \sum_i v_i^2 = 1$. Therefore $\beta(v) \geq 0$.
The square $r^2$ in the second expression of Eq.~\ref{eq:rz-r2} implies
that $\alpha(v)\geq 0$. 

Maximizing $\Phi(v)$ is therefore equivalent to maximizing
\begin{equation}
\label{eq:Phiv1}
\Phi_1(v) =  \frac{\alpha(v)}{\sqrt{\beta(v)}}, \vvnorm{v} = 1.
\end{equation}
The numerator is a Rayleigh-Ritz ratio, which requires taking a vector
in the eigenspace of the top eigenvalue.  This observation motivates the
heuristic used in Algorithm Alg.~\ref{alg:estimate-circula}.

Note however, that the coupling with the
denominator makes the optimization non trivial.
\toblack

\newcommand{\vcirct}{\mathring{v}^3}
\begin{remark}
In the subsequent calculations, we use the notation  $\vcirct = \latrans{(v_1^3,\dots,v_d^3)}$
for the cubic-wise coordinate vector associated with vector $v$.
\end{remark}

%%ii-%-%-%-%-%-%-%-%-%-%-%-%-%-%-%-%-%-%-%-%-%-%-%-%-%-%-%-%-%-%-%-%-%-%-%-%-%-%-%
\paragraph{Step 3: maximizing $\Phi$ using Lagrange multipliers.}
To solve this problem, we use Lagrange multipliers and introduce
\begin{equation}
\calL =   \Phi(v) - \lambda (\latrans{v}v-1).
\end{equation}
Writing $\nabladec[v]\calL = 0$ yields the following colinearity between the gradient of the objective function
and that of the constraint:
\begin{equation}
\label{eq:Lag-colinearity}
\nabladec[v]{\Phi(v)} = \nabladec[v]{ \frac{\alpha(v)^2}{\beta(v)} } = 2\lambda v. 
\end{equation}

\noindent Let us compute partial derivatives. For $\alpha(v)$, we get:
\begin{equation}
\alpha(v) = \latrans{v}Av \Rightarrow   \nabladec[v]{\alpha(v)} = 2 A v.\\
\end{equation}
For $\beta(b)$, we get:
\begin{equation}
\partiald*{\sum_j v_j^4}{v_i} = 4v_i^3 \Rightarrow   \nabladec[v]{\beta(v)} = -4\vcirct
\end{equation}
\noindent Differentiating $\Phi(v)$ yields
\begin{align}
\nabladec[v]{\Phi(v)} 
&= \frac{2\alpha \nabladec[v] \alpha(v)\ \beta(v) - \alpha(v)^2\beta(v)}{\beta(v)^2}\\
&= \frac{4\alpha(v)}{\beta(v)^2} (\beta(v) A v + \alpha(v) \vcirct).
\end{align}

\noindent Plugging the previous expression into Eq.~\ref{eq:Lag-colinearity}:
\begin{equation}
4\alpha(v) \bigl[ \beta(v) A v + \alpha(v) \vcirct \bigr] = 2\lambda v \beta(v)^2,
\end{equation}
which expresses the fact that $\beta(v) A v + \alpha(v) \vcirct$ is colinear to $v$.
Denoting $\mu$ the proportionality coefficient, we get:
\begin{equation}
\begin{cases}
\beta(v) A v + \alpha(v) \vcirct = \mu v\\
\sum_i v_i^2 =1
\end{cases}
\end{equation}
or in fully delineated:
\begin{equation}
\begin{cases}
(1-\sum_i v_i^4) A v + (\latrans{v}Av) \vcirct = \mu v\\
\sum_i v_i^2 =1
\end{cases}
\end{equation}
This is a polynomial system of $d+1$ equation of degree five in the $d$ coordinates $v_i$ and $\mu$.
\end{proof-not-ams}

By Bezout's theorem, the number of solution is bounded by the product
of the degrees of the individual equations, namely $2\times 5^d$.

\subsection{Heuristics}
\label{sec:Rc-heuristics}

\paragraph{Via Rayleigh-Ritz ratio.}
As a heuristic, consider again the definition of the function $\Phi$ to be maximized:
\begin{equation*}
\Phi(v) =  \frac{ (\latrans{v}Av)^2}{(1 -\sum_i v_i^4)}, \vvnorm{v} = 1.
\end{equation*}
The numerator is a Rayleigh-Ritz ratio which is maximized for the eigenvector $v=v_1$ of matrix $A$ -- Eq.~\ref{eq:def-A}.
From Eq.~\ref{eq:rz-r2}, we have $r=\sqrt{\alpha(v)/\beta(v)}$ and $w = r v$.
We finally get
\begin{equation}
\label{eq:heuristic1}
w^* = \sqrt{\frac{\lambda_1}{1-\sum_i \vij[4]{1i}}} v_1.  
\end{equation}
While we have not minimized the denominator, unless $v_1$ is isotropic, the sum of quartic terms will be small.

\paragraph{Via direct optimization upon choosing a unit vector.}
The previous expression was obtained by maximizing $\Phi$, given the
optimal value of $r$. But if the direction $v=v_1$ is chosen, we may
proceed vice-versa and optimize $r$.  Rewriting Eq.~\ref{eq:frv}
yields
\begin{equation}
g(r) = \sum_{i\neq j}  (\Aij - r^2 v_iv_j)^2,
\end{equation}
which is linear least squares problem in relation to $r$. The solution is then given by setting the derivative to zero :

\begin{align}
g(r) &= \sum_{i\neq j} (\Aij[2] - 2\Aij r^2 v_i v_j + r^4 v_i^2 v_j^2)\\
\partiald{ g(r)}{\delta r}  &= \sum_{i\neq j} (- 4\Aij r v_i v_j + 4r^3 v_i^2 v_j^2) = 0,
\end{align}
whence the final expression
\begin{equation}
\label{eq:heuristic2}
r = \sqrt{\frac{\sum_{i\neq j} \Aij v_i v_j}{\sum_{i\neq j} v_i^2 v_j^2}}.
\end{equation}
Eq.~\ref{eq:heuristic2} is the expression used in Algorithm \ref{alg:estimate-circula}.

\begin{remark}
\label{rmk:R-RI}
In Algorithm \ref{alg:estimate-circula}, 
the maximization of the numerator of Eq.~\ref{eq:Phiv2} 
commands using the largest eigenvector (algebraically)
of $R-\idmatrix$.
We note that this is equivalent to taking the eigenvector associated
to the leading eigenvalue of $R$. Indeed, 
$R-\idmatrix$ have the same eigenvectors, and the eigenvalues of the latter
are those of the former minus one, so that their ordering is preserved.
%%
%% we may start from the vector $v_1$ given by
%% \Call{SingularValueDecomposition}{\rred{$R$}} or that given by
%% \Call{SingularValueDecomposition}{\rred{$R-\idmatrix$}}.  The
%% subtraction of $\idmatrix$ does not change eigenvectors, does not
%% change the ordering of eigenvalues either, but possibly changes their
%% ordering in magnitude (absolute value).
%% %%
%% As seen from the analysis of our optimization problem, we should therefore start from 
%% the eigenvector associated to the leading eigenvalue of $R-\idmatrix$.
\end{remark}

\section{Algorithms}
%%i%%%%%%%%%%%%%%%%%%%%%%%%%%%%%%%%%%%%%%%%%%%%%%%%%%%%%%%%%%%%%%%%%%%%%%%%%%%%%%%
\label{sec:algorithms}
\subsection{Circula estimation}
\label{sec:circula-estimation}
%%ii-%-%-%-%-%-%-%-%-%-%-%-%-%-%-%-%-%-%-%-%-%-%-%-%-%-%-%-%-%-%-%-%-%-%-%-%-%-%-%

Alg.~\ref{alg:estimate-circula} describes our circula estimation algorithm.
Note the initialization of $\rho_{init}$
whose entries are used to estimate the  $\rho_{kl}$ from Eq.~\ref{eq:rhokl-dependence}.
This initialization combines two ingredients:
\begin{itemize}
\item the approximation scheme developed in Sec.~\ref{sec:rank-one-approx};
\item the clipping of the vector $r w$ to $[-1,1]$, which suffices to guarantee that the generated matrix is PSD and compliant with the construction of $R_c$ (Eq.~\ref{eq:Rc-structure}).
\end{itemize}
The resulting $w_{\text{init}}$ yields estimates for the discrete sign
parameters $q$, and provides initial estimates for correlation
magnitude parameters $\rho_{\text{init}}$ which are then refined numerically.
\toblack

\begin{algorithm}
\begin{algorithmic}[1]
\Procedure{EstimateCircula}{$\Theta, \Lambda_m$}
\State $R \gets$ \Call{JSCorrMatrix}{$\Theta$}
\State $\{\lambda_i, v_i\} \gets$ \Call{EigenValueDecomposition}{$R-\idmatrix$} \Comment{Of $R$ or $R-\idmatrix$, see Rmk \ref{rmk:R-RI}}
\State $v \gets v_1$
\State $r \gets \sqrt{\frac{\sum_{i\neq j}  R_{ij}  v_i  v_j}{\sum_{i\neq j} v_i^2  v_j^2}}$ \Comment{Eq.~\ref{eq:heuristic2}}
\State $w_{\text{init}} \gets r v$
\State $q \gets \mathrm{sign}(w_{\text{init}})$ 
\State $\rho_{\text{init}} \gets |w_{\text{init}}|$ clipped to $[-1,1]$
\State $\rho \gets $ \Call{LBFGS-B}{$\Theta,\Lambda_m, \rho_{\text{init}}$}
\State \Return $\{\Lambda_m, q, \rho\}$
\EndProcedure
\end{algorithmic}
\vspace{1em}
\begin{algorithmic}[1]
\Procedure{EstimateDistribution}{$\Theta$}
\State $\Lambda_m \gets $ \Call{EstimateMarginals}{$\Theta$}
\State $\Lambda \gets $ \Call{EstimateCircula}{$\Theta, \Lambda_m$}\\
\State \Return $\Lambda$
\EndProcedure
\end{algorithmic}
\caption{Estimation procedure of a circula-based multivariate distribution (\ref{eq:full-density})}
\label{alg:estimate-circula}
\end{algorithm}

\subsection{Fitting mixtures of circulae}
\label{sec:fitting-mixtures}
%%ii-%-%-%-%-%-%-%-%-%-%-%-%-%-%-%-%-%-%-%-%-%-%-%-%-%-%-%-%-%-%-%-%-%-%-%-%-%-%-%

\subsubsection{Initialization using k-means++}

To the best of our knowledge, there is no comparative work on the initialization of EM for mixture models with von Mises component marginals. However, the problem of initializing Gaussian mixture models has been well studied, which is the closest analogue to von Mises in euclidian space.

The main approaches to tackle this problem include starting with simpler variants of the EM algorithm (CEM, SEM)\cite{biernacki2003choosing}, relying on an initial clustering of the data \cite{mclachlan2000finite}, or even adapting seeding strategies used for clustering initialization \cite{kwedlo2013new, blomer2016adaptive, you2023new}.

Focusing on toroidal mixture models, \cite{mardia2012mixtures} proposed a mixture of multivariate von Mises sine distributions (normalized through numerical approximations) which was initialized from a data partitioning obtained through the PAM algorithm. The mixtures of wrapped normal distributions studies in \cite{greco2023finite} are also initialized from a partitioning, obtaining with the Ward's method.

Our proposal is to use a partitioning of the data, obtained through the toroidal k-means++ seeding,
%%~\cite{carriere2026clustering},
 to initialize the mixture model as described in Alg.~\ref{alg:means-to-cmm}. This is motivated by the observation that careful seeding has proved efficient in initializing gaussian mixture models \cite{blomer2016adaptive}.

\begin{algorithm}
\caption{Initialization scheme for circula mixture models}
\label{alg:means-to-cmm}
\begin{algorithmic}[1]
\Procedure{meanstocmm}{$X$}

\Statex
\State $\{\mu_k, C_k\} \gets$ \Call{toroidalKMplusplus}{$\Theta$} \Comment{cluster centers and associated partitions}
\For{$k \gets 1$ to $K$}
\State $w_k \gets |C_k|\  /\  |\Theta|$
\State $\lambda_k \gets $ \Call{EstimateDistribution}{$C_k$} \Comment{Estimate the remaining parameters using the partition}
\EndFor
\EndProcedure
\end{algorithmic}
\end{algorithm}

\subsubsection{Unsupervised learning of mixture models}

Selecting the correct amount of components when estimating a mixture model cannot rely entirely on maximizing likelihood. Indeed, increasing the number of components will always increase likelihood. However, such models are overly complex and overfit, capturing noise instead of the true underlying structure of the data, and unable to generalize. Instead, the appropriate amount of components balances between maximizing goodness of fit and minimizing model complexity. 

In the general case, this trade-off is assessed using a variety of techniques, including the Bayesian Information Criterion (BIC) \cite{schwarz1978estimating}, the Akaike Information Criterion (AIC) \cite{akaike1974new}, the Minimum Description Length (MDL) framework \cite{rissanen1978modeling}, cross-validation methods \cite{bishop2006pattern} or likelihood ratio tests \cite{lehmann2005testing} amongst others. In the following we consider the Minimum Message Length (MML) information-theoretic framework \cite{wallace1987estimation}, which quantifies the total encoding cost of both model and data as a two-part message, with length expressed as :
\begin{equation}
\MMLlength{\theta}{X} = \text{Length}(\theta) + \text{Length}(X | \theta)
\end{equation}

For statistical models, this equates to
\begin{equation}
\label{eq:mml-length}
\MMLlength{\theta}{X} = \frac{p}{2} \log q_p - \log \left( \frac{u(\theta)}{\sqrt{|\calF (\theta)|}} \right) - \calL (X | \theta) + \frac{p}{2}
\end{equation}
with $p=\size{\theta}$ the number of free parameters in the model, $q_p$ the $p$-dimensional lattice quantization constant, $u(\theta)$ the prior on the model's parameters, $\calF (\theta)$ the Fisher information matrix and $\calL (X | \theta)$ the log-likelihood function.
In mixture models, the Fisher information matrix $F(\theta)$ generally does not admit an analytical expression, such that an exact computation of the Message Length is intractable. However, an efficient approximation was proposed in Figueiredo and Jain \cite{figueiredo2002unsupervised} by replacing $F(\theta$) with the complete-data Fisher information matrix. Combined with an appropriate handling of empty components ($w_k = 0$) and a careful choice of parameter prior, this approximation yields the following form :
\begin{equation}
\label{eq:mml-length-ours}
\text{Length}(\theta, X) = \frac{N}{2} \sum_{k: w_k > 0} \log(\frac{n w_k}{12}) + \frac{K_{nz}}{2} \log(\frac{n}{12}) + \frac{K_{nz}(N+1)}{2} - \calL (X | \theta)
\end{equation}
where $K_{nz}$ denotes the number of components with non-zero weights, $n=|X|$ is the sample size and $N$ is the number of free parameters in a single component ($N=3d$ for CBMD components from Eq.~\ref{eq:full-density} \footnote{The correlation sign and intensity vectors can be jointly represented as a continuous vector in $[-1,1]^d$. This remains consistent with the Figueiredo and Jain ML approximation, which depends only on the parameter dimension rather than the parameter space}).
Figueiredo and Jain \cite{figueiredo2002unsupervised} also introduce an estimation procedure based on EM, designed to minimize $\text{Length}(\theta, X)$ over the component parameters $\theta$ as well as the number of components $K$ within a specified range $[k_{min}, k_{max}]$. This is the approach adopted in this work, which we denote as MML-EM in the following.

\begin{remark}
\label{rmk:mml-length}
Eq.~\ref{eq:mml-length} corresponds to \cite[Eq. 4]{amarasinghe2023getting}, minus one term, namely
$\size{X} d \log (\epsilon)$, with $\epsilon$ the \quoteen{\em uncertainty of each datum}.
In practice, it is taken as $0.0873$ radians ($\sim 5\deg$ per torsion angle) and corresponds to a resolution
of $\sim 0.1$\ang on the Cartesian coordinates.

We omit this additive constant which  is useless in comparing models on the same data.

Note however that as a consequence, the {\em length} may be negative
so that the result  of Eq.~\ref{eq:mml-length} should be called {\em relative MML}.

The notion of bit size for the difference between two models is strictly speaking correct in any case.
\end{remark}

%%===Inclusion end for file cmm-algorithms.tex

\clearpage
%\input{cmm-results.tex}
%%===Inclusion starts for file cmm-results.tex

\section{Results}
%%i%%%%%%%%%%%%%%%%%%%%%%%%%%%%%%%%%%%%%%%%%%%%%%%%%%%%%%%%%%%%%%%%%%%%%%%%%%%%%%%

\subsection{Protocol and data availability}
\label{sec:protocal-availability}
%%ii-%-%-%-%-%-%-%-%-%-%-%-%-%-%-%-%-%-%-%-%-%-%-%-%-%-%-%-%-%-%-%-%-%-%-%-%-%-%-%

Experiments were run on a 4-node HPC cluster (2× Intel Xeon Gold 6240 CPUs per node, 18 cores/CPU, 2.60 GHz, totaling 144 CPU cores) and two workstations equipped with an Intel Xeon W-2155 CPU (10 cores, 20 threads, 3.30 GHz) and an Intel Core i9-13900K CPU (24 cores, 32 threads, 5.80 GHz).

All algorithms are implemented in C++ in [LIBRARY OMITTED FOR ANONYMITY], a well known / carefully documented library
which currently features circa 150 packages.

The code used to produce the results presented in this work uses several 
such packages to:
\begin{itemize}
\item Parse PDB files,
\item Manipulate covalent structures and the associated internal coordinates,
\item Run \kmeanspp[black] on the flat torus,
\item Represent periodic functions on the flat torus,
\item Compute  message lengths,
\item Run the MML-EM algorithm.
\end{itemize}

This complexity prevents the diffusion of a stand-alone version of the C++ code,
which will be integrated to the release of the library upon publication of the paper.

Meanwhile, as proof-of-concept, the attached archive xxx provides 
(i) all  models in CSV format,
(ii) the full numerical values used for the Table \ref{fig:aa-pairs-results-matrix},
(iii)  python code to load the models and  generate samples.
(iv) a basic notebook to demonstrate the offered functionalities.
\toblack

\subsection{Shape analysis}
\label{sec:shape-analysis}
%%ii-%-%-%-%-%-%-%-%-%-%-%-%-%-%-%-%-%-%-%-%-%-%-%-%-%-%-%-%-%-%-%-%-%-%-%-%-%-%-%

\paragraph{Modes and critical points.}
Circula components and the CMM are real valued functions on the
flat torus $\torusd$. As such, they obey Morse
theory~\cite{milnor1963morse}. In particular, consider the critical
points $\{c_i\}$ of such a function. Assuming the function is a Morse
function (isolated critical points and non degenerate Hessian at each
of them), the sum of the indices of critical points is equal to the
Euler characteristic of the flat torus $\chi(\torusd)$.  (The index of
a critical point is the number of negative eigenvalues of the
Hessian.) By the product property~\cite{hatcher2002algebraic}, one has
$\chi(\torusd) = \prod_i \chi(S^1) = 0$. This has the following
implications:
\begin{itemize}
\item In 2D: the number of local minima/saddle points/maxima, respectively denoted $a,b,c$,
satisfy $a-b+c=~0$.

\item In 3D: the number of local minima/saddle points of index
  1/saddle points of index 2/maxima, respectively denoted $a,b,c,d$,
  satisfy $a-b+c-d=0$.
\end{itemize}

\paragraph{Illustrations.}

\begin{figure}[htbp]
\centerline{\includegraphics[width=\linewidth]{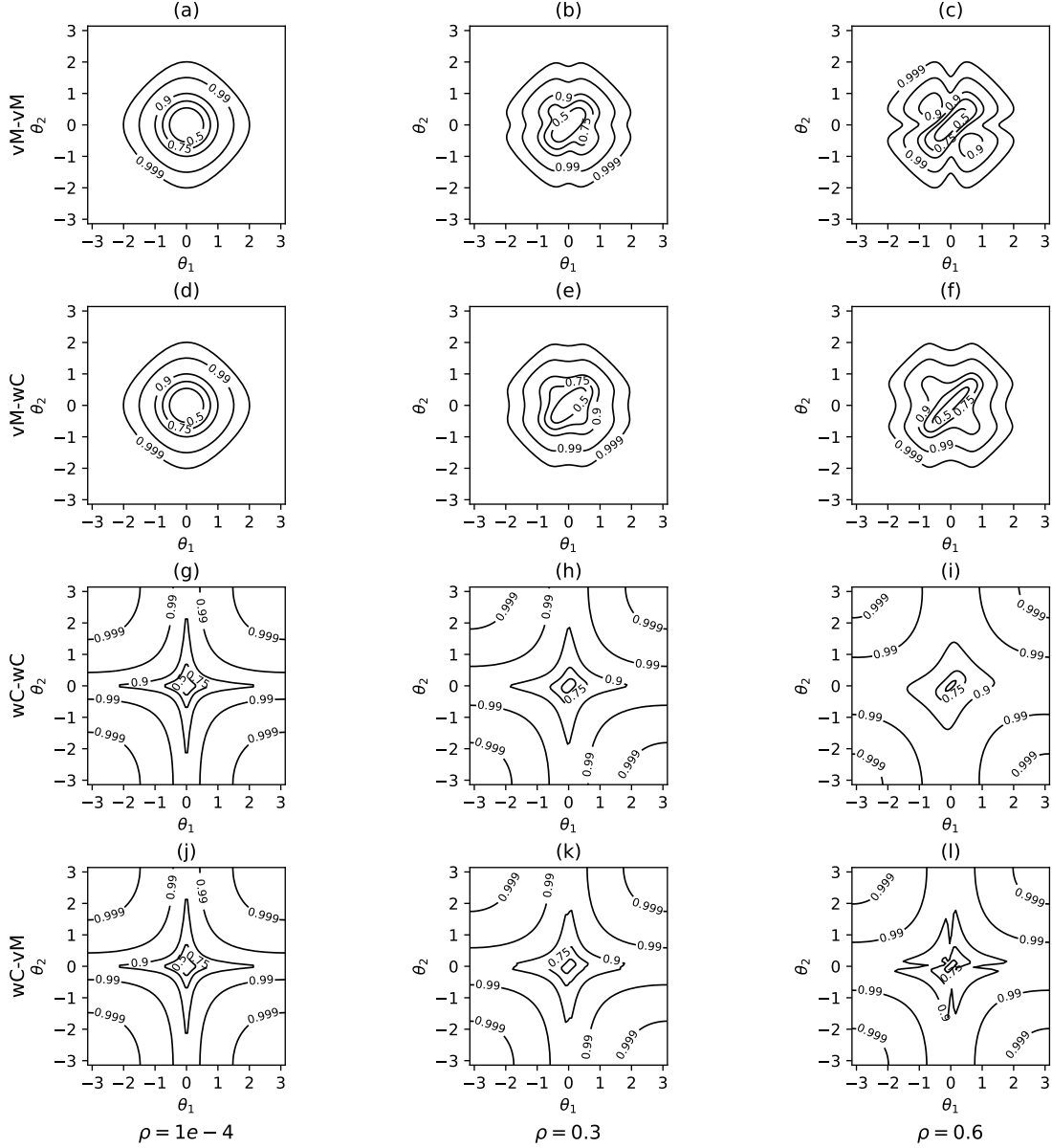}}
\caption{{\bf Comparison of circulae: 2D.}
The \vMvM (a-c), \vMwC (d-e), \wCwC (g-i) and \wCvM (j-l) distributions with $d=2$, for increasing values of mean resultant length $\rho_g$ in the binding densities, such that the correlation is equal to $\rho_{1,2} = 1e-4$ (first column), $\rho_{1,2} = 0.3$ (second column) and  $\rho_{1,2} = 0.6$ (third column). All marginals have $\mu = 0$ and $\rho = 0.9$ and all $q_i=1$ to simplify the comparison.}
\label{fig:circula-variants-2D}
\end{figure}

To assess the practical utility of the different possible instances of CBMDs yielded by the von Mises and wrapped Cauchy circulae, we compare four CBMDs models : \vMvM, \vMwC, \wCwC, and wC-vM. The following observations are made (Fig.~\ref{fig:circula-variants-2D}, Fig.~\ref{fig:circula-variants-3D} and Fig.~\ref{fig:circula-variants-3D-2}):
\begin{itemize}
\item The \vMwC and \vMvM exhibit density bumps in the direction orthogonal to that imposed by the correlation. This artifact a side effect of the circula construction, and is undesirable in the majority of practical applications. Because the correlation pattern induced by the circula is periodic, it correlates points in the circular uniform space that are distant in the original space. For the \vMwC and \vMvM distributions, this effect is not entirely compensated by the marginal distributions, such that a small portion of the probability mass is drawn towards these values (Fig.~\ref{fig:circula-decomposition}, Fig.~\ref{fig:circula-variants-2D}). However, in our experiments with correlated data, the reduction in likelihood caused by these bumps is generally smaller than the increase in likelihood brought by modeling correlation -- see Sec.~\ref{sec:experiments-results}.
\item The \vMvM distribution can become multimodal when covariance increases. With $d=2$, the distribution becomes trimodal, but when the dimension increases the number of modes increases accordingly. Fig.~\ref{fig:circula-variants-3D-2} shows this behavior with $d=3$ where the \vMvM distribution has $7$ modes.
\item The \wCwC distribution does not show multiple modes in our experiments. The bivariate wrapped Cauchy \cite{kato2015mobius}, which correspond to a bivariate circula-based distribution with wrapped cauchy marginals and binding density, is proven to be unimodal. We conjecture this property holds in the multivariate extension. In contrast, the wrapped Cauchy marginals, unlike von Mises, are not analogues of gaussian distribution on the circle. This is not desirable in many practical applications, where the data is marginally "gaussian-like".
\item The \vMwC distribution seems to share characteristics of both the \vMvM and \wCwC distributions: desirable von Mises marginals while keeping the distribution unimodal in higher covariance. 
\end{itemize}

\ifLONG
\begin{conjecture}
\vMwC : the distribution stays unimodal when the covariance is high. 
\end{conjecture}
\fi

\begin{figure}[htbp]
\centerline{\includegraphics[width=\linewidth]{\wdir/circula-variants-3d.png}}
\caption{{\bf Comparison of circulae: 3D.} {$\mu = 0$ and $\rho=0.6$ for all marginals. $\rho_g$ of each binding densities is $\sqrt{0.3}$ such that any bivariate marginal of the circula has $\rho_c = 0.3$ as in the second column of Fig.~\ref{fig:circula-variants-2D}}}
\label{fig:circula-variants-3D}
\end{figure}

\begin{figure}[htbp]
\centerline{\includegraphics[width=\linewidth]{\wdir/circula-variants-3d2.png}}
\caption{{\bf Comparison of circulae: 3D.} {$\mu = 0$ and $\rho=0.6$ for all marginals. $\rho_g$ of each binding densities is $\sqrt{0.6}$ such that any bivariate marginal of the circula has $\rho_c = 0.6$ as in the third column of Fig.~\ref{fig:circula-variants-2D}}}
\label{fig:circula-variants-3D-2}
\end{figure}

\begin{remark}
\label{rmk:mu-zero}
The CBMD instances as presented use circulae constructed with centered binding densities, however this is not required in the multivariate extension of the circula model. In fact, \cite[Section 3.1]{jones2015class} studies the behavior of bivariate circulae when varying the mean parameter of the binding density, and concludes that setting $\mu_g=0$ is preferable. This insight carries to the multivariate extension of the circula, such that we keep the focus on centered binding densities in this work.
\end{remark}

\subsection{Evaluation of the 1-factor approximation heuristic}

\label{sec:parameter-estimation-results}

We evaluate the parameter estimation heuristic based on 1-factor approximation of the JS correlation matrix, as proposed in Sec.~\ref{sec:1-factor-approx}. To this end, we compare the runtime and resulting log-likelihood when estimating \vMwC distributions using either the 1-factor approximation heuristic or an exploration over all the $q$ configurations. 

Our experiments involve datasets generated by sampling multivariate normal distributions with random parameters. The samples from these distributions are then mapped to the hypertorus. Specifically, the multivariate normal parameters are generated as follows : 
\begin{itemize}
\item The correlation matrix is sampled uniformly from an LKJ distribution with $\eta=1$
\cite{lewandowski2009generating};
\item The variance vector is sampled uniformly in $[0,\pi/2]^d$;
\item The mean vector is sampled uniformly in $[0,2\pi]^d$
\end{itemize}

We perform this process $100$ times, each time sampling $1000$ points from the resulting multivariate normal distribution. Finally this is repeated for dimensions $d=\{3,5,10\}$. Importantly, because the correlation matrices are drawn uniformly from the LKJ distribution, the samples may exhibit dependency structures not attainable by the \vMwC distribution, providing a fair assessment of the 1-factor approximation heuristic.

Results are reported in Tab.~\ref{tab:parameter-comparison}, and also include the log-likelihood of an independent joint von Mises distribution to serve as reference. The 1-factor approximation shows a small decrease in mean log-likelihood, corresponding to datasets where the one-factor approximation does not align with the direction that maximizes the likelihood for the \vMwC distribution. However, this reduction is negligible compared to the substantial gain in runtime, which grows exponentially with the dimension due to avoiding the $2^d$ exhaustive search over all $q$ configurations.
%\input{fig/table_heuristic_comparison.tex}
%%===Inclusion starts for file fig/table_heuristic_comparison.tex
\begin{table}[h!]
\centering
\begin{tabular}{l|lll}
 & Joint Von Mises & Full exploration & Rank-1 approximation \\
\hline
d = 3 & -3761.458 & -3743.046 (0.445 s) & -3744.386 (0.053 s)\\ 
d = 5 & -6252.759 & -6222.610 (4.257 s) & -6225.876 (0.156 s)\\ 
d = 10 & -12489.852 & -12431.538 (520.587 s) & -12438.071 (1.036 s)\\ 
\end{tabular}
\caption{\textbf{Mean log-likelihood (Runtime) of vMwC depending on the parameter estimation heuristic}. Joint von mises model log-likelihood is included to serve as reference}
\label{tab:parameter-comparison}
\end{table}%%===Inclusion end for file fig/table_heuristic_comparison.tex

\subsection{Single amino acid torsion angles}
%%ii-%-%-%-%-%-%-%-%-%-%-%-%-%-%-%-%-%-%-%-%-%-%-%-%-%-%-%-%-%-%-%-%-%-%-%-%-%-%-%
\label{sec:phisical-results}
%\input{fig/table_phisical_bits.tex}
%%===Inclusion starts for file fig/table_phisical_bits.tex
\begin{table}[ht]
\center
\resizebox{1.0\textwidth}{!}{
\begin{tabular}{crrrrrrrrrrrrrrr}
\hline
&   & \multicolumn{4}{c}{PhiSiCal ML} && \multicolumn{4}{c}{VMVM mixture ML} && \multicolumn{4}{c}{VMWC mixture ML} \\  \cline{3-6}\cline{8-11}\cline{13-16} 
(aa) & N & K & Model & Data & Total &&  K & Model & Data & Total &&  K & Model & Data & Total  \\ \hline
ALA & 1861359 & 25 & 886 & 1749479 & 1750365 && 25 & 1220 & 1746097 & \textbf{1747318} && 25 & 1217 & 1746310 & 1747527 \\
&&&  &&&&&  \tored + 334 &  \togreen -3382 &&&&  \tored + 331 &  \togreen -3169 &  \\
ARG & 1130448 & 250 & 19136 & 5143922 & 5163059 && 250 & 27435 & 5084387 & 5111823 && 250 & 27492 & 5076165 & \textbf{5103657} \\
&&&  &&&&&  \tored + 8299 &  \togreen -59535 &&&&  \tored + 8356 &  \togreen -67757 &  \\
ASN & 948274 & 180 & 8583 & 4509280 & 4517863 && 180 & 12101 & 4492836 & 4504938 && 180 & 12100 & 4492581 & \textbf{4504681} \\
&&&  &&&&&  \tored + 3518 &  \togreen -16444 &&&&  \tored + 3517 &  \togreen -16699 &  \\
ASP & 1279567 & 170 & 8509 & 5214276 & 5222785 && 170 & 11979 & 5191094 & 5203074 && 170 & 11982 & 5190845 & \textbf{5202828} \\
&&&  &&&&&  \tored + 3470 &  \togreen -23182 &&&&  \tored + 3473 &  \togreen -23431 &  \\
CYS & 296547 & 96 & 3170 & 812834 & 816004 && 96 & 4361 & 810657 & 815018 && 96 & 4366 & 810269 & \textbf{814635} \\
&&&  &&&&&  \tored + 1191 &  \togreen -2177 &&&&  \tored + 1196 &  \togreen -2565 &  \\
GLN & 820871 & 239 & 12645 & 4478925 & 4491570 && 239 & 17908 & 4457030 & 4474939 && 239 & 17952 & 4454364 & \textbf{4472316} \\
&&&  &&&&&  \tored + 5263 &  \togreen -21895 &&&&  \tored + 5307 &  \togreen -24561 &  \\
GLU & 1446860 & 262 & 14906 & 7779832 & 7794738 && 262 & 21209 & 7738832 & 7760042 && 262 & 21211 & 7737649 & \textbf{7758861} \\
&&&  &&&&&  \tored + 6303 &  \togreen -41000 &&&&  \tored + 6305 &  \togreen -42183 &  \\
GLY & 1588115 & 30 & 1073 & 4789895 & 4790969 && 30 & 1472 & 4783378 & 4784850 && 30 & 1472 & 4783370 & \textbf{4784842} \\
&&&  &&&&&  \tored + 399 &  \togreen -6517 &&&&  \tored + 399 &  \togreen -6525 &  \\
HIS & 515611 & 163 & 7229 & 2345505 & 2352734 && 163 & 10155 & 2336201 & 2346356 && 163 & 10154 & 2336040 & \textbf{2346194} \\
&&&  &&&&&  \tored + 2926 &  \togreen -9304 &&&&  \tored + 2925 &  \togreen -9465 &  \\
ILE & 1333508 & 172 & 8275 & 1706984 & 1715259 && 172 & 11603 & 1694565 & 1706169 && 172 & 11602 & 1694373 & \textbf{1705975} \\
&&&  &&&&&  \tored + 3328 &  \togreen -12419 &&&&  \tored + 3327 &  \togreen -12611 &  \\
LEU & 2171630 & 165 & 8561 & 3977004 & 3985566 && 165 & 12068 & 3956273 & 3968342 && 165 & 12065 & 3955610 & \textbf{3967675} \\
&&&  &&&&&  \tored + 3507 &  \togreen -20731 &&&&  \tored + 3504 &  \togreen -21394 &  \\
LYS & 1176395 & 266 & 17570 & 7170847 & 7188417 && 266 & 25038 & 7096896 & 7121934 && 266 & 25147 & 7083546 & \textbf{7108693} \\
&&&  &&&&&  \tored + 7468 &  \togreen -73951 &&&&  \tored + 7577 &  \togreen -87301 &  \\
MET & 417170 & 270 & 12781 & 1966502 & 1979283 && 270 & 17838 & 1950840 & 1968679 && 270 & 18049 & 1943786 & \textbf{1961835} \\
&&&  &&&&&  \tored + 5057 &  \togreen -15662 &&&&  \tored + 5268 &  \togreen -22716 &  \\
PHE & 927298 & 226 & 10345 & 2899050 & 2909396 && 226 & 14500 & 2888168 & 2902669 && 226 & 14502 & 2888116 & \textbf{2902618} \\
&&&  &&&&&  \tored + 4155 &  \togreen -10882 &&&&  \tored + 4157 &  \togreen -10934 &  \\
PRO & 1004859 & 231 & 10458 & -2332991 & -2322533 && 231 & 14860 & -2430198 & \textbf{-2415338} && 231 & 14815 & -2428849 & -2414033 \\
&&&  &&&&&  \tored + 4402 &  \togreen -97207 &&&&  \tored + 4357 &  \togreen -95858 &  \\
SER & 1337273 & 114 & 4646 & 4173766 & 4178412 && 114 & 6463 & 4167916 & 4174380 && 114 & 6464 & 4167873 & \textbf{4174338} \\
&&&  &&&&&  \tored + 1817 &  \togreen -5850 &&&&  \tored + 1818 &  \togreen -5893 &  \\
THR & 1221604 & 90 & 3692 & 2845817 & 2849510 && 90 & 5126 & 2840289 & 2845415 && 90 & 5126 & 2840219 & \textbf{2845345} \\
&&&  &&&&&  \tored + 1434 &  \togreen -5528 &&&&  \tored + 1434 &  \togreen -5598 &  \\
TRP & 310470 & 212 & 8261 & 1026964 & 1035226 && 212 & 11512 & 1021321 & 1032834 && 212 & 11512 & 1021230 & \textbf{1032742} \\
&&&  &&&&&  \tored + 3251 &  \togreen -5643 &&&&  \tored + 3251 &  \togreen -5734 &  \\
TYR & 788176 & 192 & 8750 & 2503299 & 2512049 && 192 & 12258 & 2492770 & \textbf{2505028} && 192 & 12258 & 2492789 & 2505048 \\
&&&  &&&&&  \tored + 3508 &  \togreen -10529 &&&&  \tored + 3508 &  \togreen -10510 &  \\
VAL & 1601058 & 96 & 3989 & 1895908 & 1899898 && 96 & 5517 & 1889285 & 1894802 && 96 & 5516 & 1889137 & \textbf{1894653} \\
&&&  &&&&&  \tored + 1528 &  \togreen -6623 &&&&  \tored + 1527 &  \togreen -6771 &  \\
\hline
\end{tabular}}
\caption{\textbf{Message lengths (model length + data length) of \vMwC and \vMvM mixture models obtained by re-estimating the PhiSiCal von Mises mixture models with added correlation}. $N$ is the dataset size ang $K$ is the number of components}\label{tab:phisical-results}\end{table}
%%===Inclusion end for file fig/table_phisical_bits.tex

\subsection{Amino acid pairs torsion angles}
%%ii-%-%-%-%-%-%-%-%-%-%-%-%-%-%-%-%-%-%-%-%-%-%-%-%-%-%-%-%-%-%-%-%-%-%-%-%-%-%-%

\label{sec:cross-landscape-analysis}
%%===Inclusion end for file cmm-results.tex

%% For journals like Proteins etc
%\renewcommand{\figurename}{Figure S\hspace{-.05cm}~}
%\renewcommand{\tablename}{Table S~}
%\setcounter{figure}{0}
%\setcounter{table}{0}
%\makeatletter %% package algorithm
%\renewcommand{\ALG@name}{Algorithm S\hspace{-.1cm}~}
%\makeatother

\clearpage

{\scriptsize
\tableofcontents
}
\end{document}